\documentclass[]{article}

\usepackage{graphicx}
\usepackage{natbib}
\usepackage{natbibspacing}

\setkeys{Gin}{draft=false}
\begin{document}

%
%

\centerline{\bf Planetary gyre, time-dependent eddies, torsional waves}
\centerline{\bf and equatorial jets at the Earth's core surface}
\vspace*{.5cm}
%
%

%
%









\centerline{N. Gillet$^{1,2}$, D. Jault$^{1,2}$
and C. C. Finlay$^{3}$ }
\vspace*{.5cm}

$^{1}${University Grenoble Alpes, ISTerre,  BP 53, F-38041 Grenoble, France}

$^{2}${CNRS, ISTerre,  BP 53, F-38041 Grenoble, France}

$^{3}${Division of Geomagnetism, DTU Space, National Space Institute, Technical University of Denmark, Elektrovej,   2800 Kgs. Lyngby, Denmark}

\vspace*{.5cm}

Reference: Gillet, N., Jault, D., \& Finlay, C. C. (2015). Planetary gyre, time‐dependent eddies, torsional waves, and equatorial jets at the Earth's core surface. {\it J. Geophys. Res.: Solid Earth}, 120(6), 3991-4013.
%
%


\begin{abstract}

We report a calculation of time-dependent quasi-geostrophic core flows for 1940-2010. Inverting recursively for an ensemble of solutions, we evaluate the main source of uncertainties, namely the model errors arising from interactions between unresolved core surface motions and magnetic fields. 
Temporal correlations of these uncertainties are accounted for.
The covariance matrix for the flow coefficients is also obtained recursively from the dispersion of an ensemble of solutions.
Maps of the flow at the core surface show, upon a planetary-scale gyre, time-dependent large-scale eddies at mid-latitudes and vigorous azimuthal jets in the equatorial belt. 
The stationary part of the flow predominates on all the spatial scales that we can resolve.
We retrieve torsional waves that explain the length-of-day changes at 4 to 9.5 years periods. 
These waves may be triggered by the nonlinear interaction between the magnetic field and sub-decadal non-zonal motions within the fluid outer core.
Both the zonal and the more energetic non-zonal interannual motions were particularly intense close to the equator (below 10 degrees latitude) between 1995 and 2010. 
We revise down the amplitude of the decade fluctuations of the planetary scale circulation and find that electromagnetic core-mantle coupling is not the main mechanism for angular momentum exchanges on decadal time scales if mantle conductance is $3 \times 10^8$ S or lower.

\end{abstract}

%
%

%


\section{Introduction}
\label{intro}

\subsection{On the resolution of core motions}
\label{intro-res}

Some consensus has emerged about the possibility of estimating core surface flows ${\bf u}$ from the secular variation (defined as the time variation of the magnetic field and henceforth abbreviated to SV), and about their geometry for the most recent epochs. 
This picture is obtained by inverting for $\mathbf{u}$ using the radial component of induction equation at the core surface,
\begin{equation}
\label{eq:ind_rad}
\displaystyle\frac{\partial B_r}{\partial t} = - \nabla_H \cdot \left( \mathbf{u} B_r \right)
\end{equation}
  ($B_r$ is the radial magnetic field, $\nabla_H\cdot$ is the horizontal divergence operator).
Here magnetic diffusion has been neglected.
While stressing the significance of magnetic diffusion as a source for SV, \cite{holme06} concluded that it does not contribute a larger fraction of the high degree SV than of the low degree SV. A recent study based on geodynamo modeling supports this statement \citep{aubert2014earth}.
Meanwhile, the SV associated with the unresolved part of the radial magnetic field has been diagnosed as the main 
source of uncertainty in single-epoch analyses of equation (\ref{eq:ind_rad}) \citep{eymin05,pais08,baerenzung2013bayesian}. 
Consequently, our ability to reconstruct core flows is reduced.

There is nevertheless growing evidence of the equatorial symmetry (ES) of large scale core surface flows \citep[but see][for an alternative view]{whaler2015derivation}.
\cite{gillet11} find an increase of the percentage of ES flow as better SV data are available.
\cite{baerenzung2013bayesian} observe from satellite SV models that without imposing any topological constraints, more than 80\% of the kinetic energy of core surface motions for the single epoch 2005.0 is stored into their ES component \citep[see also][]{Wardinski:2008qy}.  
Furthermore, \cite{aubert2013bottom} have recently shown that the eccentric gyre proposed by \cite{pais08} can also emerge from three-dimensional geodynamo simulations as a columnar structure that persists over centuries. 
With this in mind, flows have been continued within the core assuming quasi-geostrophy \citep{pais08,aubert13}.
Our knowledge of time variations of the flow is less advanced, partly because time series of satellite data are relatively short.

Early calculations of time-dependent flow models were largely conducted in order to estimate the time changes of core angular momentum \citep[e.g.][]{jackson97,pais00}. 
Domination of the steady component of the large length-scale  flow in the kinetic energy budget was suggested by \cite{bloxham88}. Conversely, emergence of large scale vortices on 50 year time scales \citep{amit2006time} as well as
sudden local accelerations over a few months have been reported \citep{olsen08}.   
Interpretations of core flow variations have however been hampered because most previous studies assumed that the magnetic field entering the forward problem was perfectly known and relied on ad-hoc regularizations of the core surface velocity.
A first attempt at the estimation of model uncertainties (on which we try to improve here) gave confidence in the detection of  fast torsional waves \citep[][see also \cite{Asari:2011jk} for a description of these waves from 2000 to 2010]{gillet10}.  Here, we attempt to go further using more consistent forms of prior information concerning the core flow.

\subsection{Deterministic versus stochastic modeling}

One attractive solution to prescribing prior information is to use a dynamical model i.e. resorting to geomagnetic data assimilation \citep{fournier2010introduction}. 
Relying on the dynamics of a 3D numerical geodynamo simulation, \cite{fournier2011} and \cite{aubert11} made inferences about the core surface flow.
Dynamics entered their linear estimation in the form of covariances between the velocity and the magnetic field, the interior and the surface, all at a single epoch.
\cite{aubert13} remarked thereafter that linear estimation may have limited applicability because it works well only if the initial guess is not too far from the final solution.
Hence, he resorted to a classical frozen-flux inversion of the core surface flow and used the first and second statistical moments obtained from geodynamo simulations to build prior information about the core surface flow and to determine the error covariance matrix, including contributions from model uncertainties that arise as a consequence of truncations. 
Along the same lines, \cite{aubert2014earth,Aubert:2014ys} has recently presented images of core flows using statistics provided by the magnetic field model {\it COV-OBS} \citep{gillet13}, which is also our source of data here.
Remarkably, these calculations do not require ad-hoc penalization of small-scale flows, which is an awkward feature of classical core surface flow estimation. 
On the other hand, the important issue of temporal correlations in unresolved scales is not addressed.
Furthermore, the present generation of three-dimensional geodynamo simulations probably lacks part of the short time scale variability observed in geomagnetic series \citep{aubert2014earth}.

As an alternative, we propose here that a stochastic framework can be employed, following a strategy similar to that  used during the derivation of the {\it COV-OBS} magnetic field model \citep{gillet13}. 
The radial magnetic field at the core surface $B_r (t)$ was then considered to be a realization of a stationary stochastic (i.e. random) process ${\xi}(t)$.
Knowledge about the spectral density of $B_r (t)$ (in the frequency interval $\sim \left[ 10^{-2}, 1\right] \mbox{ yr}^{-1}$) was translated into prior information about the time correlation properties of the process ${\xi}$ sampled by $B_r$ via a correlation function of the form:
\begin{equation}
\rho(t,t')=\left(1+\sqrt{3} \frac{|t-t'|}{\tau_0} \right) 
\exp\left(-\sqrt{3}\frac{|t-t'|}{\tau_0} \right),
\label{corr:AR2}
\end{equation}
where $\tau_0$ is a characteristic time. The process $\xi$ with correlation function (\ref{corr:AR2}) is the solution of 
 the stochastic differential equation \citep[equations 2.153 and 2.155']{Yaglom:1962kx}
\begin{equation}
d \frac{d {\xi}}{dt} + \frac{2\sqrt{3}}{\tau_0} d{\xi} + \frac{3}{\tau_0^2} \xi  dt = d \zeta (t) \, ,
\label{eq:stoch-AR2}
\end{equation}
where $\zeta(t)$ is the Brownian motion (or Wiener) process.
Note that equation (\ref{eq:stoch-AR2}) corrects the equation (12) of \cite{gillet13}, which is wrong and corresponds to a process that is not stationary (the mistake in writing this equation did not affect the results presented by \citet{gillet13} because the expression for their correlation function (\ref{corr:AR2}) was correct).

Here, we incorporate pre-existing knowledge about the time evolution of the core surface velocity field $\mathbf{u}$ via a time correlation function and consider the time series of the flow coefficients to be realizations of a stochastic process.
Our specific choice of process for $\mathbf{u}$, which is defined by the stochastic differential equation
\begin{equation}
 {d{\xi}} + {\xi}\frac{dt}{\tau_u} = {d\zeta(t)} \, ,
\label{eq:stoch-AR1}
\end{equation}
follows from our selection of  (\ref{eq:stoch-AR2}) as the process for $B_r$, assuming the same differentiability properties for $\mathbf{u}$ and $\partial B_r/\partial t$.
This choice of prior stochastic process implies that the flow coefficients series are continuous, albeit non differentiable  (their increments during a time interval of length $\tau$ are defined but  are not proportional to $\tau$ as $\tau$ is decreased within the range of time-scales that we consider). 
Realizations of the process (\ref{eq:stoch-AR1}) have a simple exponential time correlation function of the form
\begin{equation}
\rho(t,t')= \exp\left(-\frac{|t-t'|}{\tau_u} \right)\,,
\label{eq:corr-AR1}
\end{equation}
where $\tau_u$ is the characteristic time scale for flow changes.  The crucial point here is that the choice of prior for $\mathbf{u}$ impacts the calculation of SV model errors arising due to unresolved small scales.

\subsection{Ensemble calculation of SV model errors}
\label{sec: intro SV}

SV model errors arise when estimating core surface flows because the radial magnetic field that enters the right hand side of (\ref{eq:ind_rad}) is an uncertain parameter  \citep{Jackson:1995zl}. 
These SV model errors should be carefully distinguished from the SV observation errors that are provided by {\it COV-OBS} in our case. 
The foremost example of this difficulty is our ignorance of the small-scale core surface magnetic field $B_r$ with harmonic degree above 14. 
Workers have traditionally circumvented this problem by searching for flows $\mathbf{u}$ presenting  rapidly converging spectra \citep[e.g.,][]{hulot92}, for which the interaction with the small-scale magnetic field is artificially reduced. 
This strategy was tempting as long as only the very largest scales of the secular variation were known (up to degree about 8 from ground observations). 
However, magnetic field models obtained from satellite data now display resolved SV up to degree 14 \citep[e.g.][]{olsen10,lesur62second,finlay12}.
This observation has made it necessary to relax the large-scale flow hypothesis, placing the SV model error problem foremost. \cite{eymin05} stressed this point and dealt with the problem in the framework of regularized single epoch core surface flow inversions, tuning the trade-off parameter such that the SV residuals have similar amplitude to the estimated SV modeling errors. 
\cite{pais08} later recursively estimated a diagonal approximation (i.e. for a single epoch) of the SV model error covariance matrix, with elements being a function of the harmonic degree only, adding this to the observation error covariance matrix to obtain the required covariance matrix \citep[see also][]{aubert13}.
Alternatively, \cite{baerenzung2013bayesian} represented the subgrid processes in the induction equation (\ref{eq:ind_rad}) as a function of the large scale fields.
However, none of these procedures account for time correlations of model uncertainties, which turns out to be essential if one wishes to resolve flow time variations.
Consideration of such temporal correlations is a crucial issue because of the high coherence, on decadal time scales, of both the flow and the unresolved magnetic field that enter the induction equation (\ref{eq:ind_rad}). 
Neglecting the temporal correlation of SV model errors should actually result in an  underweighting of the information in geomagnetic data concerning the rapid flow changes, and may lead to a biased solution for the core flow.

Here, we adopt a new ensemble estimation of SV model errors, taking due account of their time correlation.
\cite{gillet09} made a first attempt at an ensemble estimation of time-dependent surface core flows from a magnetic field model and its covariance. 
They generated an ensemble $\left\{{B_r}^p(t)\right\}$ of core surface magnetic fields (including its unresolved component at small length-scales), substituted these for $B_r$ in equation (\ref{eq:ind_rad}) and estimated an ensemble of flow solutions $\left\{\mathbf{u}^p(t)\right\}$ from SV data associated with observation errors only. 
The most probable solution was then calculated as the ensemble average of flows  $\left<{\mathbf{u}}\right>$. 
Considering each realization of the unresolved field as perfectly known when inverting equation (\ref{eq:ind_rad}), SV model errors were estimated a posteriori once and for all. 
This assumption keeps the inverse problem linear, thus avoiding an iterative estimate of SV model errors as in \cite{pais08}.
However, there was a major drawback in the approach of \cite{gillet09}:
substituting  $\left<{\mathbf{u}}\right>$  for ${\mathbf{u}}$ in equation (\ref{eq:ind_rad}) systematically underpredicted the SV monitored in ground observatories even though each flow member ${\mathbf{u}}^p$ of the ensemble adequately predicted the observed SV. 
The cause of this deficiency was that the SV error covariance used during the estimation of the flow contained only the contribution from observation errors.  
The correct way to calculate ensemble statistics is to generate an ensemble of forward models and to build in a consistent fashion the error covariance matrix by adding both the SV model error and observational error covariances \citep{evensen09}. 
In agreement with this statement, \cite{baerenzung2013bayesian} remark in their synthetic tests that omitting SV model error covariances, as in the study of \cite{gillet09}, causes a degradation in performance compared to the iterative scheme initiated by \cite{pais08}.
Note that SV model error covariance matrix has to be calculated recursively because  it involves the quantity $\mathbf{u}$ that is estimated.  
Furthermore, in order to properly account for time correlation in the SV model error, we must estimate the flow over the entire time interval in a single batch.  For the SV observation error covariance matrix, we rely on the  {\it COV-OBS} magnetic field model. 

Section \S \ref{modeling} is devoted to methodology, and constitutes the heart of the present work. 
We develop a time-dependent ensemble approach to calculate a covariance matrix for the SV model errors, following the conclusions of the above discussion. We describe how this transforms the kinematic inversion of core flows into a non-linear problem. 
We furthermore extend the ensemble method to the calculation of the covariance matrix for the uncertainties in the model parameters. 
It is at this stage that prior knowledge about time-correlations is incorporated. 
In section \S\ref{sec:results obs} we analyse the predictions of the resulting flow for geophysical observations, namely length-of-day and SV changes at observatories. 
Then, the geometry and the time dependence of the flow are discussed (section \S\ref{sec:results flow}), together with their uncertainties. 
This section ends with considerations about torsional waves, electro-magnetic core-mantle coupling, and a focus on the dynamics in the equatorial region.
We conclude this paper (section \S\ref{sec: discussion}) with discussions of possible methodological improvements and perspectives concerning core physics. 

\section{Method}
\label{modeling}

We first set out the notations employed throughout this paper (\S\ref{sec: notations}) before formulating the QG topological constraints on the flow that we use (\S\ref{sec:def-QG}). 
Next in sections \S\ref{ensemble}--\ref{iterative} we present how we implement a recursive ensemble method to obtain information about both the variances of the parameters describing the core motions and the temporal cross-covariances of SV model errors.
The use of the QG assumption, motivated in \S\ref{intro-res}, offers a significant shrinkage of the parameter space that greatly reduces the numerical cost.
Limiting the spatial complexity enables us to investigate non-zero time-correlations and dense covariance matrices.
The results we obtain concerning the dynamics should be considered within this QG approximation, which could in principle be replaced by any other hypothesis. 
The method derived to account for time-correlated SV errors is however general, and the conclusions about the use of cross-covariances are independent of the topological framework. 
We also present in Appendix \ref{sec: tutorial} a tutorial example that illustrates the impact of considering temporal cross-covariances.

\subsection{Notations}
\label{sec: notations}

We expand all quantities ($B_r$, $\partial B_r/\partial t$ and the poloidal and toroidal scalars $S$ and $T$ describing the surface flow 
$\mathbf{u}={\nabla} \times (T \mathbf{r}) +  \nabla_H (rS)$, where $\mathbf{r}$ is the position vector) in spherical harmonics:
\begin{eqnarray}
\displaystyle
\left\{
\begin{array}{rl}
\displaystyle S=&\displaystyle\sum_{\ell=1}^{L_{\sf x}} \sum_{m=-\ell}^{\ell} s_{\ell m} Y_{\ell m}(\theta,\phi), \; \mbox{with} \; s_{\ell,{-m}} = \overline{s_{\ell m}} \\
\displaystyle T=&\displaystyle\sum_{\ell=1}^{L_{\sf x}} \sum_{m=-\ell}^{\ell} t_{\ell m} Y_{\ell m}(\theta,\phi), \; \mbox{with} \; t_{\ell,{-m}} = \overline{t_{\ell m}}
\end{array}
\right.\,.
\end{eqnarray}
($(\theta,\phi)$ are spherical coordinates, $Y_{\ell m}$ are
Schmidt quasi-normalized functions, and $ \overline{x}$ is the complex conjugate of $x$). The expansions of the magnetic field and its time derivative are truncated respectively at degrees $L_{\sf b}$ (for $B_r$), and $L_{\sf y}$ (for $\partial B_r / \partial t$). 
Contrary to \cite{jackson97} and \cite{gillet09}, we do not expand in time in terms of a B-spline basis and consider instead the spherical harmonic coefficients of the time variable fields as discrete-time parameter sets $\left\{t_n\right\}_{n\in[1,N]}$ regularly sampling $[t_1\dots t_N]=[1940\dots 2010]$ every $\delta t=(t_N-t_1)/(N-1)=1$ year (N=71).
At each epoch $t_n$ we store $B_r$, $\partial B_r/\partial t$ and core flow spherical harmonic coefficients in vectors ${\bf b}(t_n)$, ${\bf y}(t_n)$  and ${\bf x}(t_n)$ respectively. 
From these we build vectors ${\bf Y}=\left[{\bf y}(t_1)\dots{\bf y}(t_N)\right]^T$, 
${\bf B}=\left[{\bf b}(t_1)\dots{\bf b}(t_N)\right]^T$ and ${\bf X}=\left[{\bf x}(t_1)\dots{\bf x}(t_N)\right]^T$, which are linked through the forward problem
\begin{equation} 
\displaystyle
{\bf Y}={\cal H}(\mathbf{B}){\bf X}+{\bf E}\,,
\label{eq:forward}
\end{equation}
with ${\bf E}=\left[{\bf e}(t_1)\dots{\bf e}(t_N)\right]$ the SV error vector and ${\cal H}$ the operator  
\begin{eqnarray}
\displaystyle
{\cal H} (\mathbf{B})= 
\left[
\begin{array}{cccc}
{\sf H}(\mathbf{b}(t_1)) & {\sf 0} & \dots  & {\sf 0} \\
{\sf 0} & {\sf H}(\mathbf{b}(t_2))  & \ddots & \vdots \\
\vdots & \ddots & \ddots & {\sf 0} \\
{\sf 0}  & \dots & {\sf 0} & {\sf H}(\mathbf{b}(t_N))  
\end{array}
\right]\,.
\label{forward op}
\end{eqnarray}
${\sf H}({\bf b}(t_n))$ results from the transformation of the snapshot equation (\ref{eq:ind_rad}) at epoch $t_n$ in matrix form. 

\subsection{Formulation of the quasi-geostrophic topological constraint}
\label{sec:def-QG}

We assume quasi-geostrophy and incompressibility in the outer core volume, which results in the columnar flow constraint at the core surface \citep{amit04,amit13},
\begin{equation} 
\displaystyle
\nabla_H\cdot\left({\bf u}\cos^2\theta\right)=0\,,
\label{helical hyp}
\end{equation}
together with the equatorial symmetry constraint \citep[eq. (27)]{pais08}.
In contrast with our previous attempts at calculating core flows \citep{pais08,gillet09}
we do not impose no penetration across the cylindrical surface tangent to the solid inner core.
We note below that our method makes it easy to incorporate this constraint at a later stage if desired.

The above hypotheses yield a set of linear constraints that can be formally written as 
\begin{equation} 
\displaystyle
\forall n, \qquad
{\sf Q}\,{\bf x}(t_n)={\bf 0}  \, .
\label{constraint matrix}
\end{equation}
Following \cite{jackson97}, we use the matrix ${\sf G}$ obtained from the QR decomposition of  ${\sf Q}$ in order to project the vector of unknowns onto a reduced basis :
\begin{equation} 
\displaystyle
{\bf x}(t)={\sf G}{\bf w}(t) \, .
\label{basis change}
\end{equation}
The solution ${\bf W}=\left[{\bf w}(t_1)\dots{\bf w}(t_N)\right]$ for which we invert is composed of vectors whose size, $L_{\sf x}(L_{\sf x}+1)/2$ for $L_{\sf x}$ even, is reduced by a factor about 4 compared to the size $2L_{\sf x}(L_{\sf x}+2)$ of the original unknown vectors. 

\subsection{Ensemble core flow estimation}
\label{ensemble}

We use an ensemble approach, following the example of previous studies in oceanic and atmospheric dynamics \citep{evensen2003ensemble}, to recursively estimate stationary second-order statistics for the flow coefficients and the model errors.
We derive an ensemble of $P$ solutions (typically here $P=20$) to the forward problem (\ref{eq:forward}) under the constraint (\ref{constraint matrix}) from the ensemble of replications $\left\{{\bf B}^p,{\bf Y}^p\right\}_{p\in[1,P]}$  drawn from {\it COV-OBS}. 
Pre-existing knowledge about $\mathbf{w}$ is represented by the covariance matrix ${\sf C}_{\bf w}$ while SV error covariances are described by ${\sf C}_{\sf e}$.
Then, for each replication $\left\{{\bf B}^p,{\bf Y}^p\right\}$, the least-square solution ${\bf W}^p$ minimizes the cost function 
\begin{equation} 
\displaystyle
J^p({\bf W})=\left[{\bf Y}^p-{\cal A}({\bf B}^p){\bf W}\right]^T{\sf C}_{\sf e}^{-1}\left[{\bf Y}^p-{\cal A}({\bf B}^p){\bf W}\right]
+{\bf W}^T{\sf C}_{\sf w}^{-1}{\bf W}\,.
\label{cost function}
\end{equation}
Here ${\cal A}(\mathbf{B})$ is the forward operator of equation (\ref{forward op}) rotated into the reduced basis using (\ref{basis change}).
Rather than define and fix the two matrices ${\sf C}_{\sf e}$ and ${\sf C}_{\sf w}$ prior to the inversion, here we update these at each iteration using the current ensemble of flow solutions $\left\{{\bf W}^p\right\}_{p\in[1,P]}$. 
Thus the minimization of the functional (\ref{cost function}) becomes a nonlinear problem, and we calculate the ensemble of models recursively, with at each iteration $k$ 
\begin{equation} 
\displaystyle
{\bf W}^{p,k+1} = 
\left[{\cal A}({\bf B}^p)^T \left({\sf C}_{\sf e}^k\right)^{-1}{\cal A}({\bf B}^p) + \left({\sf C}_{\sf w}^k\right)^{-1}\right]^{-1}
{\cal A}({\bf B}^p)^T\left({\sf C}_{\sf e}^k\right)^{-1}{\bf Y}^p \,.
\label{solution}
\end{equation}
We keep the same ensemble of replications $({\bf B}^p,{\bf Y}^p)$ from one iteration to the next. 
 
Concerning the construction of ${\sf C}_{\sf w}^k$, we estimate the a priori covariances between the coefficients ${\sf w}_{i}(t_n)$ and ${\sf w}_{i'}(t_{n'})$ at iteration $k+1$ from time and ensemble averages of the variances of the coefficients of the flow models $\left\{{\bf W}^p\right\}_{p\in[1,P]}$, as calculated at iteration $k$.   
In addition, for the required temporal correlation function, we simply adopt  the exponential function (\ref{eq:corr-AR1}).  Combining these gives
\begin{equation} 
\displaystyle
E\left({\sf w}_{i}(t_n){\sf w}_{i'}(t_{n'})\right)=\delta_{ii'}\left<
\frac{1}{N}\sum_{n=1}^N{{\sf w}_i}(t_n)^2
\right>
\exp\left(-\frac{|t_n-t_{n'}|}{\tau_u}\right)\,.
\label{E(wwt)}
\end{equation}
Here the notation $\displaystyle\left<X\right>$ means ensemble averaging:
\begin{equation} 
\displaystyle
\left<X\right> = \frac{1}{P}\sum_{p=1}^P X^p\,.
\label{eq:ens ave}
\end{equation}
The choice of a common value for all coefficients of the characteristic flow time scale $\tau_u$ is made for the sake of simplicity.
Spatial cross-covariances (excepted those associated with the QG assumption) are ignored. 
To summarize, equation (\ref{E(wwt)}) prescribes the elements of the flow model covariance matrix ${\sf C}_{\sf w}^{k}$ that is used as prior information at iteration $k+1$. 
In order to initialize the calculation,  ${\sf C}_{\sf w}^0$ is built from a flat CMB spatial power spectrum ($\forall m\in[-\ell,\ell],E(t_{\ell m})^2=10^2/\ell(\ell+1)$) and the time auto-correlation function (\ref{eq:corr-AR1}).
The final flow solutions are found to be insensitive to small changes of ${\sf C}_{\sf w}^0$.

Use of equation (\ref{E(wwt)}) means that we build the a priori probability distribution of the flow from the expected values obtained from the existing ensemble of flow models.
This approximation is made to reduce the numerical cost; indeed, one should in principle also account for the posterior uncertainties associated with each flow realization \citep[see][]{baerenzung2013bayesian}. 
Tests on problems of small dimensions demonstrate that with this simplification we only slightly under-estimate the posterior uncertainties on the flow model and the related SV model errors. 

We have no prior knowledge of the covariance matrices that enter the cost function (\ref{cost function}) to be minimized. 
The iterative process yields (i) the flow model, (ii) the SV model errors (as described in \S\ref{iterative}), and (iii) the flow covariances. 
We seek to exhibit possible solutions within this framework. 
On the other hand, the inversion process is inherently nonlinear, and we do not claim unicity of the flow solution.

\subsection{Accounting for time-correlated SV model errors}
\label{iterative}

The SV model error covariance matrix ${\sf C}_{\sf e}^{k}$ is also estimated recursively using an ensemble approach.  Recall that the error vector ${\bf E}$ in equation (\ref{eq:forward}) contains both the SV observation errors ${\bf E}^o$ and the SV model errors ${\bf E}^m$ that arise from unresolved interactions between the core flow and the magnetic field. 
SV model errors ${\bf E}^m$ result from the unresolved flow $\delta{\bf W}$ interacting with the entire (resolved or not) magnetic field, plus the resolved flow interacting with the unresolved field $\delta{\bf B}$.
We omit model errors at the first iteration.  We calculate the resolved flow and magnetic field as the ensemble averages of the flow solutions, $\left<{\bf W}\right>$ (calculated at iteration $k$), and of the {\it COV-OBS} field models, $\left<{\bf B}\right>$, respectively. 

An ensemble of $Q=40$ realizations of $\delta{\bf B}$ is obtained from the product of normally distributed random vectors with the Choleski decomposition of the covariance matrix ${\sf C_b}=E\left(\delta{\bf B}\delta{\bf B}^T\right)$.
For $\ell\le14$, we simply take the {\it COV-OBS} error covariance matrix as ${\sf C_b}$. 
Note these are potentially under-estimated at degrees $\ell\sim 14$ due to the signature of unmodeled lithospheric field at large length-scales, \citep[see][]{jackson90,thebault15}.
For $\ell>14$, ${\sf C_b}$ is constructed from the correlation function (\ref{corr:AR2}) proposed in \cite{gillet13}, using correlation times and variances extrapolated with power laws from those obtained in satellite field models for $\ell\le12$. 
We truncate the unresolved field at $L_{\sf b}=30$ since increasing $L_{\sf b}$ further does not significantly affect the results.
The ensemble of $Q$ realizations is calculated independently for the $P$ replications $\mathbf{B}^p$ that are used in (\ref{solution}).
The dispersion within the ensemble of flow solutions defines the unresolved flow, i.e. $\delta{\bf W}^p={\bf W}^p-\left<{\bf W}\right>$.
It should be accounted for when estimating the covariances of SV model errors because these depend on the flow \citep[the reason why our problem is nonlinear, see][]{pais08}. 
We thus estimate $PQ = 800$ realizations of the SV errors from all possible pairs $(\delta{\bf W}^p, \delta{\bf B}^{q})$:
\begin{equation} 
\displaystyle
\delta{\bf Y}^{pq}={\cal A}\left(\delta{\bf B}^{q}\right)\left(\left<{\bf W}\right>+\delta{\bf W}^p\right) + {\cal A}\left(\left<{\bf B}\right>\right)\delta{\bf W}^p\,,
\label{SV mod errs}
\end{equation}
from which we build the matrix 
\begin{equation} 
\displaystyle
{\sf S}_{\sf e}^m = \frac{1}{PQ-1}\sum_{p=1}^{P}\sum_{q=1}^{Q}\left(\delta{\bf Y}^{pq}-\left<\delta{\bf Y}\right>\right)\left(\delta{\bf Y}^{pq}-\left<\delta{\bf Y}\right>\right)^T\,,
\label{SV mod errs Cmat}
\end{equation}
using the ensemble average
$\displaystyle\left<\delta{\bf Y}\right> = \frac{1}{PQ}\sum_{p=1}^{P}\sum_{q=1}^{Q} \delta{\bf Y}^{pq}$.
We have found it important that the realizations of $\delta{\bf B}$ that enter equation (\ref{SV mod errs}) are independent of the $P$ realizations used in (\ref{solution}). 
This yields an unbiased estimate of $\delta{\bf Y}$ (i.e. in practice $\forall i, \left|\left<\delta{\sf Y}_i\right>\right| \ll \left| \delta{\sf Y}_i \right|$), and thus of ${\sf S}_{\sf e}^m$. 
Otherwise, SV model errors tend to contribute constructively to the observed SV (their sign is predominantly that of the SV data), and variances in (\ref{SV mod errs Cmat}) tend to be dramatically under-estimated.

The matrix ${\sf S}_{\sf e}^m$ is positive semi-definite but unfortunately most of its eigenvectors correspond to null eigenvalues because the size $PQ$ of the ensemble is smaller than the number of rows $R_{\sf y}=N L_{\sf y}(L_{\sf y}+2)$. 
We have
\begin{equation}
 \forall (p,q), \qquad\mathbf{V} \cdot \left(\delta{\bf Y}^{pq}-\left<\delta{\bf Y}\right>\right) = 0 \qquad  \Longrightarrow \qquad\mathbf{V}^T {\sf S}_{\sf e}^m \mathbf{V} = 0\,,
\end{equation}
where $\mathbf{V}$ is any non-zero column vector.  As a result, ${\sf S}_{\sf e}^m$ is not a valid covariance matrix and is not invertible.
It would require $PQ\gg R_{\sf y}$ in order to have a converged estimate of all cross-covariances in ${\sf S}_{\sf e}^m$; their accuracy is limited by the size of the ensemble. 
To avoid considering spurious cross-covariances that would bias the measure of the SV model errors, we have to modify ${\sf S}_{\sf e}^m$.
First, spatial correlations, in the spherical harmonic domain, of the flow model uncertainties are neglected in order to avoid overestimates of these quantities, and we consider only temporal correlations. 
Tests on a smaller problem show that ignoring these spatial correlations does not significantly affect the time changes of the output flow models.   
Second, we guard against artificial correlations between SV model errors at distant epochs by using a covariance localization approach \citep{gaspari1999construction}. This consists in taking the Hadamard product (element-by-element multiplication) of ${\sf S}_{\sf e}^m$ and ${\sf L}$,
\begin{equation} 
\displaystyle
{\sf C}_{\sf e}^m = {\sf L}\circ{\sf S}_{\sf e}^m\,,
\label{Ce loc}
\end{equation}
where ${\sf L}$ is a covariance matrix constructed from a correlation function $\rho_{loc}$.
The matrix ${\sf C}_{\sf e}^m$ is a valid covariance matrix because the Hadamard product of a positive definite matrix (such as ${\sf L}$) and of a positive semi-definite matrix with all its diagonal elements strictly positive (such as ${\sf S}_{\sf e}^m$) is positive definite \citep[Schur's theorem, see][]{horn1990hadamard}.
This heuristic approach has been extensively used in data assimilation studies of the dynamics in the atmosphere and the ocean, to filter the background covariance matrix as a function of distance \citep{hamill2001distance}. 
Instead we use it here to filter the SV model error covariance matrix as a function of time separation.

\begin{figure}
\centerline{
\noindent\includegraphics[width=30pc]{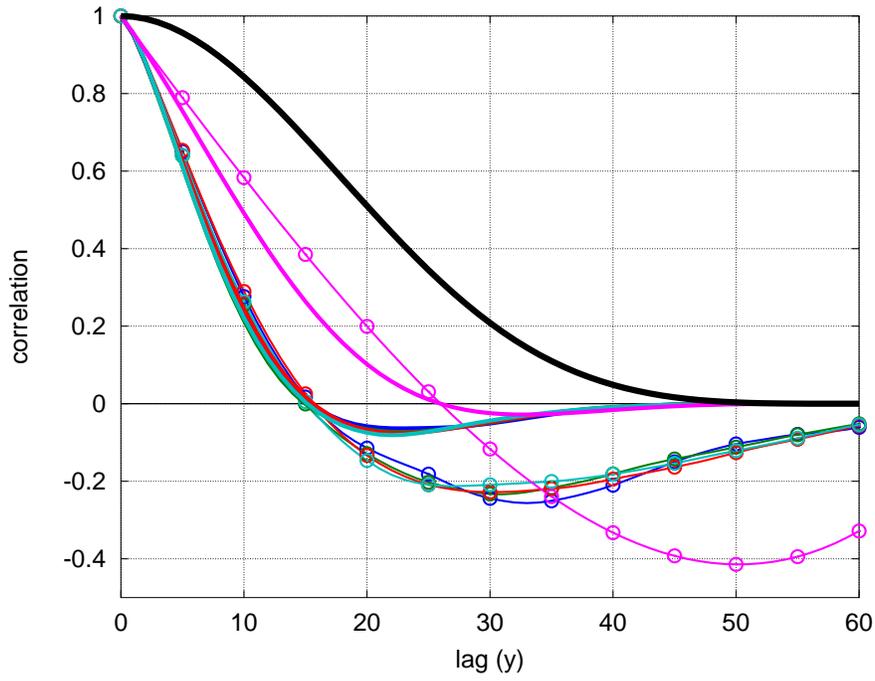}}
 \caption{Time-correlations of the empirical SV model errors determined for the coefficients $dg_n^m/dt$: for the axial dipole (magenta), and averages over all spherical harmonic orders of degrees 1 (blue, excepted for the axial dipole), 3 (green), 5 (red), and 7 (cyan). 
In thick (resp. thin circle) lines the correlations after (resp. before) localization, as implemented using the function $\rho_{gc}(\tau)$ from \cite{gaspari1999construction} (in black).}
\label{fig: localization}
 \end{figure}

As in oceanic applications, ${\sf L}$ is constructed from a correlation function $\rho_{gc}$ defined by equation (4.10) in \cite{gaspari1999construction}, which involves a cut-off period $\tau_{loc}$. 
As illustrated in figure \ref{fig: localization}, this allows time differentiability properties compatible with the SV spectrum (see the smooth behavior of $\rho_{gc}$ at zero lag). 
Furthermore, being defined on a compact support prevents over-estimation of cross-covariances at large lag. 
The cut-off period should be slightly larger than empirical estimates of the correlation length \citep{oke2007impacts}.  
We find through the analysis of ${\sf S}_{\sf e}^m$ that the SV model error typically have a decay time between 10 and 20 years (depending on the SV coefficients). 
We therefore adopt a cut-off period $\tau_{loc}=30$ years. 
Figure \ref{fig: localization} illustrates the impact of the localization process on the ensemble estimate of time correlations. 
The time-correlation functions at zero lag is less sharp than the Laplacian function obtained for a stochastic process such as that defined by equation (\ref{eq:stoch-AR1}):
this corresponds to a case intermediate between the two tutorial examples illustrated in Appendix \ref{sec: tutorial}.  
Note that the most extreme localization would involve keeping only the diagonal elements of ${\sf S}_{\sf e}^m$, i.e. $\rho_{loc}(\tau)=\delta(\tau)$, which we also test for comparison purposes. 

Finally, the combined SV data error covariance matrix is then ${\sf C_e}={\sf C}_{\sf e}^o+{\sf C}_{\sf e}^m$. 
Covariances ${\sf C}_{\sf e}^o$ for ${\bf E}^o$ are derived from the error covariance matrix ${\sf C}_{spl}$ for the {\it COV-OBS} spline model coefficients \citep{gillet13}:
if ${\sf D}(t)$ is the operator relating the SV coefficients at epoch $t$ to the spline coefficients, then the matrix ${{\sf C}_{\sf e}^o}$ is built from $N\times N$ blocs 
${{\sf C}_{\sf e}^o}_{nn'} = E\left({\bf e}^o(t_n){{\bf e}^o(t_{n'})}^T\right)={\sf D}(t_n) {\sf C}_{spl} {\sf D}(t_{n'})^T$ that contain the covariances for the observation errors between all SV spherical harmonic coefficients at epochs $t_n$ and $t_{n'}$.

Note that \cite{gillet09,gillet11} assumed ${\sf C_e}={\sf C}_{\sf e}^o$.
Our procedure to obtain ${\sf C_e}$ also differs from the iterative method of \cite{pais08} inasmuch as they needed only one realization $({\bf B}^1,{\bf Y}^1)$ to estimate how the diagonal elements of ${\sf C_e}$ vary with the degree.
In practice less than 10 iterations suffice to obtain a converged ensemble of solutions and thus converged covariance matrices ${\sf C_e}$ and ${\sf C_w}$.
For all frequencies and length-scales, the rate of change of the average solution, from one iteration to the next, is smaller than the dispersion within the ensemble of solutions (about 5\%~of the kinetic energy), and we observe no drift of the solution through the iterative process. 
The variances that enter ${\sf C_w}$ in equation (\ref{E(wwt)}), averaged per degree $\ell$ and over time, change by about 4\%~ from one iteration to the next (similarly, with no drift). 
 
\section{Flows accounting for rapid magnetic field and length-of-day changes}
\label{sec:results obs}

We first discuss in section \S\ref{sec:choice} the importance of accounting for time correlated SV model errors in order to consistently recover geophysical observations, especially on sub-decadal time scales.  This guides our choice for the free parameter $\tau_u$ that enters the model prior covariance matrix. 
In section \S\ref{sec: SV pred} we analyse more closely the fit to the SV observations, comparing the flow predictions to time series of spherical harmonic coefficients, and also, more directly, to observatory series. 

\subsection{Importance of time correlations of flow coefficients and SV model errors}
\label{sec:choice}

In order to measure the impact of the methodology choices (for example the time covariances in ${\sf C_e}$, the choice of $\tau_u$), we compare different flow solutions, presenting in Table \ref{tab:diagnostics} the following statistics for the solutions:
\begin{itemize}
\item 
the ensemble average of the normalized misfits to the COV-OBS field model SV,
\begin{equation} 
\displaystyle
\chi^2=
\left<\frac{1}{R_{\sf y}}
\left[{\bf Y}-{\cal A}({\bf B}){\bf W}\right]^T{\sf C_e}^{-1}
\left[{\bf Y}-{\cal A}({\bf B}){\bf W}\right]\right>\,;
\label{chi2}
\end{equation}
This measure contains information from all frequencies when cross-covariances, due to temporal error correlation, are considered in ${\sf C_e}$ (cases `$A_{\tau_u}$' with $\rho_{loc}(\tau)=\rho_{gc}(\tau)$).
When instead $\rho_{loc}(\tau)=\delta(\tau)$ (cases `$D_{\tau_u}$'), this measure is mainly sensitive to the longer periods where the SV signal amplitude is the largest, and is indifferent to the high frequency SV. 
As a consequence, $\chi^2$ can be smaller in cases $D_{\tau_u}$ even though rapid changes are not well reproduced (cf Figure \ref{fig: SV g10}).

\item 
the ensemble average of the ratio between the r.m.s. of the recorded ($\varphi^o_{obs}$) and predicted ($\varphi^p_{obs}$) SV series at ground-based stations, calculated over $[t_i,t_e]=[1940,2010]$, 
\begin{equation} 
\displaystyle
{\cal L}_{SV} =  \frac{1}{N_{obs}}\sum_{obs=1}^{Nobs}\frac{1}{3}\sum_{\varphi=X,Y,Z}\left(\int_{t_i}^{t_e}(\varphi^o_{obs}-\hat{\varphi}^o_{obs})^2dt\right)^{-1}\left<
\int_{t_i}^{t_e}(\varphi^p_{obs}-\hat{\varphi}^p_{obs})^2dt
\right>\,,
\label{ratio SV}
\end{equation}
averaged over all three components and over four test observatories (Kakioka, Hermanus, Niemegk and Alibag). 
We use the notation $\displaystyle \hat{x}=\frac{1}{t_e-t_i}\int_{t_i}^{t_e}x(t)dt$ for the time-averaging. 
A similar quantity ${\cal L}_{SV}^f$ is defined for the signals filtered between 4 and 9.5 years.

\item
the ensemble average of the ratio between the rms values for length-of-day (LOD) geodetic data $\gamma^{o}(t)$ and their predictions $\gamma^p(t)$, calculated over $[t_i,t_e]=[1940,2010]$, 
\begin{equation} 
\displaystyle
{\cal L}_{\gamma} =  \left(\int_{t_i}^{t_e}(\gamma^o-\hat{\gamma}^o)^2dt\right)^{-1}
\left<
\int_{t_i}^{t_e}(\gamma^p-\hat{\gamma}^p)^2dt\right>\,.
\label{ratio LOD}
\end{equation}
Here $\gamma^p(t)$ is calculated using equation (103) in \cite{TOG8Jault13}.
We use a definition similar to (\ref{ratio LOD}) for $\displaystyle {\cal L}_{\gamma}^f$ the ratio for the LOD series filtered at periods between 4 and 9.5 years. 
\end{itemize}

We find that ignoring temporal covariances of model errors (cases `$D$' with $\rho_{loc}(\tau)=\delta(\tau)$) leads to a loss of information. Table (\ref{tab:diagnostics}) shows that this omission leads us to either  over-predict LOD changes on decadal time scales (see $D_{10}$) or under-predict LOD changes on inter annual time scales (see $D_{100}$).

We also find that the amplitude of LOD predictions is a very useful diagnostic in assessing the appropriate flow correlation time $\tau_u$. For $\tau_u=10$ years (or less), the predicted LOD fluctuations are significantly more intense than the actual LOD fluctuations (see $A_{10}$). On the other hand, it appears impossible to account for rapid SV fluctuations with $\tau_u=300$ years, or more (see $A_{300}$).
We thus consider the solution obtained for $\tau_u=100$ years as providing an acceptable compromise, producing reasonable amplitudes of the SV and LOD changes at both decadal and interannual periods. 
Henceforth we focus on our preferred flow ensemble $A_{100}$, and discuss it in detail. 

\begin{table}
\caption{Statistics of the derived flow models for several values of $\tau_u$, and several localization functions $\rho_{loc}$. $\rho_{gc}$ stands for the localization function from \cite{gaspari1999construction}, used with a cut-off frequency $\tau_{loc}=30$ years. $r_{\gamma}$ is the correlation coefficient between the LOD data and the LOD prediction from the ensemble average solution (similarly $r_{\gamma}^f$ is defined for the series filtered at periods between 4 and 9.5 years).  
Other symbols are defined in the text (section \ref{sec:choice}). }
\centering
\begin{tabular}{|c|c|c|r|rr|rr|rr|rr|rr|}
\hline
case & $\rho_{loc}(\tau)$  & $\tau_u$ (y) & $\chi^2$ & ${\cal L}_{SV}$ & ${\cal L}_{SV}^{f}$ & ${\cal L}_{\gamma}$ & ${\cal L}_{\gamma}^{f}$ & $r_{\gamma}$ & $r_{\gamma}^f$ \\
\hline
$A_{10}$   & $\rho_{gc}(\tau)$   & 10  & 0.213 & 0.86 & 1.02 & 1.72 & 1.56 & 0.87 & 0.77\\	
$A_{100}$  & $\rho_{gc}(\tau)$   & 100 & 0.406 & 0.81 & 0.86 & 1.34 & 1.03 & 0.86 & 0.80\\	
$A_{300}$  & $\rho_{gc}(\tau)$   & 300 & 0.664 & 0.75 & 0.74 & 1.19 & 0.86 & 0.85 & 0.69\\	
$D_{10}$   & $\delta(\tau)$      & 10  & 0.045 & 0.93 & 0.75 & 1.91 & 0.88 & 0.92 & 0.79\\	
$D_{100}$  & $\delta(\tau)$      & 100 & 0.110 & 0.88 & 0.44 & 1.49 & 0.37 & 0.89 & 0.75\\	
\hline
\end{tabular}
\label{tab:diagnostics}
\end{table}

\subsection{Fit to SV observations}
\label{sec: SV pred}

\begin{figure}
\centerline{
 \noindent\includegraphics[width=30pc]{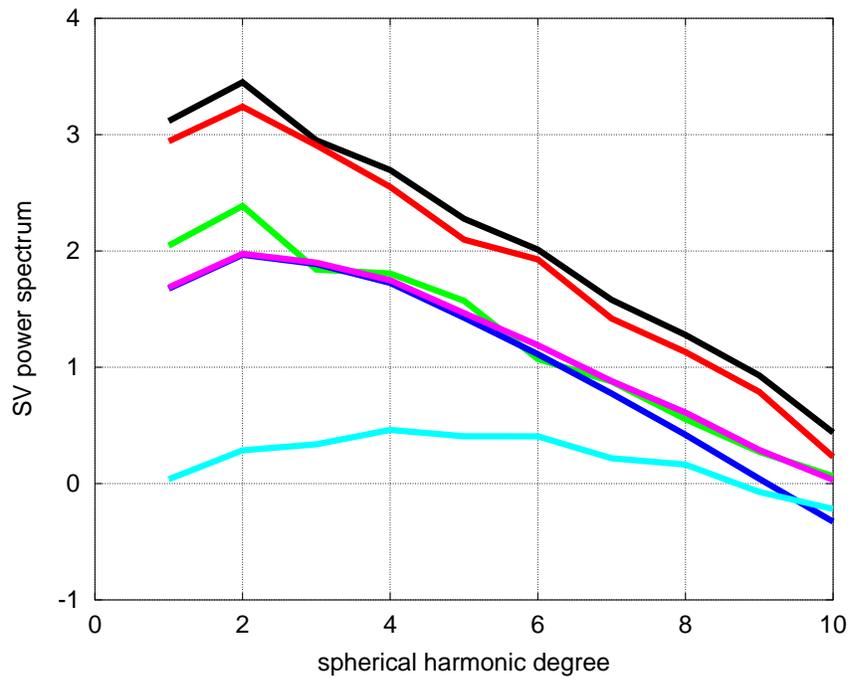}}
 \caption{SV power spectra at the Earth's surface, time-averaged over 1940--2010 (scale is $\log_{10}$, in units of (nT/yr)$^2$):
SV spectrum from COV-OBS (black) and its associated observation errors (cyan);
ensemble average of the SV spectra for the model predictions (red), the model prediction errors (green), the SV model errors due to unresolved scales (dark blue), and the SV model plus observation errors (magenta).
}
\label{fig: SV spec}
 \end{figure}

In this section we analyse how predictions from our ensemble of flow solutions fit the SV data. 
We consider both the misfit to COV-OBS Gauss coefficients and comparisons to annual differences of observatory annual means.
The former are analyzed in terms of time-averaged SV spatial power spectra, presented in Figure \ref{fig: SV spec}. 
Overall, the SV spectra of the a posteriori and a priori errors calculated through our ensemble scheme  superimpose. 
The latter is similar to that found by \cite{pais08} for a snapshot problem.
We obtain a posteriori errors slightly larger (smaller) than the prior errors at long (short) length-scales, which translates into a misfit less than unity (see Table \ref{tab:diagnostics}). 

\begin{figure}
\centerline{
 \noindent\includegraphics[width=20pc]{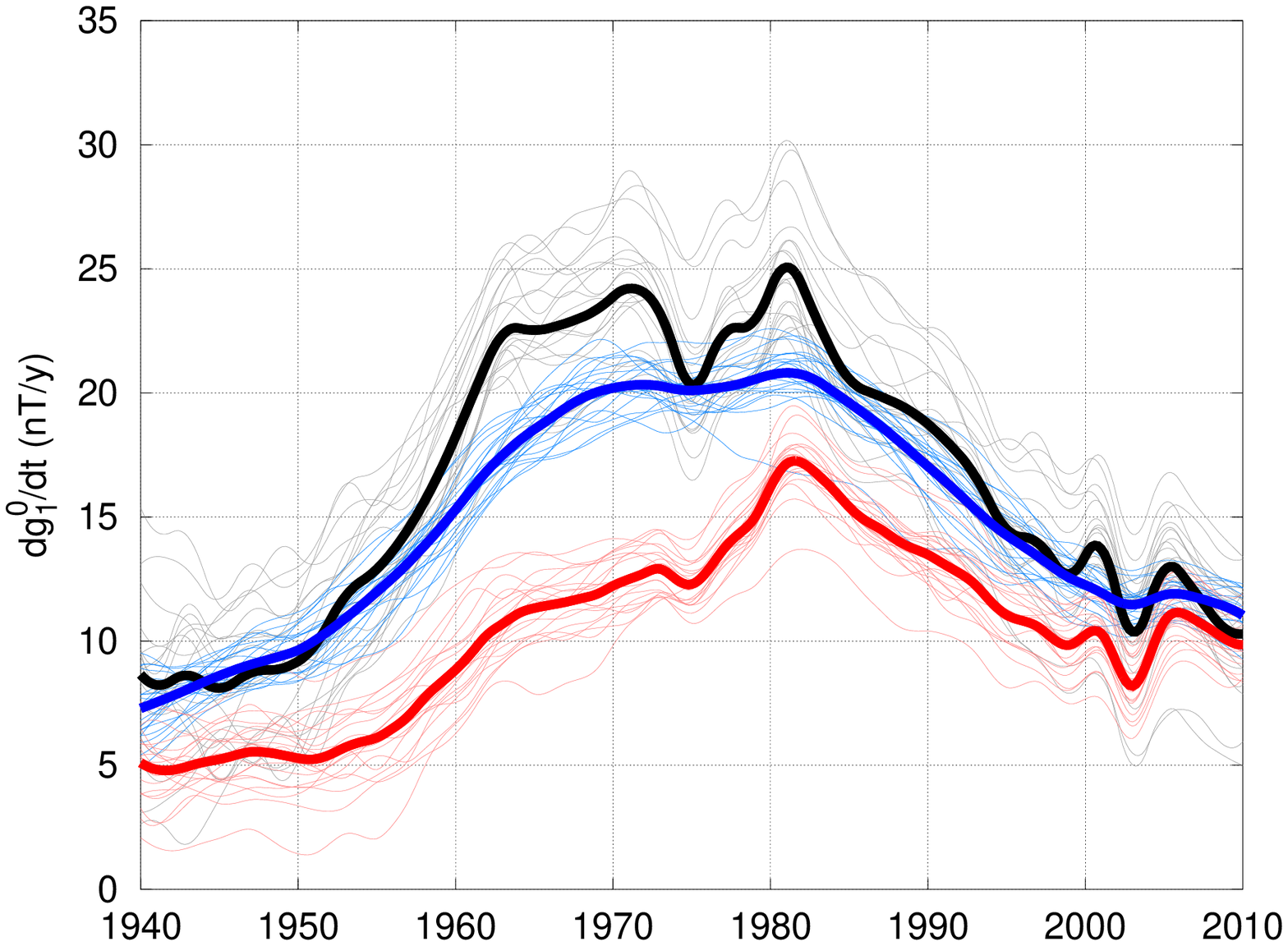}
 \noindent\includegraphics[width=20pc]{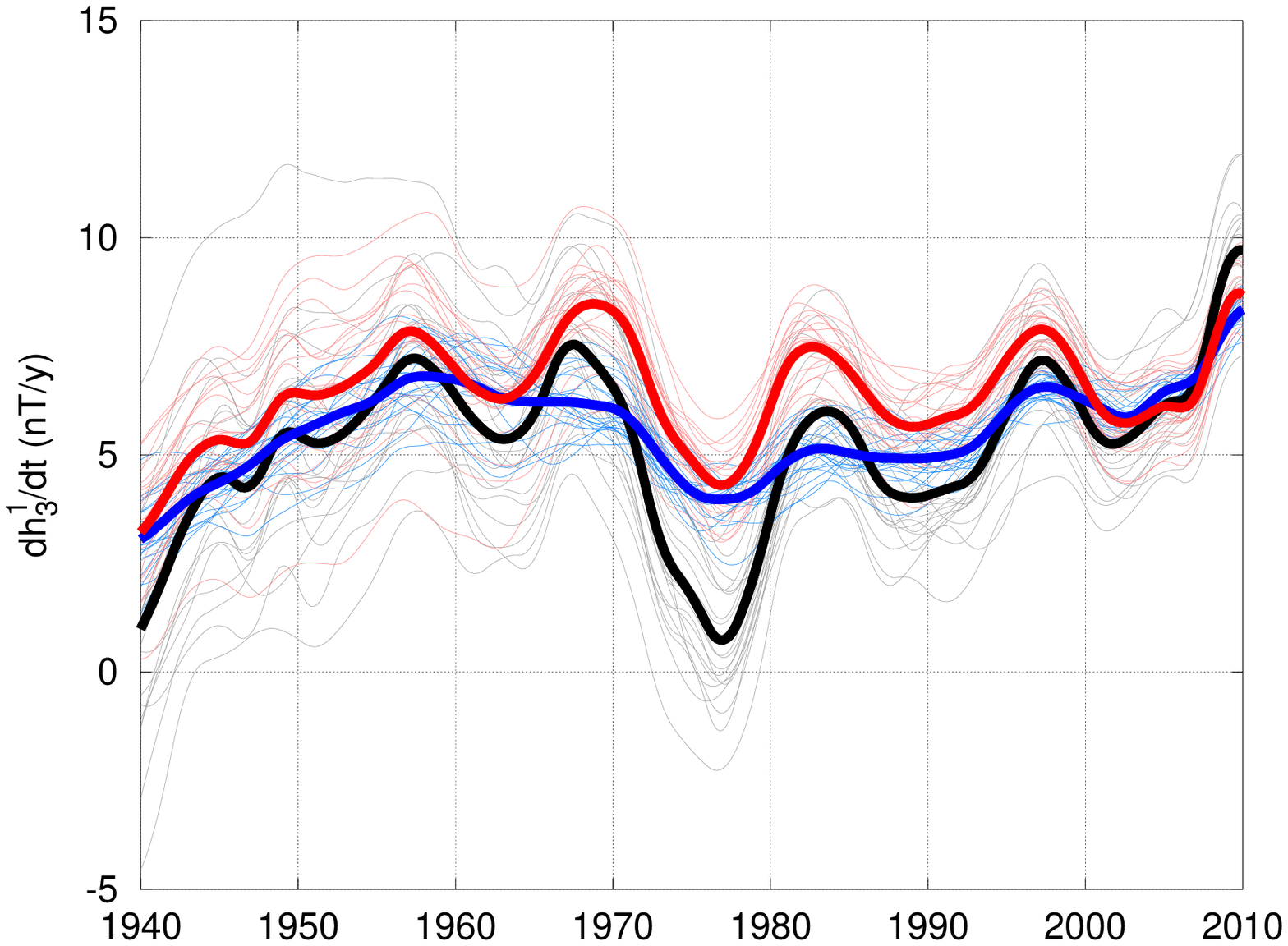}}
\centerline{
 \noindent\includegraphics[width=20pc]{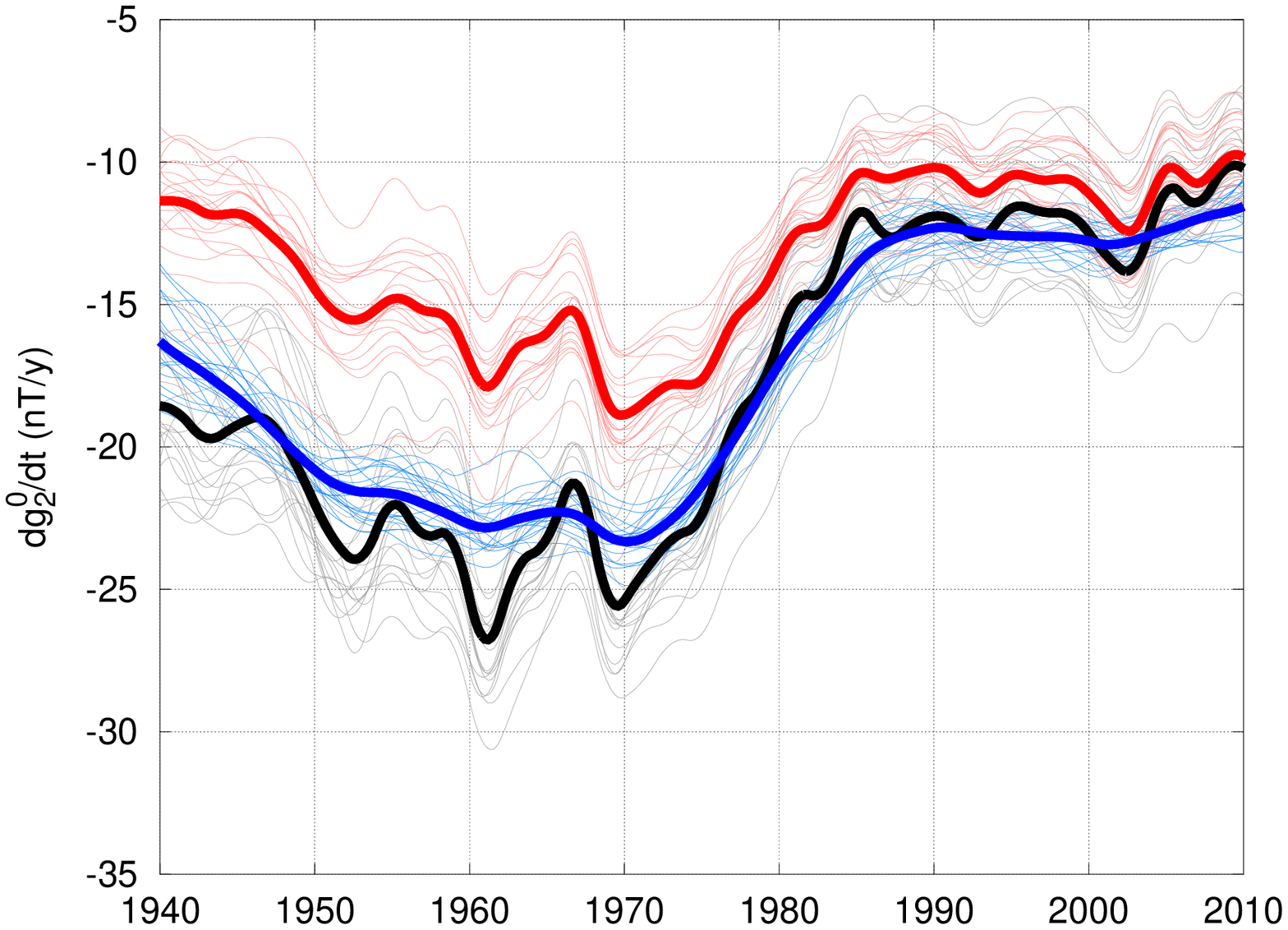}
 \noindent\includegraphics[width=20pc]{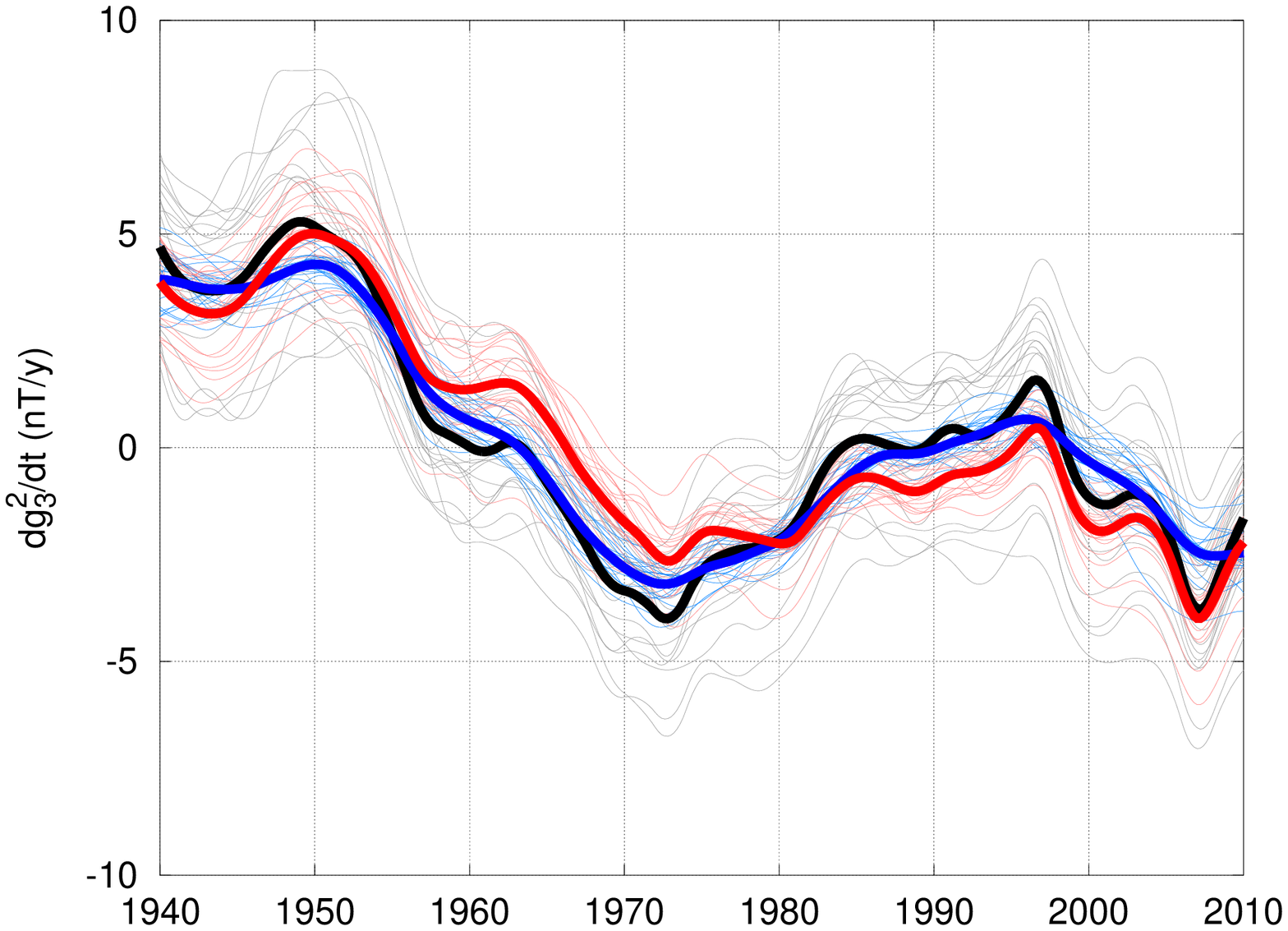}}
 \caption{Time changes of the SV spherical harmonic coefficients $dg_n^m/dt$ (in nT/yr).  
SV from the {\it COV-OBS} model truncated at degree $n=10$ 
(thick black line: average model; grey thin lines: ensemble of realizations including the impact of SV model errors), 
and SV predictions from the ensemble of flows: 
accounting for time correlations of the SV model errors 
(thick red line: ensemble average; thin pink lines: ensemble of realizations), 
or neglecting it 
(thick blue line: ensemble average; thin blue lines: ensemble of realizations). 
In both cases for $\tau_u=100$ years. 
Our preferred flow ensemble corresponds to the red curves.}
\label{fig: SV g10}
 \end{figure}

\begin{figure}
\centerline{
 \noindent\includegraphics[width=20pc]{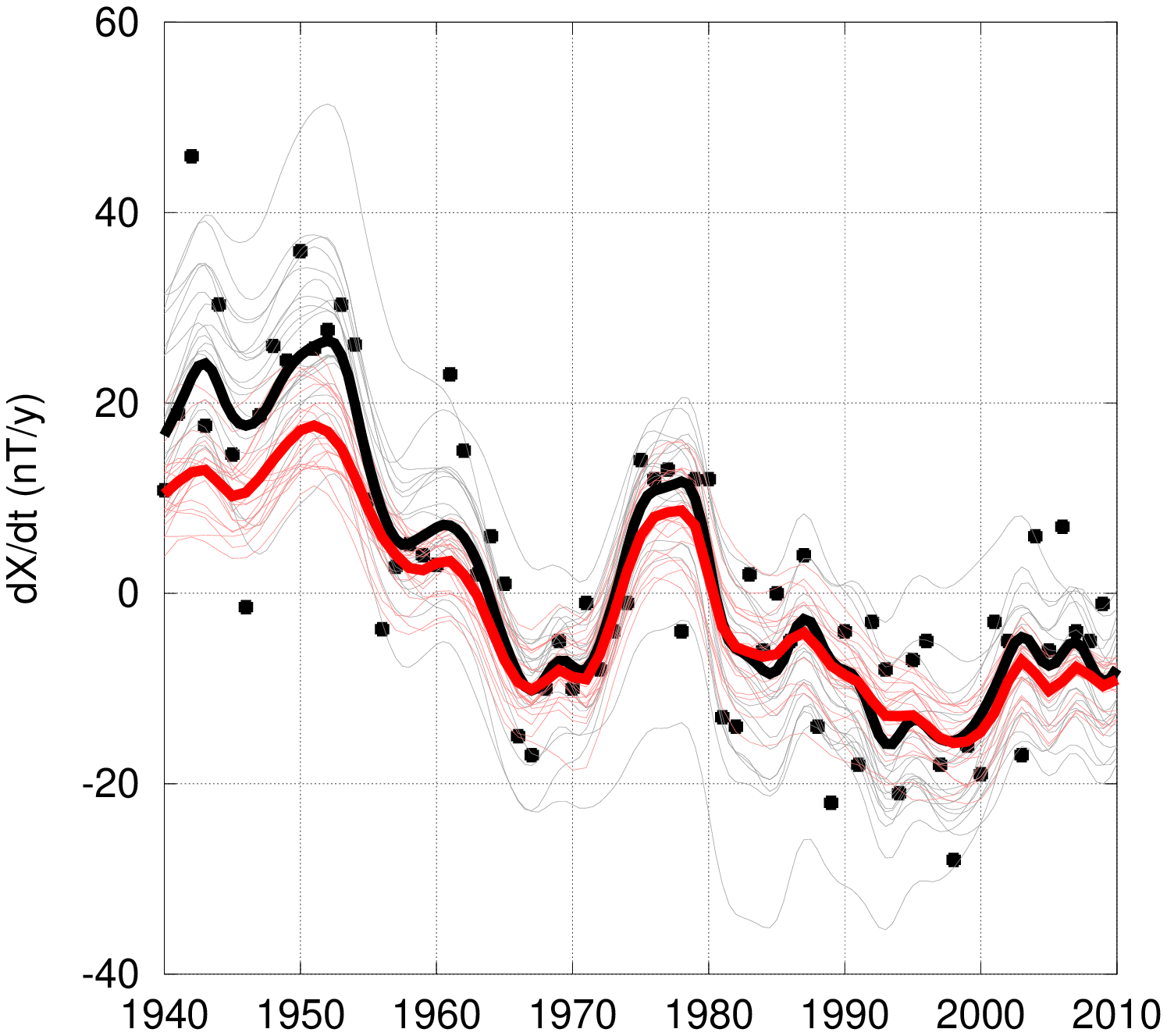}
\hspace*{-2cm}
 \noindent\includegraphics[width=20pc]{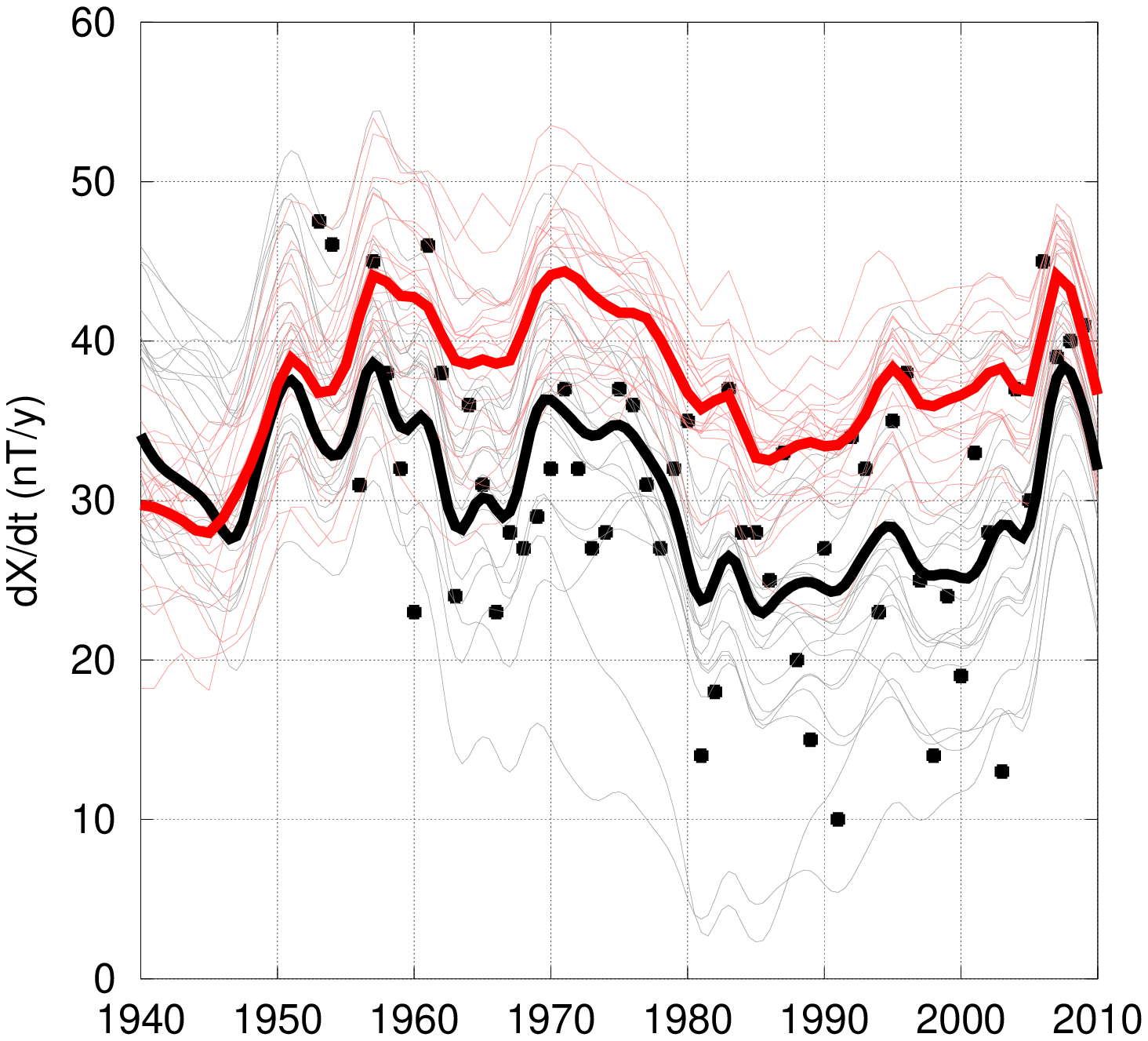}
\hspace*{-2cm}
 \noindent\includegraphics[width=20pc]{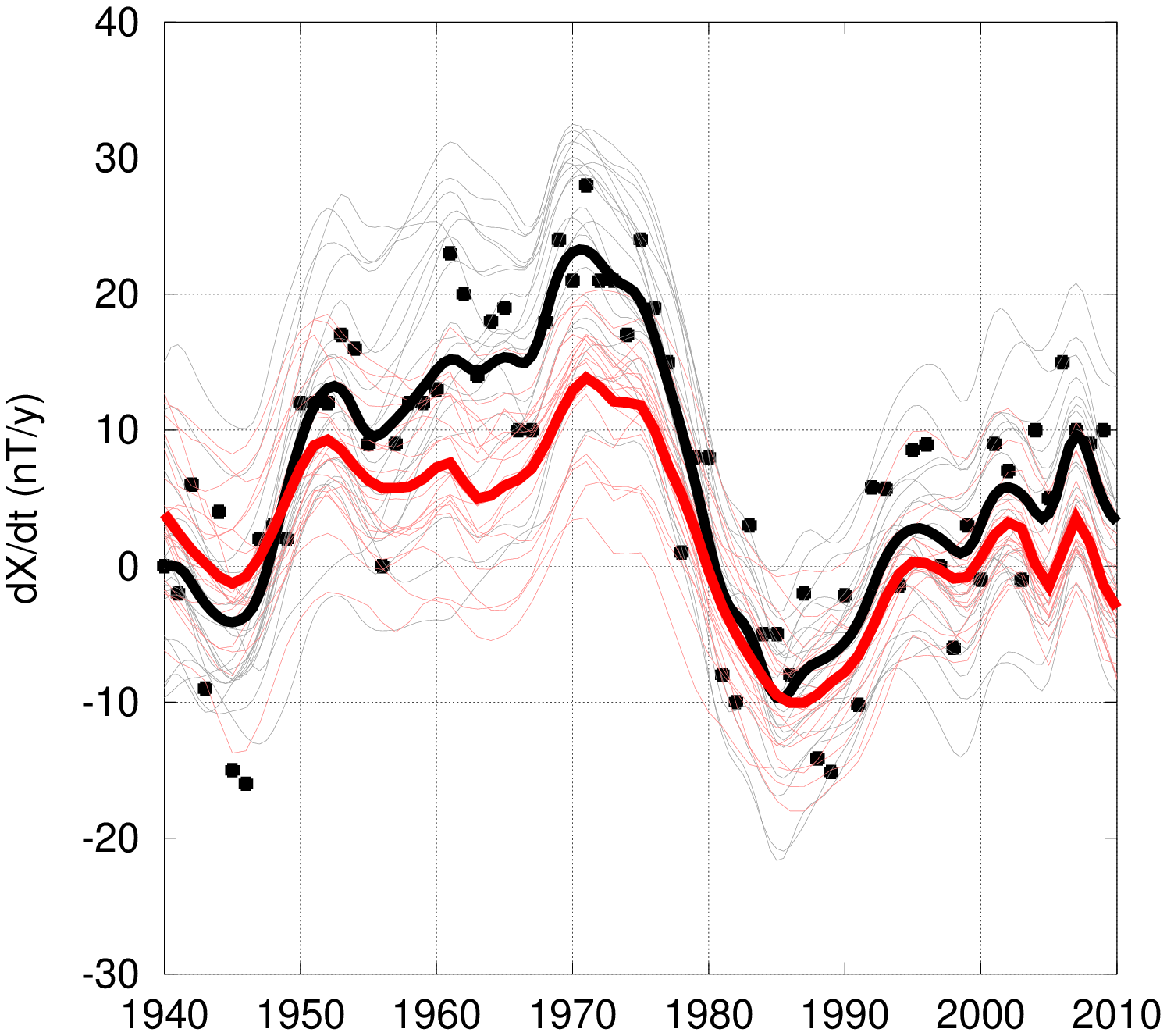}}
\centerline{
 \noindent\includegraphics[width=20pc]{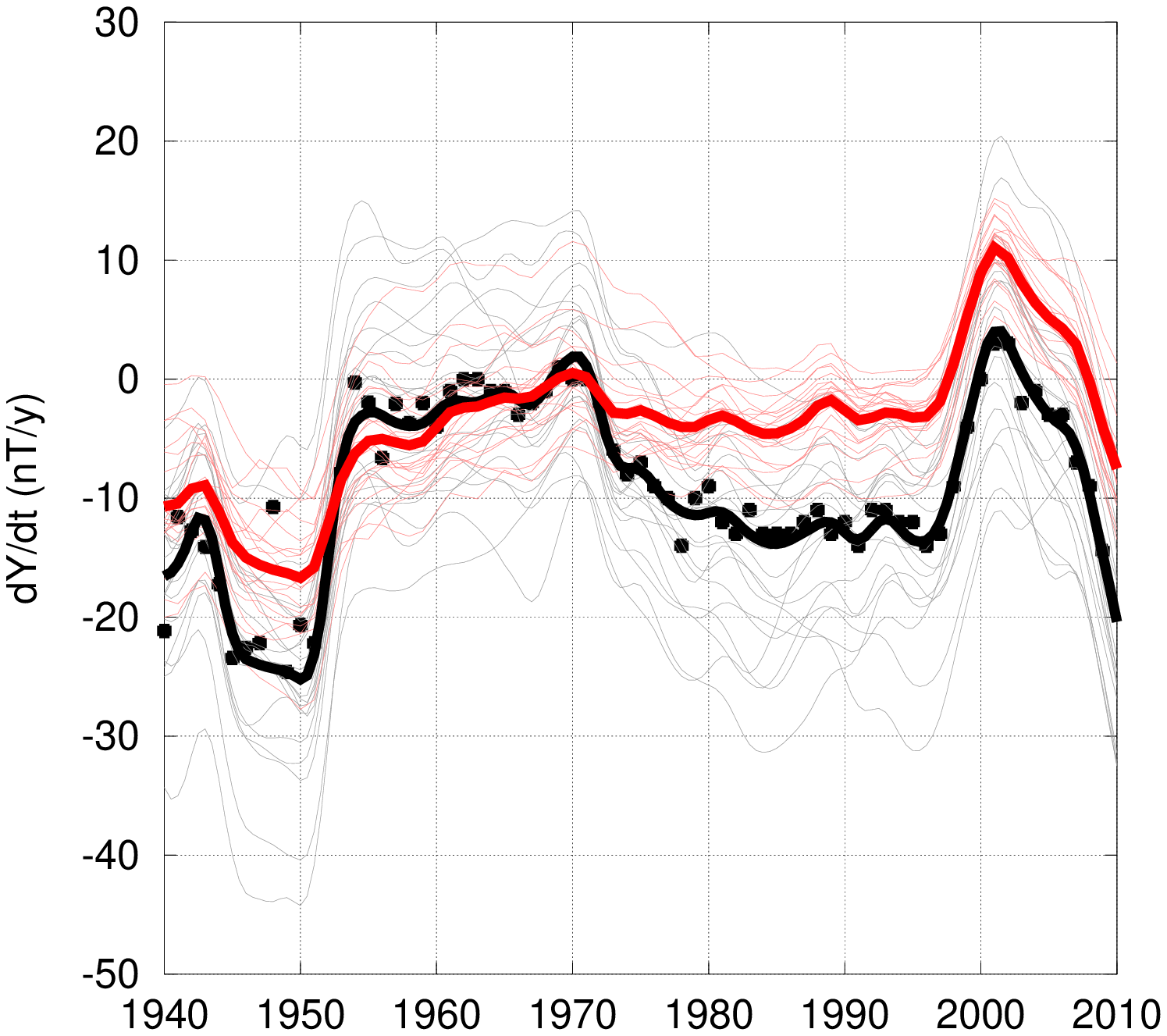}
\hspace*{-2cm}
 \noindent\includegraphics[width=20pc]{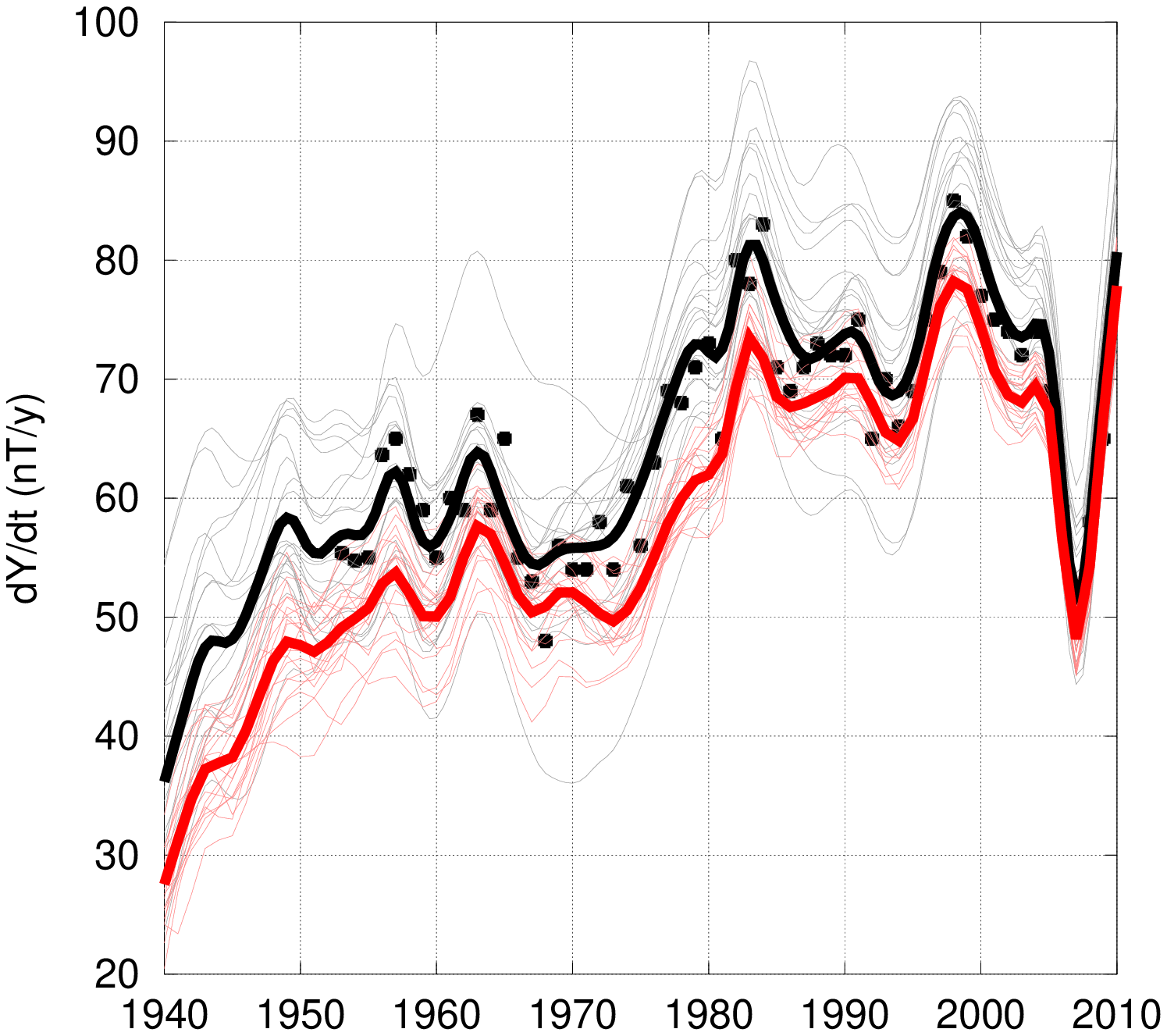}
\hspace*{-2cm}
 \noindent\includegraphics[width=20pc]{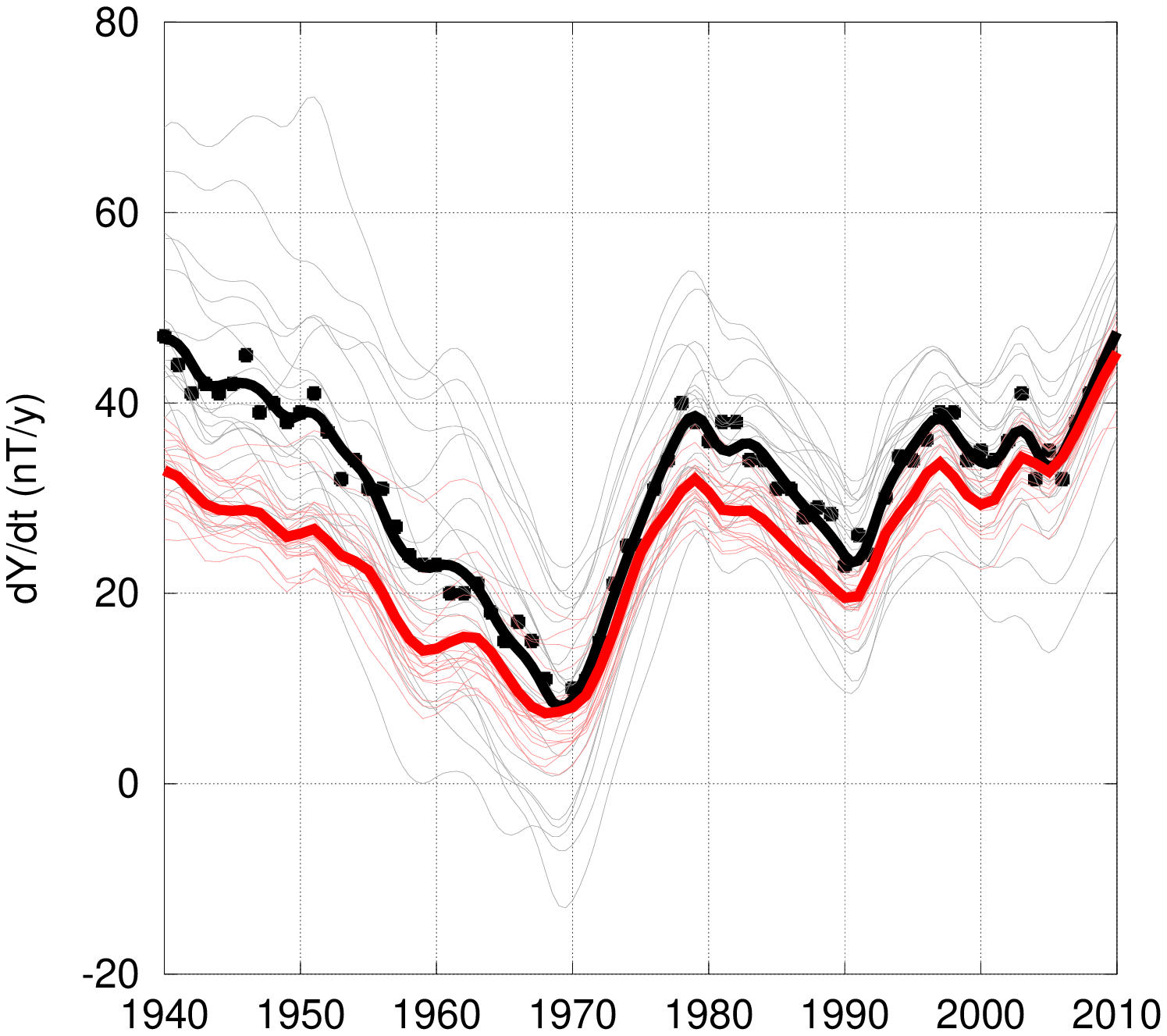}}
\centerline{
 \noindent\includegraphics[width=20pc]{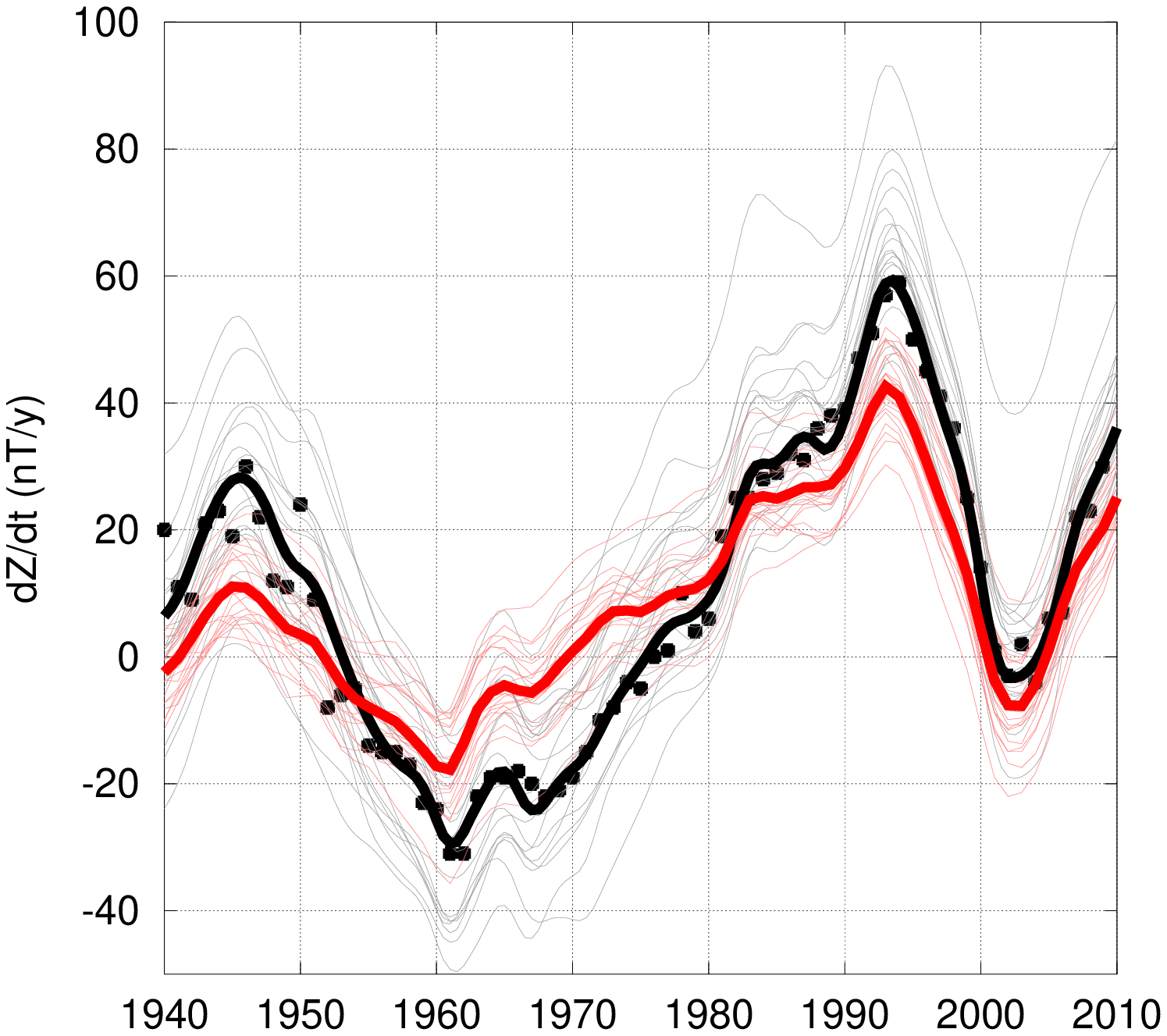}
\hspace*{-2cm}
 \noindent\includegraphics[width=20pc]{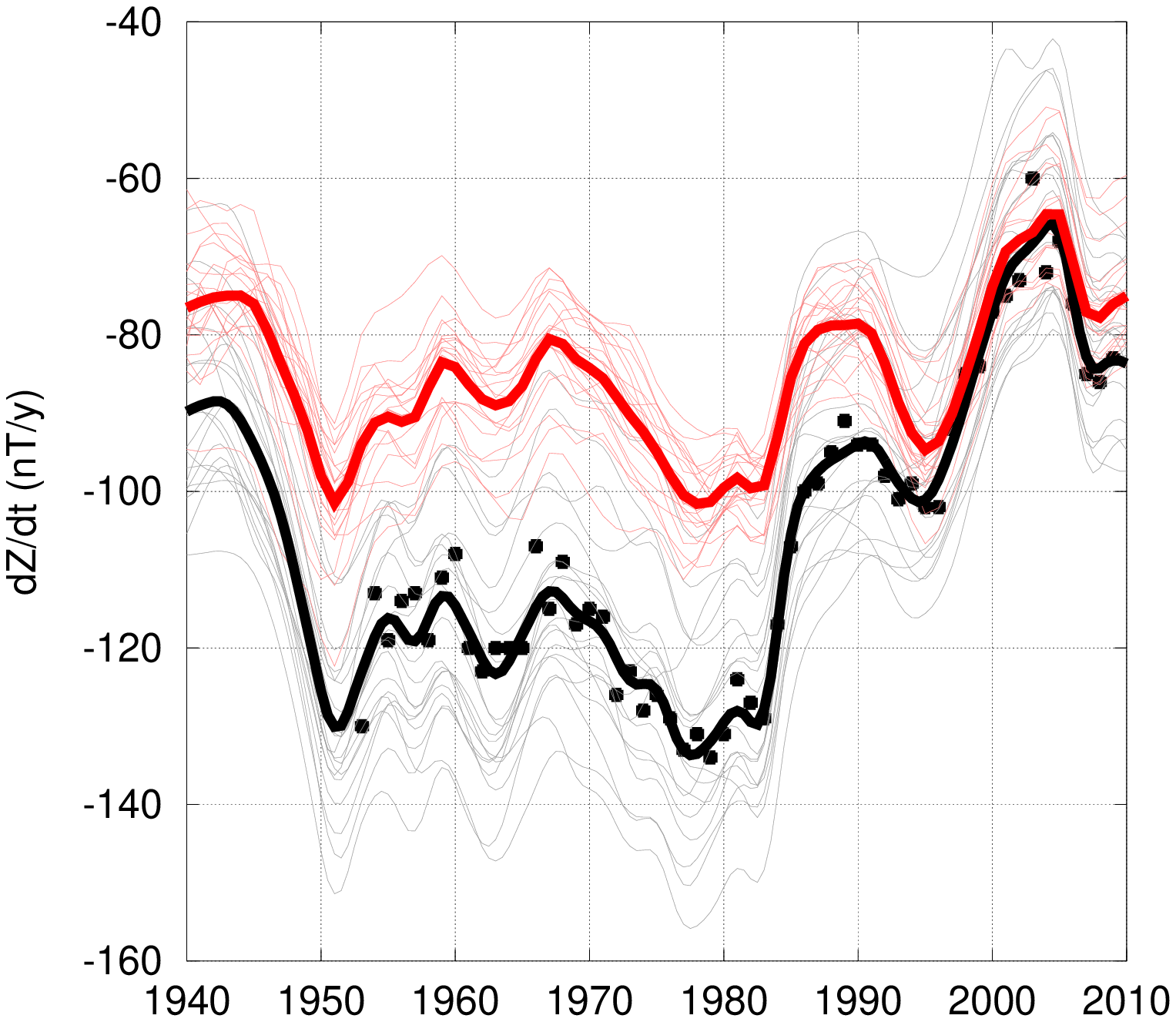}
\hspace*{-2cm}
 \noindent\includegraphics[width=20pc]{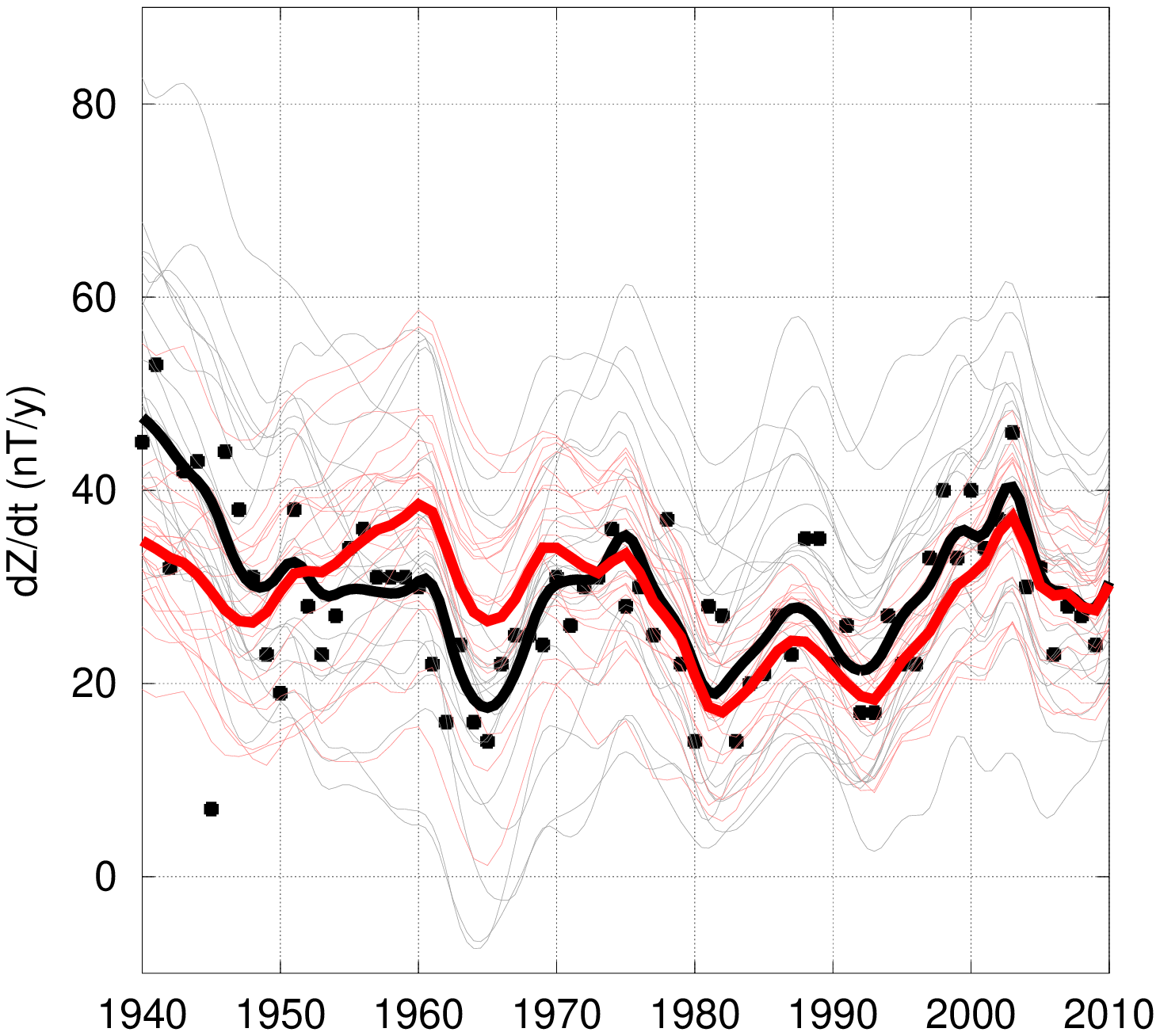}}
 \caption{$X$, $Y$ and $Z$ components (from top to bottom) of the SV observed at the Kakioka (36$^{\circ}$N, 140$^{\circ}$E, left), M'Bour (14$^{\circ}$N, 17$^{\circ}$W, middle) and Niemegk (52$^{\circ}$N, 13$^{\circ}$E, right) observatories (in nT/yr): predictions from our ensemble of flow models (average in red, ensemble in pink), superimposed with SV time series from the {\it COV-OBS} field (in black the ensemble average field model, truncated at degree $n=10$, in grey including the impact of SV model errors) and annual differences of observatory annual means (black symbols).
Flow predictions are again for the preferred case of $\tau_u=100$ years, and accounting for the time correlation of SV model errors.}
\label{fig: SV obs}
 \end{figure}

We focus on degrees 1 and 2 of the SV that show large posterior errors compared to the prior errors. 
Analyzing SV coefficients individually we find anomalously enhanced misfits for $\dot{g}_1^0,\dot{g}_2^0,\dot{h}_1^1$. 
SV predictions for these coefficients are biased toward zero as illustrated by figure \ref{fig: SV g10} for the axial dipole. Extrema of $| \dot{g}_1^0 |$ (e.g. from 1960 to 1980) are particularly poorly reproduced (see the red curves in the figure \ref{fig: SV g10}). 
The difficulty to reproduce large values for some low degree coefficients (for instance $\dot{g}_1^0$ and $\dot{g}_2^0$ in Figure \ref{fig: SV g10}) explains why SV predictions from our flows have slightly yet systematically lower amplitude than SV measured at observatories, especially when the magnitude of the SV is high (e.g., $dY/dt$ at Niemegk or $dZ/dt$ at Kakioka and M'Bour, etc. in Figure \ref{fig: SV obs}).  
The biases are less pronounced for recent epochs (cf Figures \ref{fig: SV g10} and \ref{fig: SV obs}), due to the improved accuracy of SV data.
In contrast with the slow changes, the high frequency fluctuations are very well reproduced, once we account for time-correlated SV model errors.

Note that these results are independent of the imposition of the equatorial symmetry constraint. 
 If we omit the time correlation of SV model errors (see the blue curve in figure \ref{fig: SV g10}) or if we artificially decrease these errors for $\dot{g}_1^0,\dot{g}_2^0,\dot{h}_1^1$, we are able to better fit these coefficients, which indicates that the topological constraints are not the origin of the difficulty.
Furthermore, looking at the spatial distribution of residuals at the core-mantle boundary (not shown), we find no evidence for any particular region displaying preferentially large SV misfit. 
Despite the improvements we have introduced for the calculation of the SV model errors, a bias in the distribution of SV residuals persists. To some extent, this is unavoidable if
the prior distributions of flow coefficients are all centered around 0.
The optimization scheme then preferentially selects models with SV predictions biased towards zero, within the specified error bars (see the tutorial example in Appendix \ref{sec: tutorial}). 
Unfortunately, the relatively large amplitude of the SV model errors for the low degree coefficients exacerbates this effect. 
In section \ref{sec: discussion} we discuss possible approaches for handling this issue.

In Figure \ref{fig: SV obs} we compare the SV predicted by our ensemble of flow models with annual differences of observatory annual means in Kakioka, Niemegk and M'Bour.
We observe a larger dispersion of the ensemble at earlier epochs: as noted by \cite{pais08}, the more knowledge we have about the observed SV (at recent epochs), the smaller SV model errors become. 
Nevertheless, within this dispersion, the interannual changes are rather coherent between realizations. 
Interannual changes can be recovered, despite the relatively large SV model errors, due to the  consideration of time-correlations  in ${\sf C_e}$ (see table \ref{tab:diagnostics}). 
Of course, the bias mentioned above in the spherical harmonic domain also maps into the predictions at observatory locations (see figure \ref{fig: SV obs}) albeit to a smaller extent than in the study of \cite{gillet09}. 

\section{Time evolution of the core flow}
\label{sec:results flow}

We now present in detail our preferred probabilistic ensemble of core flow models. 
We first analyse their resolution within the ensemble as a function of frequency and wave number (section \S\ref{sec: time changes}). 
Next we describe the structure of the resolved flow and its time variations (section \S\ref{sec: geometry}). 
In section \S\ref{sec: TW}, we discuss the fit to changes in the LOD and put forward an interpretation of the geostrophic circulation as the superposition of a slowly varying flow determined through the Taylor's condition and torsional waves.  
We revisit in section \S\ref{sec: EM coupling} electro-magnetic coupling between the core and the mantle. 
Finally we focus in \S\ref{sec: eq} on the dynamics of the equatorial region. 

\subsection{Resolution of the calculated core flows}
\label{sec: time changes}

\begin{figure}
\centerline{
 \noindent\includegraphics[width=30pc]{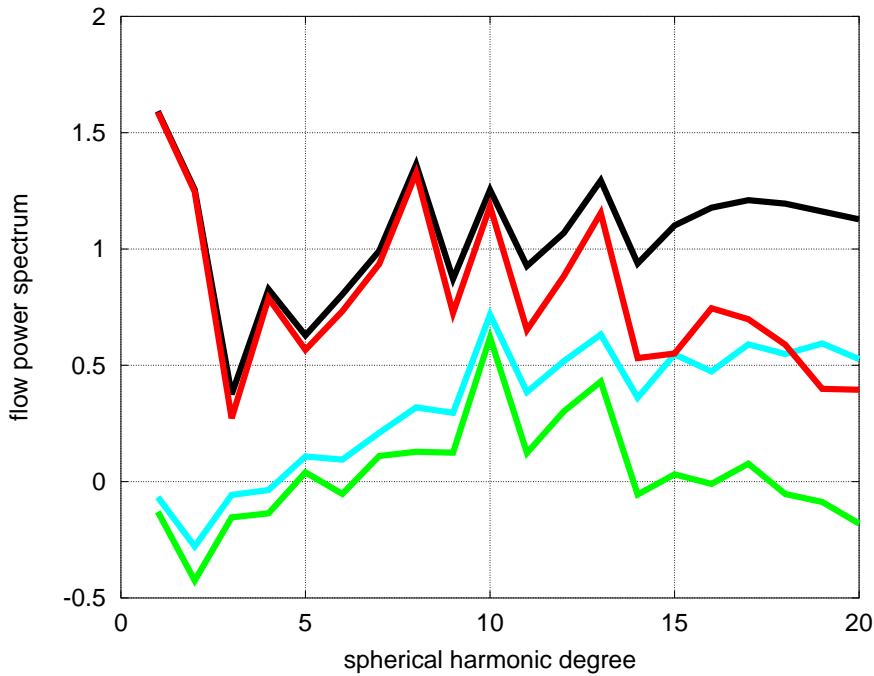}}
 \caption{Spatial power spectra of the core surface flow, time-averaged over 1940--2010 (scale in units of $\log_{10}$, in (km/yr)$^2$): 
ensemble average of the spectra for the total flow models (black), their time-dependent components only (cyan); 
spectra of the ensemble average flow model (red) and its time-dependent component (green).}
 \label{fig: flow spectra}
 \end{figure}

The spatial structure in our ensemble of flow models  is predominantly steady throughout the studied interval of 1940--2010.
By way of illustration, the correlation coefficient of core surface velocity maps calculated for pairs of epochs 10 years apart remains very close to $0.93$ during $[1945, 2005]$. Assuming that the correlation function has the form (\ref{eq:corr-AR1}),
this corresponds to a correlation time of at least $140$ years.
Remarkably, the time-averaged flow also includes significant small scale constituents, as can be seen from
the time-averaged spatial power spectrum of the flow 
\begin{equation} 
\displaystyle
{\cal S}_{{\bf u}}(\ell) = \frac{1}{t_e-t_s} \int_{t_s}^{t_e}\frac{\ell(\ell+1)}{2\ell+1} 
\sum_{m=-\ell}^\ell 
\left( {t_{\ell m}}(t)^2+{s_{\ell m}}(t)^2 \right) dt\, .
\label{spatial spectra}
\end{equation}
Similarly we define ${\cal S}_{{\bf u}'}(\ell)$ to be the spectrum for the time-dependent component of the flow, after the time-averaged part has been removed. 
In Figure \ref{fig: flow spectra} we display the ensemble average of the two power spectra, $\left<{\cal S}_{{\bf u}}\right>$ and $\left<{\cal S}_{{\bf u}'}\right>$, together with the power spectra for the ensemble averages, ${\cal S}_{\left<{\bf u}\right>}$ and ${\cal S}_{\left<{\bf u}'\right>}$. 
We find the flow coefficients are resolved within the ensemble up to degree about 10, above which $\left<{\cal S}_{{\bf u}}\right>$ departs gradually from ${\cal S}_{\left<{\bf u}\right>}$. 
There is likely some useful information on the time average flow up to degree 13.
A similar scale of spatial resolution is obtained for the time-dependent component of the flow when comparing $\left<{\cal S}_{{\bf u}'}\right>$ with ${\cal S}_{\left<{\bf u}'\right>}$, even though large length-scales seem slightly less resolved for the flow fluctuations than for its time average. 
The energy of the time-variable part of the flow peaks at degree 10, above which there are larger uncertainties on the flow coefficients. Thus, the time-variable flow is predominantly small-scale.

In order to obtain a more detailed measure of the flow resolution as a function of harmonic degree and frequency $f$, we introduce the quantity 
\begin{equation} 
\displaystyle
{\cal C}_{{\bf u}}(\ell,f) = \frac
{\displaystyle\left<\sum_{m=-\ell}^\ell 
|\tau_{\ell m}(f)-\left<\tau_{\ell m}(f)\right>|^2+|\sigma_{\ell m}(f)-\left<\sigma_{\ell m}(f)\right>|^2\right>}
{\displaystyle\left<\sum_{m=-\ell}^\ell |{\tau_{\ell m}}(f)|^2+|{\sigma_{\ell m}}(f)|^2\right>}\,,
\label{resolution}
\end{equation}
where the complex coefficients $(\tau_{\ell m},\sigma_{\ell m})$ are the Fourier transforms of the flow coefficients time series $t_{lm}(t), \, s_{lm}(t)$. 
This quantity is 1 (resp. 0) when the spread in the flow harmonic coefficients reaches 100\% (resp 0\%) of the model variances.  
A low value of ${\cal C}_{{\bf u}}$ is a necessary but not sufficient condition for the true core flow to be captured. 

\begin{figure}
\centerline{\noindent\includegraphics[width=30pc]{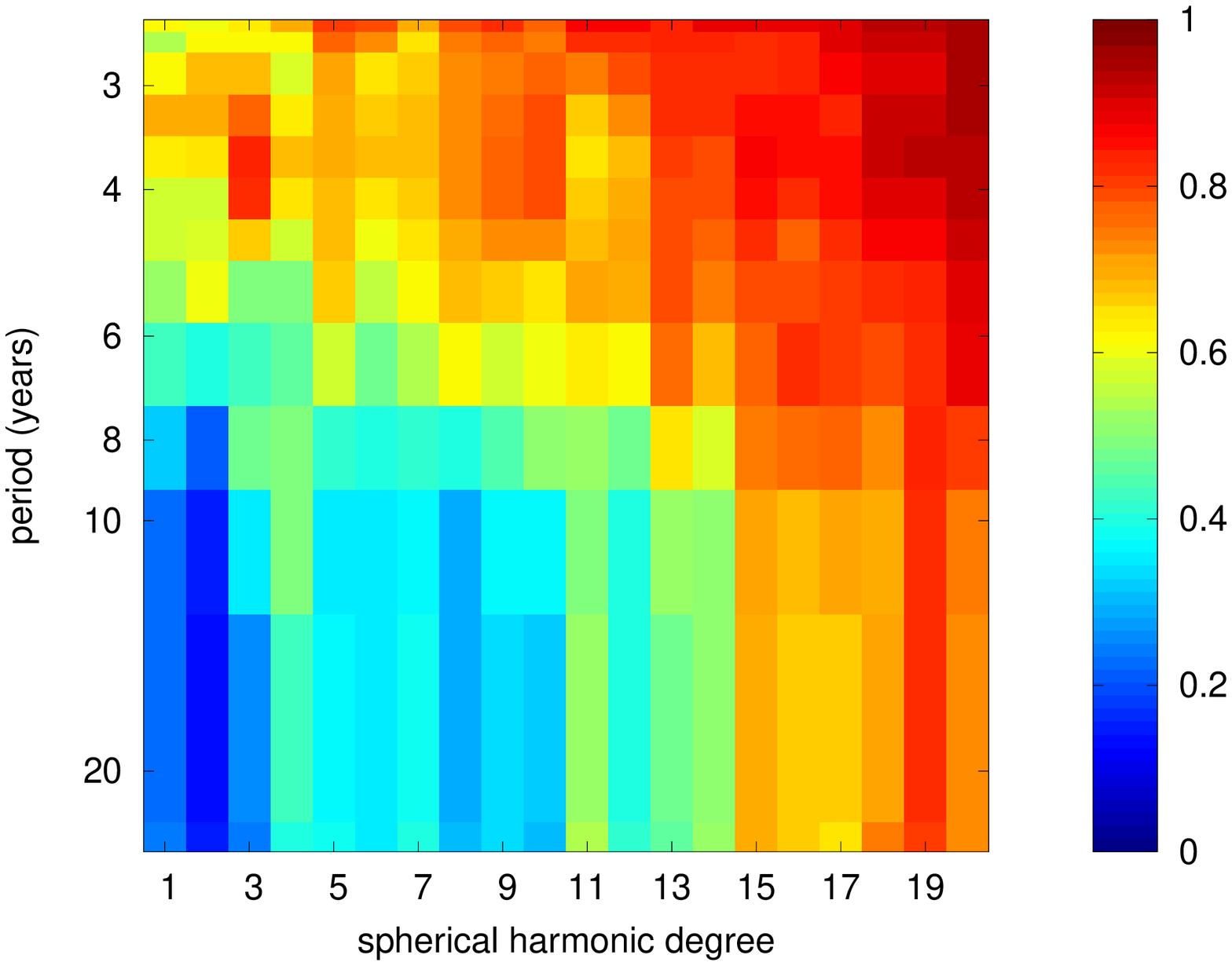}}
\centerline{\noindent\includegraphics[width=30pc]{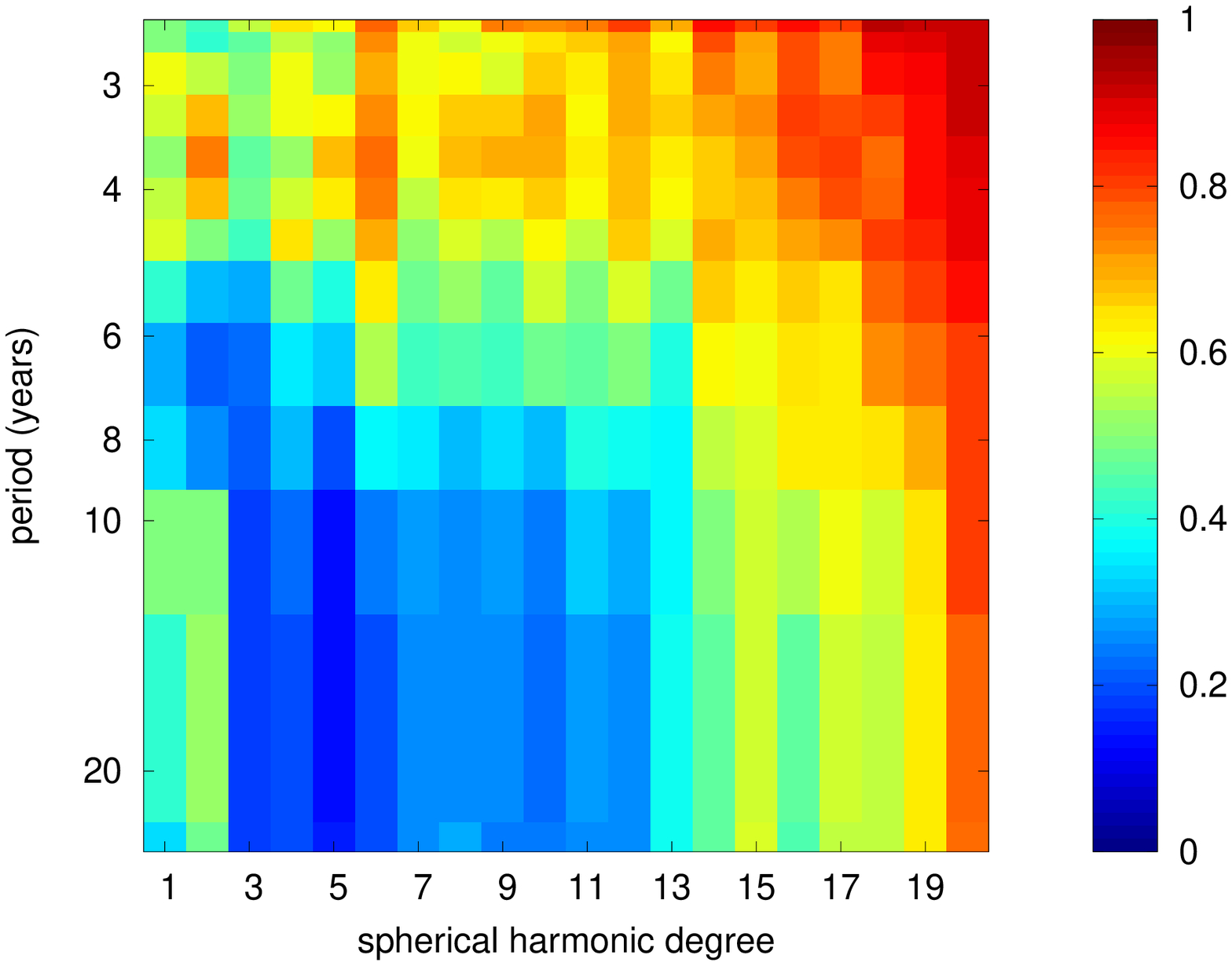}}
 \caption{Resolved features in the ensemble of flows as a function of period and spherical harmonic degree, as defined with equation (\ref{resolution}), for the two time intervals 1940--1975 (top)  and 1975--2010 (bottom).}
 \label{fig: flow resolution}
 \end{figure}

Distributions of ${\cal C}_{{\bf u}}(\ell,f)$ are shown in Figure \ref{fig: flow resolution} for the two intervals 1940--1975 and 1975--2010. 
We observe a clear improvement at recent epochs, with well resolved flows at both shorter periods and smaller length-scales.
Indeed, for the most recent time interval, flow variations with a period of about 6 years are well resolved up to degree 6, with some information provided up to degree 12, whereas slower decadal variations are well constrained until degree 12, with some information provided up to degree 16. 
Though resolution is poorer for older epochs, some interannual changes are constrained even during the first 35 years. 
Again, we wish to emphasize that our ability to reliably retrieve interannual flows is a direct consequence of accounting for time-correlated SV model errors (see Appendix \ref{sec: tutorial}).

 \subsection{Planetary gyre and and mid-latitude eddies}
\label{sec: geometry}

Under the incompressible QG hypothesis (see section \ref{sec:def-QG}), the flow in the whole volume can be represented through a stream function $\psi(s,\phi)$ \citep[eq. 14 and 15]{TOG8Jault13,canet14}:
\begin{equation}
{\bf u}(s,\phi,z)=\frac{1}{H}\nabla\times\left(\psi{\bf 1}_z\right)
-\frac{z}{H^3}\frac{\partial\psi}{\partial \phi}{\bf 1}_z\,.
\label{ue psi}
\end{equation}
(${\bf 1}_s,{\bf 1}_{\phi},{\bf 1}_z$) are the unit vectors in cylindrical polar coordinates. 
$H(s)=\sqrt{c^2-s^2}$ is the half-height of a geostrophic cylinder with $c$ the outer core radius.
We give in Appendix \ref{sec: app-psi} the relation between coefficients describing the stream function and the toroidal and poloidal coefficients. 
The fluid flow in the equatorial plane, the first term on the r.h.s. of (\ref{ue psi}), is parallel to the isolines of $\psi$, but its intensity is not proportional to the density of curves.

\begin{figure}
\centerline{
 \noindent\includegraphics[height=15pc]{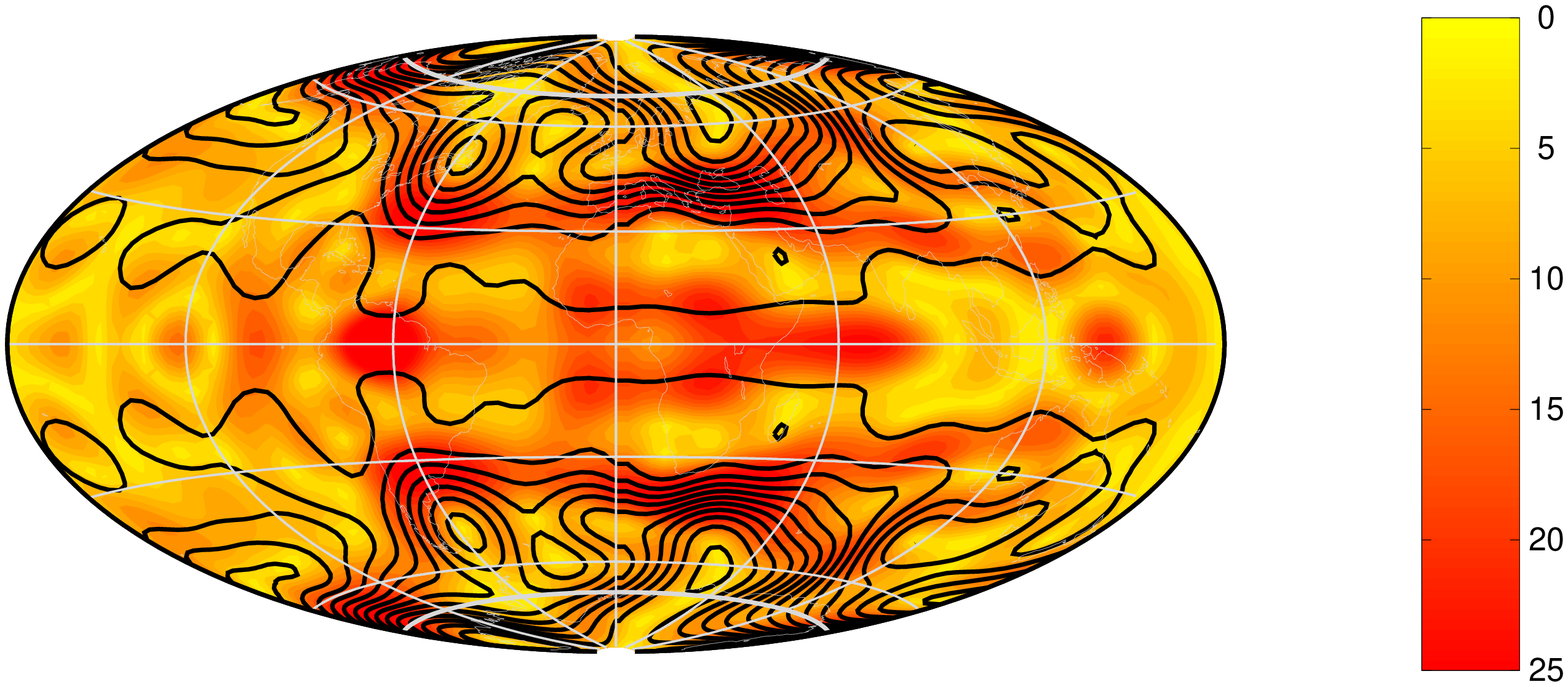}
\hspace*{-3cm}
 \noindent\includegraphics[height=15pc]{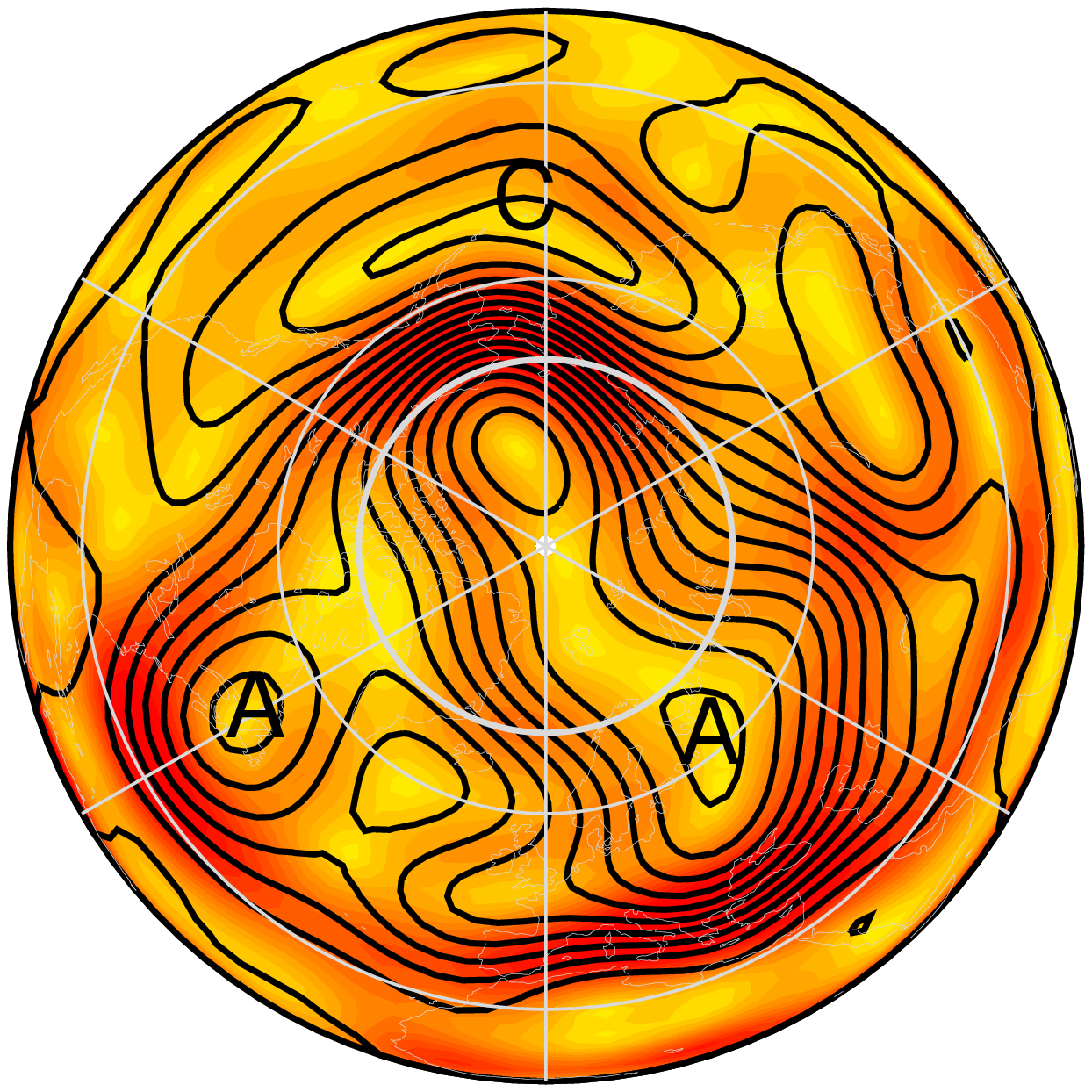}
}
\centerline{
 \noindent\includegraphics[height=15pc]{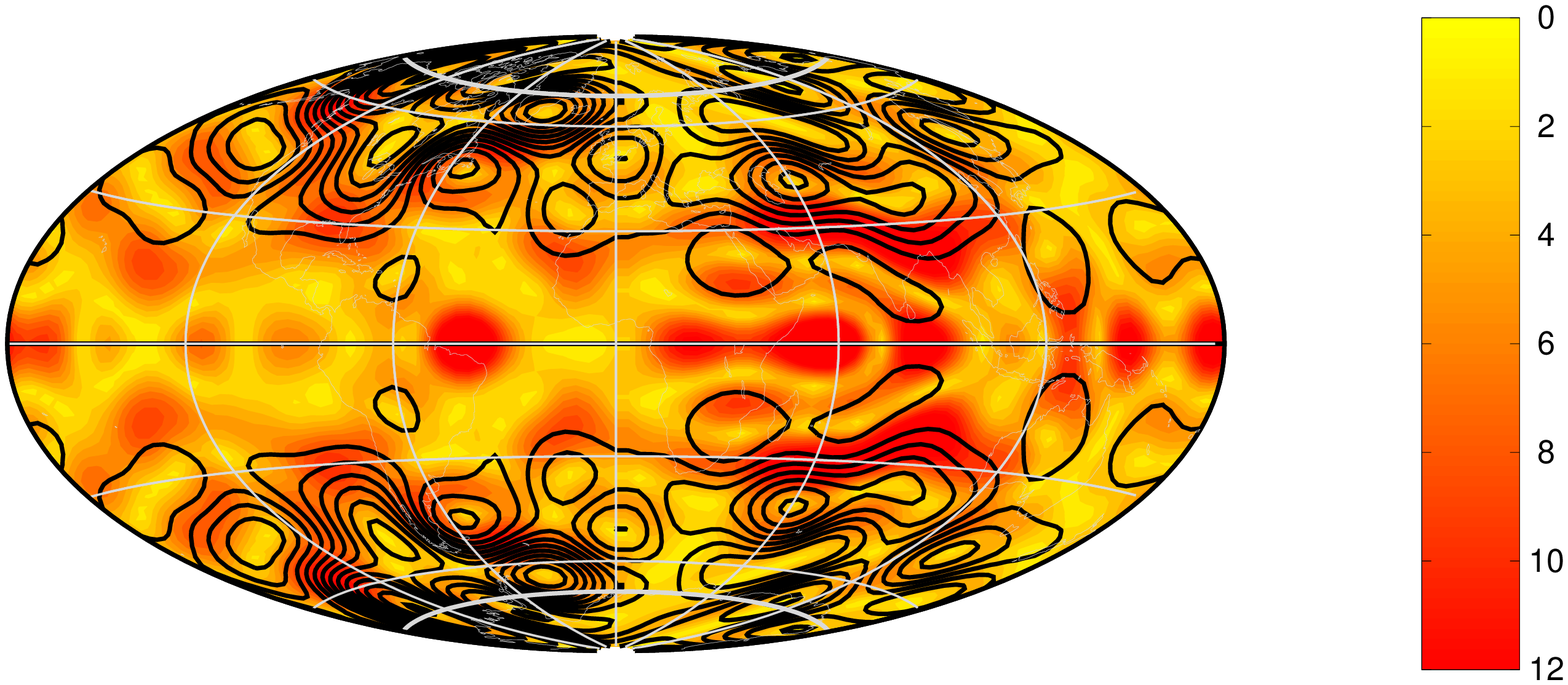}
\hspace*{-3cm}
 \noindent\includegraphics[height=15pc]{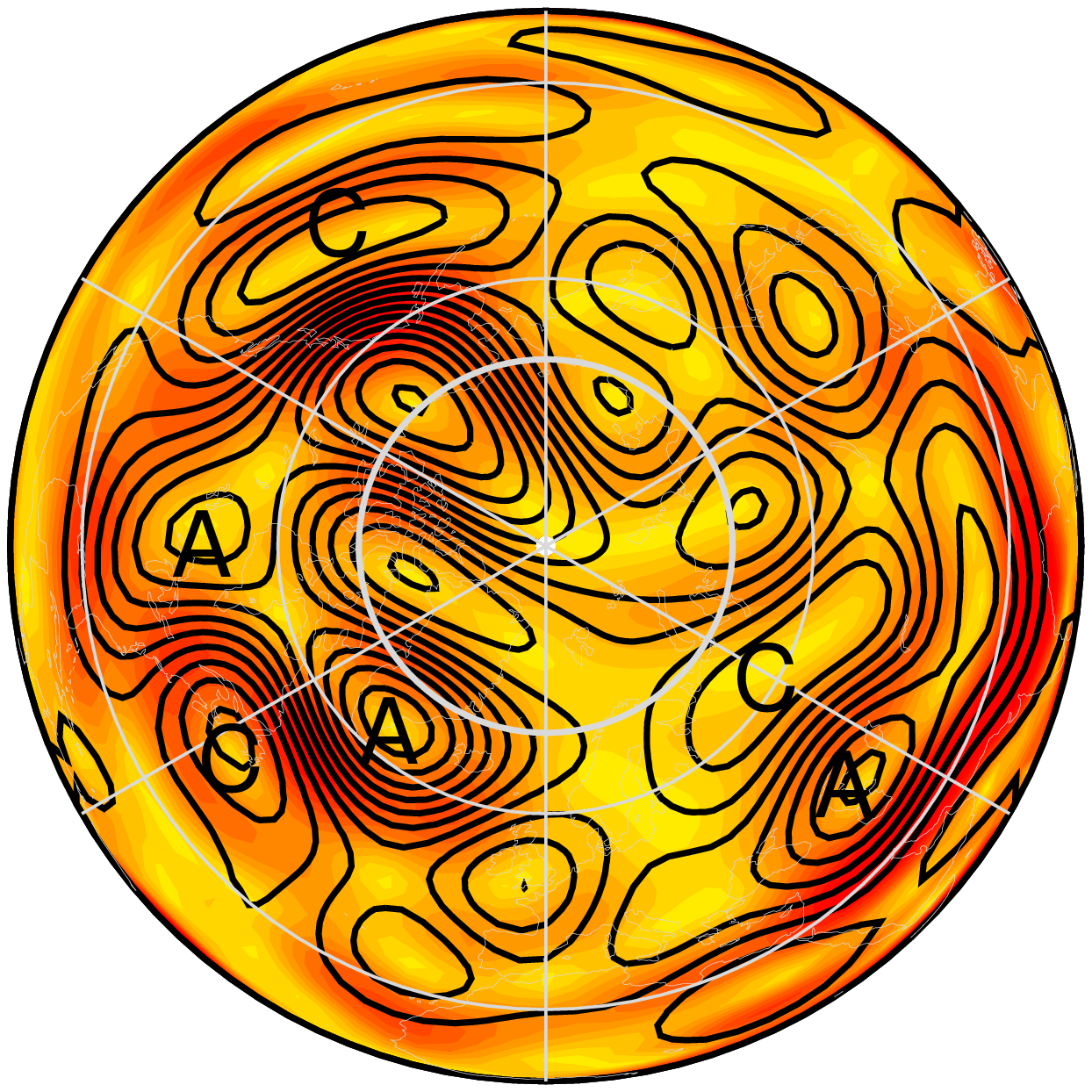}
}
 \caption{
Maps of the quasi-geostrophic stream function $\psi$ (black isolines) and norm of the velocity (colourscale, in km/y) at the CMB (left, Hammer-Aitoff projection centered on the Greenwhich meridian) and in the equatorial plane (right). 
Meridians (parallels) are marked every 60$^{\circ}$ (30$^{\circ}$).
The thick grey parallel corresponds to the projection of the tangent cylinder at the CMB. 
Top:  the time-average flow between 1940--2010. 
Bottom: an example of the flow anomaly with respect to the stationary flow in epoch 2005.  
In both cases the flow has been truncated at spherical harmonic degree 14.
All figures are for the ensemble average of the flow models. 
Blue capital letters `A' and `C' on equatorial maps stand respectively for the anticyclones and cyclones discussed in the text. 
}
 \label{fig: t-ave flow}
 \end{figure}

In Figure \ref{fig: t-ave flow} (top), we present maps at the CMB and in the equatorial plane of the time-average flow showing a planetary scale gyre similar to that described in earlier studies \citep{pais08,gillet09,aubert13}. 
We find that the gyre possesses a detailed structure, in agreement with the findings of \cite{amit13} for flows based on the incompressible QG hypothesis. 
The most conspicuous features within the gyre are two anticyclones centered at (45$^{\circ}$E, 60$^{\circ}$N,S) and (60$^{\circ}$W, 45$^{\circ}$N,S) at the core surface.
Large velocities are observed where the flow is in the azimuthal direction.
In particular, the westward flow in the Atlantic hemisphere is split into two branches, one within $\pm 10^{\circ}$ latitude at the equator, and the other at latitudes from 30 to 45$^{\circ}$. 
These are separated by a region of lower azimuthal velocity.

\begin{figure}
\centerline{
 \noindent\includegraphics[width=30pc]{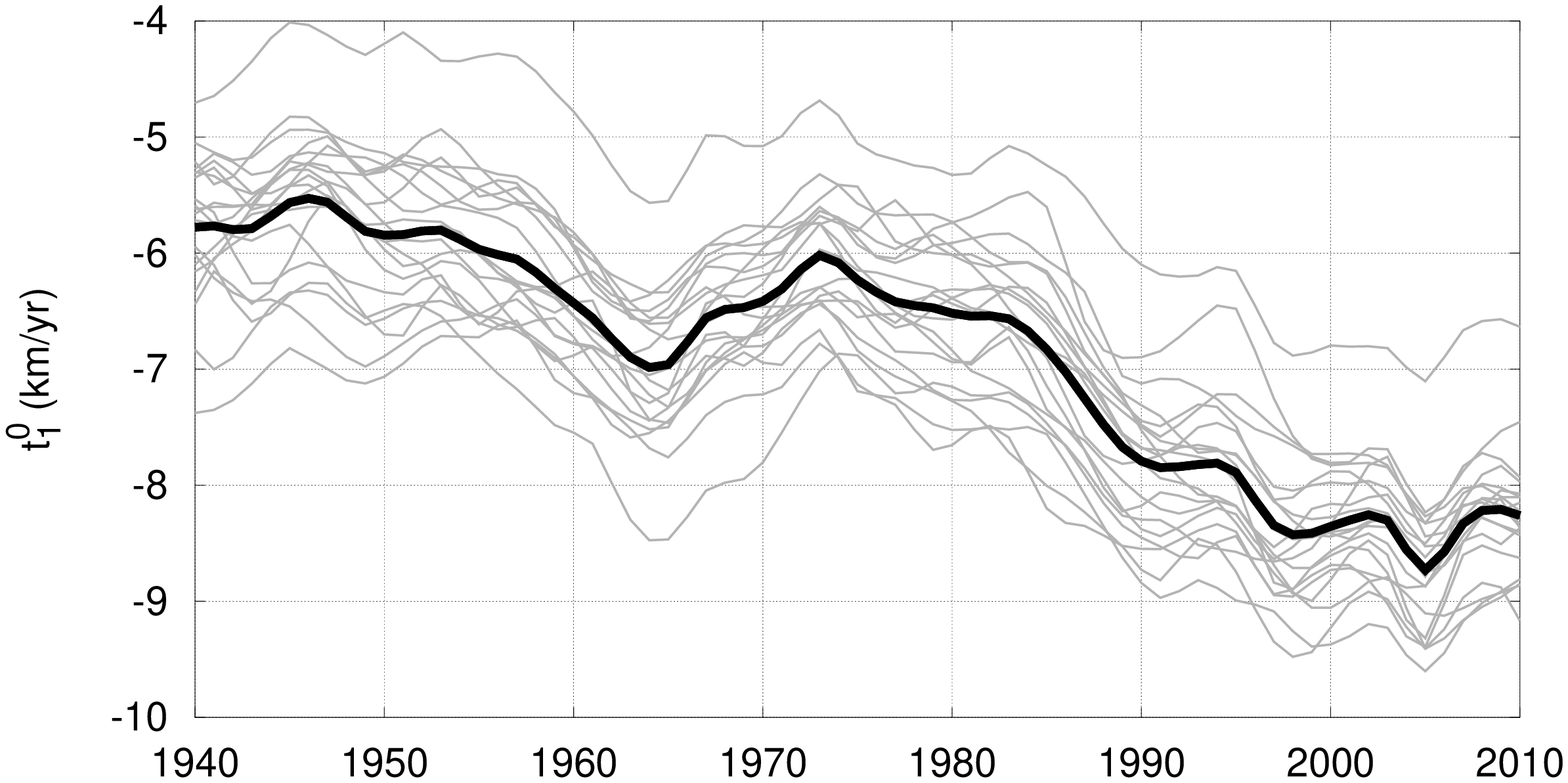}}
\centerline{
 \noindent\includegraphics[width=30pc]{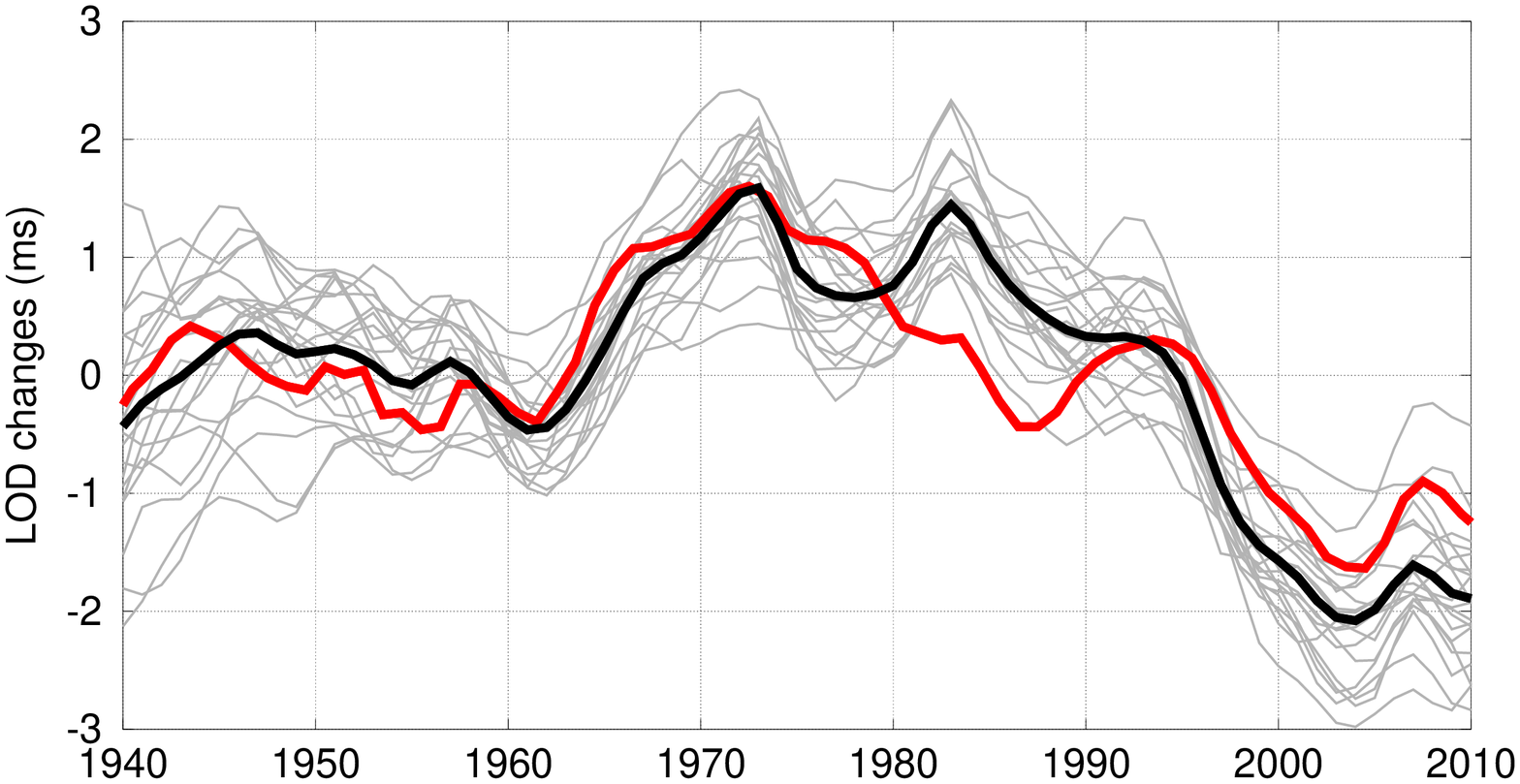}}
\centerline{
 \noindent\includegraphics[width=30pc]{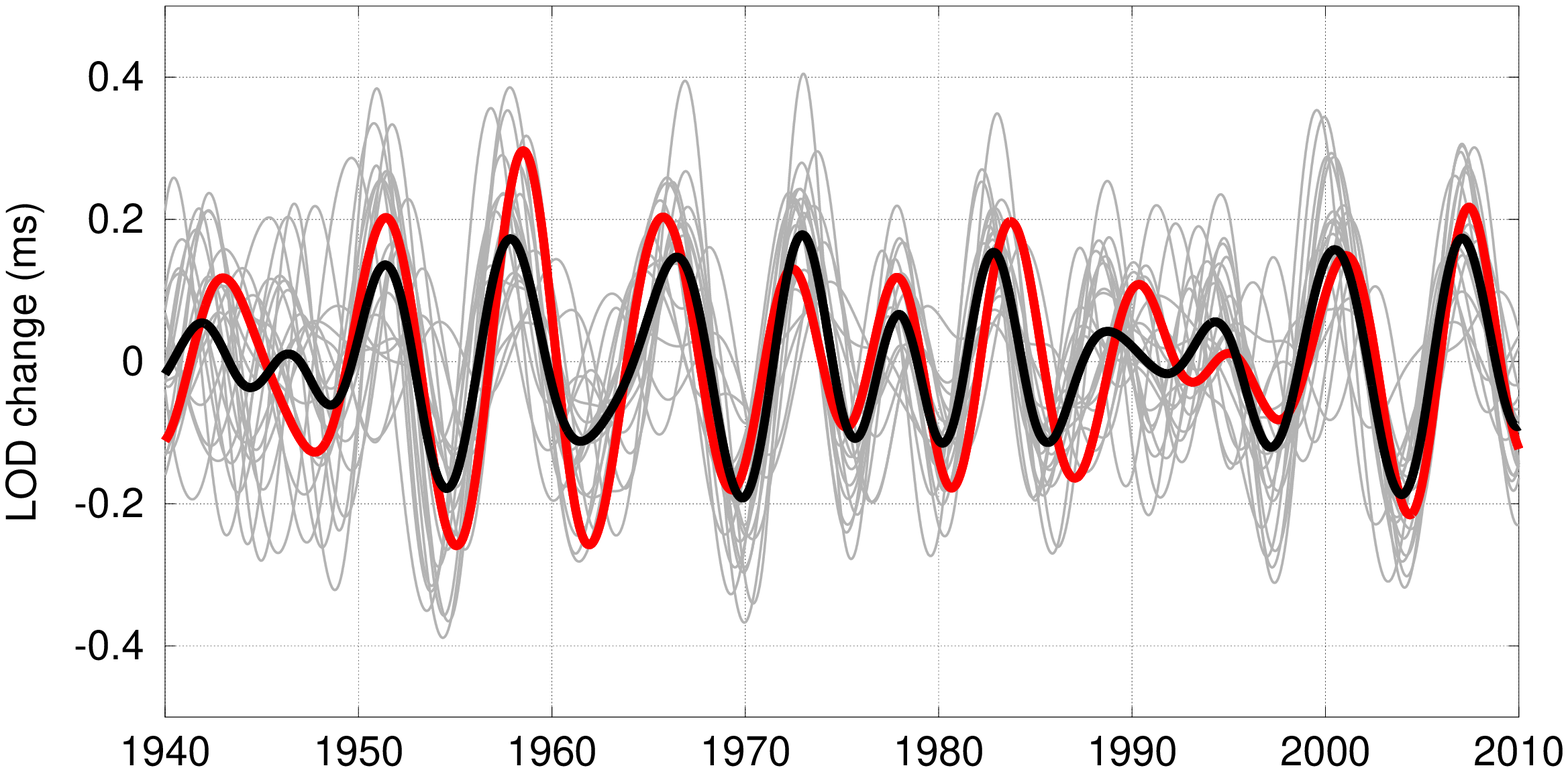}}
 \caption{Top: flow coefficient $t_1^0$ (in km/yr) for the ensemble of flow models (grey) and the ensemble average flow model (black). 
Middle and bottom: comparison between LOD predictions (in ms) from all members of the ensemble of flow models (grey), their ensemble average (black) together with the observed LOD changes (red): (middle) total LOD, with all individual LOD time-averages set to zero, and (bottom) LOD band-pass filtered between 4 and 9.5 years.}
 \label{fig: LOD}
 \end{figure}

In addition to the time-averaged gyre structure, we also observe time-dependent features at decadal time scales. 
There is a general increase in the westward solid-body rotation from 1940 onward (see Figure  \ref{fig: LOD}, top). 
Assuming that the system of solid Earth (crust, mantle and core) is closed, we find that our ensemble of geostrophic flows, of the form $u_G(\theta,t)$ at the core surface, account rather well for the observed decadal changes in the length-of-day (Figure \ref{fig: LOD}, middle). 
A correlation coefficient $r_{\gamma}$ close to 0.9 is found between the data and the predictions at decadal periods (see Table 1).

We also observe transient non-zonal circulations. 
The flow perturbation in 2005 (see Figure \ref{fig: t-ave flow}, bottom right) enhances anticyclonic eddies centered near (50$^{\circ}$E, 40$^{\circ}$N,S), (40$^{\circ}$W, 65$^{\circ}$N,S) and (90$^{\circ}$W, 45$^{\circ}$N,S), and  cyclonic eddies around ($60^{\circ}$E, 60$^{\circ}$N,S), ($60^{\circ}$W, 45$^{\circ}$N,S) and ($150^{\circ}$W, 45$^{\circ}$N,S). 
These are reminiscent of the two main time-dependent structures isolated by \cite{pais2014variability} on time-scales about 70 years and longer. 
However, the most energetic time-variable flows are non-zonal azimuthal jets located around 30$^{\circ}$ latitude and in the equatorial belt (see figure \ref{fig: t-ave flow}, bottom left). 
If the former appears related to the gyre, the latter are difficult to describe using equatorial projections of the stream function. 
The dynamics in the equatorial region is further discussed in section \ref{sec: eq}.

\subsection{Taylor's state in the Earth's core and excitation of torsional waves}
\label{sec: TW}

\begin{figure}
\centerline{
 \noindent\includegraphics[width=30pc]{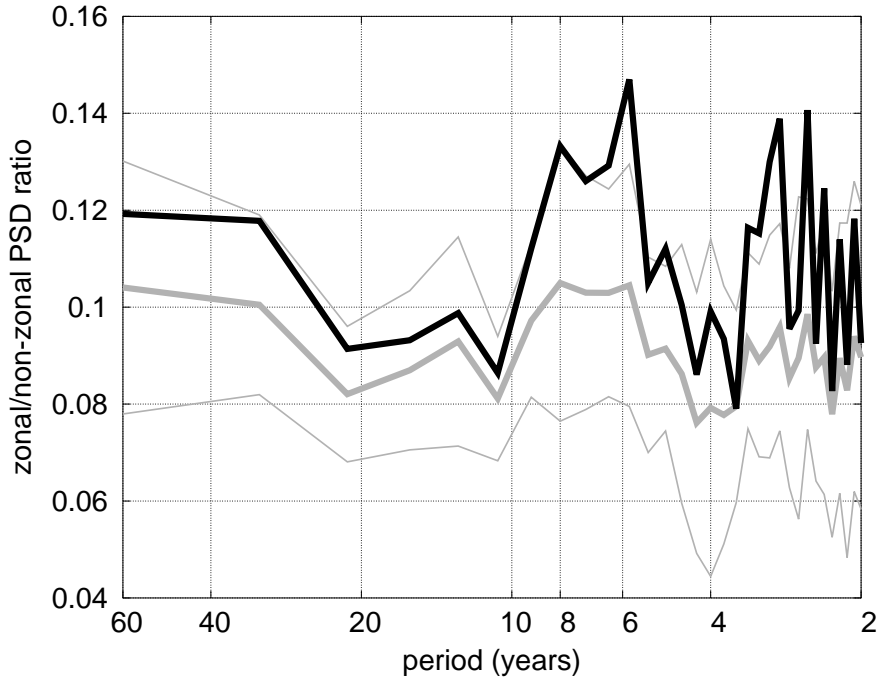}}
 \caption{Ratio ${\cal P}_Z/ {\cal P}_{NZ}$ between the power spectral densities (PSD) for the zonal and non-zonal flows as a function of period. 
In bold black line the ratio for the ensemble average of flow solutions. 
In bold grey the average ratio over the ensemble members (thin grey line: $\pm$ one standard deviation).
Flows are truncated at degree $\ell=14$.}
 \label{fig: psd ratio}
 \end{figure}

Non-zonal flows account for the majority of the energy of time variable flows, as can be inferred from the power spectral densities 
\begin{equation} 
\displaystyle
\displaystyle {\cal P}_Z(f)   =  \displaystyle\sum_{\ell}\frac{\ell(\ell+1)}{2\ell+1}{\cal P}_{t_{\ell 0}} \;\mbox{and}\;
\displaystyle {\cal P}_{NZ}(f)=  \displaystyle\sum_{\ell}\frac{\ell(\ell+1)}{2\ell+1}\sum_{m\neq0}\left({\cal P}_{t_{\ell m}}+{\cal P}_{s_{\ell m}}\right)
\label{PSD ZNZ}
\end{equation}
of respectively zonal and non-zonal motions
(${\cal P}_{t_{\ell m}}$ stands for the power spectral density of the series $t_{\ell m}(t)$, with similar definition for ${\cal P}_{s_{\ell m}}$).
Nevertheless, the ratio $\displaystyle {\cal P}_Z/{\cal P}_{NZ}$ of the zonal to non-zonal kinetic energies shows distinctive spectral bands centered on 6-8 years and around 3 years where zonal motions are relatively more intense (see figure \ref{fig: psd ratio}).   
We shall not attach much importance to the 3 years spectral band  which is not adequately resolved (see figure \ref{fig: flow resolution}), but the well-resolved peak at 6-8 years provides an indication that there may be torsional waves present in the derived core flows.

\begin{figure}
\centerline{
 \noindent\includegraphics[width=40pc]{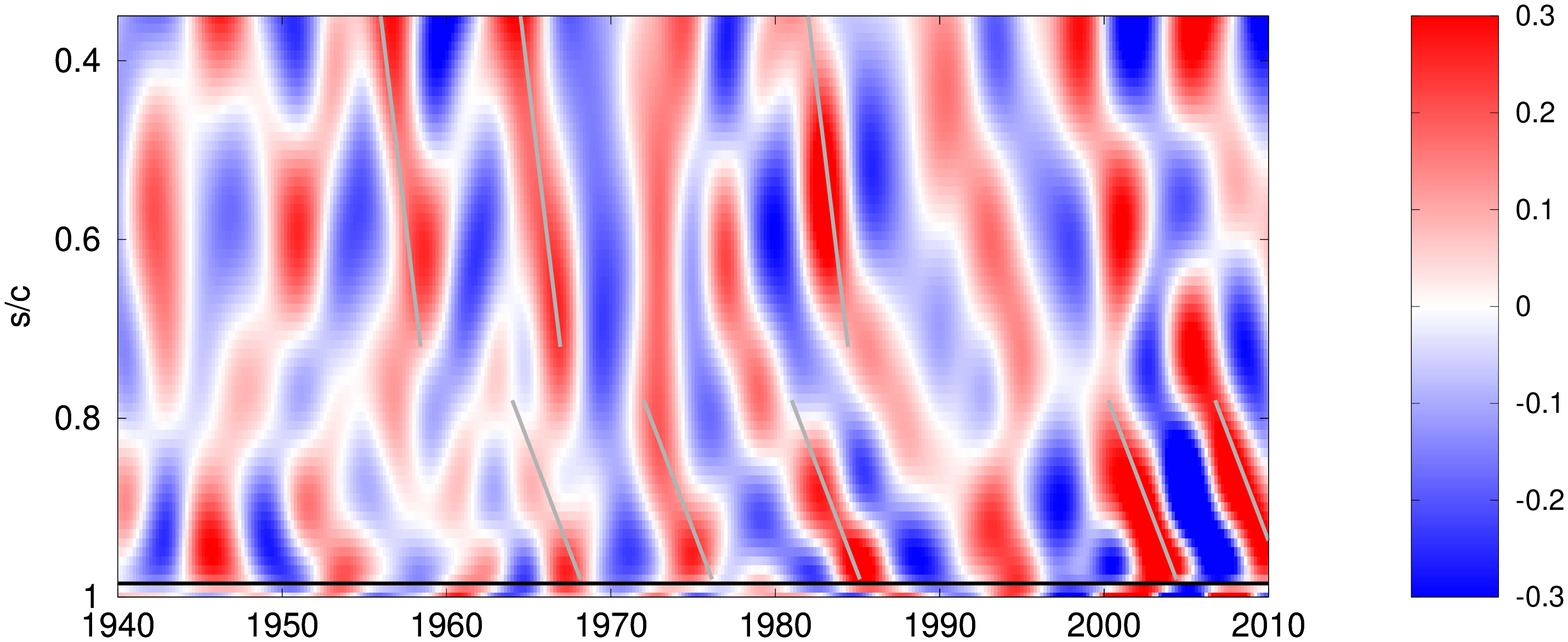}}
\centerline{
 \noindent\includegraphics[width=40pc]{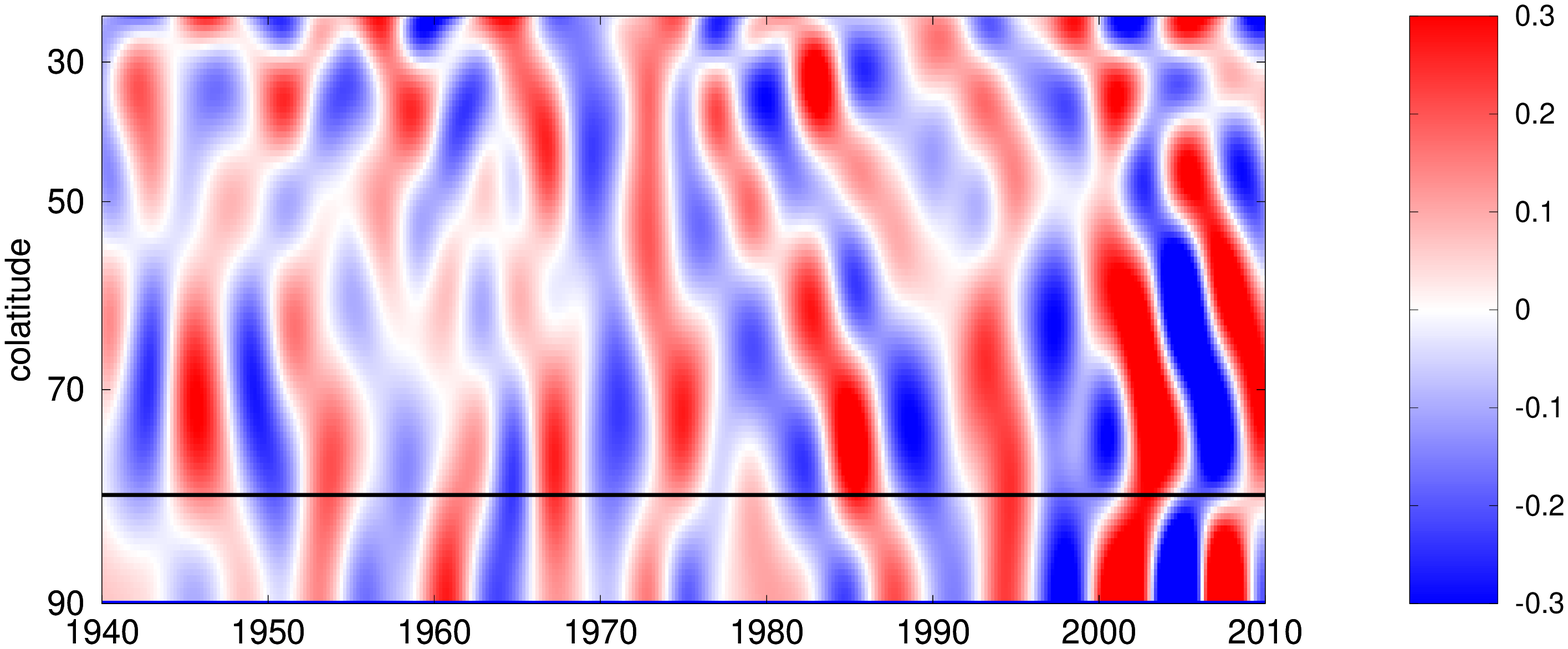}}
 \caption{Ensemble mean of the geostrophic flow (in km/yr), band-pass filtered between 4 and 9.5 years, as a function of time. The black line correspond to 10$^{\circ}$ latitude.
On the top pannel, the grey lines correspond to Alfv\'en velocities $C$ based on a rms cylindrical magnetic field of 1.9 and 0.6 mT in regions respectively close to the inner core and close to the equator.
Bottom pannel: $Y$ axis increments are proportional to the surface between $s$ and $s+\delta s$ ($dY\propto \sin\theta d\theta\Rightarrow Y\propto 1-\cos\theta$). 
}
 \label{fig: TW}
 \end{figure}

In figure \ref{fig: TW} we show a time-cylindrical radius map of the zonal velocity from 1940 to 2010, for the period band between 4 and 9.5 years where we find enhanced zonal energy in Figure \ref{fig: psd ratio}. 
The contribution of these flows to LOD changes is displayed  in the bottom panel of figure \ref{fig: LOD}, which shows the predictions from individual members of the flow ensemble and from the ensemble mean.
The contribution of external fluid envelopes (mainly the atmosphere and to a lesser extent the ocean) to angular momentum changes is known to dominate for periods up to about 3 years \citep{gross2004atmospheric}. 
However, the interannual variability of atmospheric angular momentum \citep{paek2012comparison} appears to be too small to account for a prominent signal observed with a period of around 6 years \citep{abarca00,Chao:2014mz}.  
The zonal core motions that we have isolated readily provide a explanation for this signal. 
We corroborate the results of \cite{gillet10}, which were limited to the time interval 1955--1975 (note that there is a delay compared to their figure 2$b$, due to the fact that they used a non-causal filter and omitted to shift the time-axis).
The time variability of  the observed and predicted LOD in the period range [4-9.5] years (see also \cite{Chao:2014mz}) is in apparent conflict with the finding of \cite{Holme:2013zh} who decomposed iteratively the LOD data (from 1962 to 2012) into a decadally varying signal and a 5.9 years oscillation of almost constant amplitude (compare their figure 2 with our figure \ref{fig: LOD}, bottom). 
In our opinion, the relatively small amplitude of the oscillation (in comparison with that of decadal changes) 
makes it difficult to decide whether it is long-standing or heavily damped.
Figure \ref{fig: LOD} (bottom) displays all the LOD changes produced by geostrophic flows in the frequency range [4-9.5] years.  
They need not all be attributed to the propagation of torsional waves.
In any case, the remarkable agreement between our predictions and the geodetic data encourages us in the interpretation of the flow model down to periods about 4 years.

Geostrophic motions appear very clear over 1995-2010, particularly as the torsional wave approaches the equator, at latitudes below 40$^{\circ}$, with a node of the waveform at about 10$^{\circ}$ latitude (see figure \ref{fig: TW}, bottom). 
The amplitude of the motions in this region is significantly larger than the spread in the flow ensemble (even at earlier epochs), yet the better resolution of the field model at recent epochs may have increased the sensitivity in the relatively small (in latitudinal extent) equatorial area. 
We confirm the slower propagation inferred by \cite{gillet10} as the wave gets closer to the equator, and find no evidence for reflection at the equator. 

Now, the theory of `magnetostrophic dynamos' \citep[see e.g.][]{Roberts:2015}, which has been developed to account for the Earth's magnetic field, gives us a tool to interpret the ratio $\displaystyle {\cal P}_Z/{\cal P}_{NZ}$ as a function of frequency.
We note above that this ratio remains small and does not vary much for periods larger than 8 years (see figure \ref{fig: psd ratio}).
\cite{taylor63} demonstrated that in the absence of inertia and viscosity we have
\begin{equation} 
\forall s,  \int_{\Sigma(s)} {\bf 1}_\phi\cdot \left( \left(\nabla \times {\bf B}\right) \times {\bf B} \right) d\Sigma =0\,,
\label{eq: taylor}
\end{equation}
with $\Sigma(s)$ the geostrophic cylinders (see \cite{Roberts:2012} or \cite{TOG8Jault13} for modern accounts of Taylor's theory).
Differentiating in time (\ref{eq: taylor}) and substituting ${\partial {\bf B}}/{\partial t}$ with its expression from the induction equation
\begin{equation} 
\displaystyle
\frac{\partial {\bf B}}{\partial t} = 
\nabla\times\left({\bf u}\times{\bf B}\right)+\eta\nabla^2{\bf B}\,,
\label{eq:induction}
\end{equation}
\cite{taylor63} obtained a linear relationship between the geostrophic zonal flow $u_G$ and non-geostrophic motions ${\bf u}_{NG}$, which depends on the magnetic field inside the core (see his equation 4.5, $\eta$ is the magnetic diffusivity),
\begin{equation} 
\displaystyle
 \frac{1}{s^3 H}\frac{\partial}{\partial s}\left(
s^3 H C^2 \frac{\partial}{\partial s}\left(\frac{u_G}{s}\right)\right)
 = -\int_{\Sigma(s)} \left ( {\cal F} \left({\bf B},  {\bf u}_{NG} \right) + {\cal G}_\eta \left({\bf B} \right) \right ) d\Sigma\, ,
\label{Taylor_init}
\end{equation}
with 
\begin{equation} 
\displaystyle C^2(s)={\frac{1}{4\pi sH\rho\mu}\displaystyle\int_{\Sigma(s)} B_s(s,\phi,z)^2 d\Sigma}\, .
\label{TW phase speed}
\end{equation}
The quantity $C(s)$ has the dimension of a velocity and is proportional to the rms value of the cylindrical radial field $B_s$ averaged over geostrophic cylinder
($\rho$ is the density of the outer core and $\mu$ the free space magnetic permeability). 
Both ${\cal F}$ and ${\cal G}_\eta$ are quadratic functions of the magnetic field, ${\cal F}$ depends linearly on the non-zonal velocity, and ${\cal G}_\eta$ stems from the diffusion term $\eta\nabla^2{\bf B}$.
In a quasi-geostrophy framework, we can identify ${\bf u}_{NG}$ with the non-zonal motions ${\bf u}_{NZ}$.
Equation (\ref{Taylor_init}) then determines a linear relationship between $u_G$ and ${\bf u}_{NZ}$ that may explain the uniform ratio between their energies at long periods.
There is perhaps no need to seek another mechanism for decadal variations of the geostrophic velocity and for decadal signals in the length of a day.

Restoring the inertial acceleration $\partial u_G / \partial t$ \citep[equ. 24b]{Roberts:2012},
Taylor's relationship is transformed into
\begin{equation} 
\displaystyle
\frac{\partial^2}{\partial t^2} \left(\frac{u_G}{s}\right)  - \frac{1}{s^3 H}\frac{\partial}{\partial s}\left(
s^3 H C^2 \frac{\partial}{\partial s}\left(\frac{u_G}{s}\right)\right)
 = \int_{\Sigma(s)} \left ( {\cal F} \left({\bf B},  {\bf u}_{NZ} \right) + {\cal G}_\eta \left({\bf B} \right) \right ) d\Sigma\,.
\label{Taylor ZNZ}
\end{equation}
The homogeneous part (on the left hand-side) corresponds to the torsional waves equation of \cite{braginski70}, where $C$ can now be interpreted as the torsional wave velocity.
Comparing (\ref{Taylor_init}) and (\ref{Taylor ZNZ}) makes clear that Taylor's differential equation (\ref{Taylor_init}) is valid on time-scales that are long compared to the period of the torsional waves.
The non-zonal velocities on the right-hand side of (\ref{Taylor ZNZ}) appear as a possible source term for the torsional waves provided they have the appropriate time-scale, as mentioned also by \cite{teed2014dynamics} who searched for torsional waves in numerical simulations of the geodynamo. 

Over the best resolved era (the last 15 years), 
the outward propagation of geostrophic motions, which can be interpreted as a torsional wave in most of the outer core, 
present a node 
10$^{\circ}$\, away from the equator where $u_G$ remains of small amplitude (figure \ref{fig: TW}, bottom).
The mechanism responsible for the special behavior close to the equator remains unclear.
Latitudes below 10$^{\circ}$ involve geostrophic motions only up to 50 km cylindrical depth from the equator. 
These geostrophic motions, which become important after 1995, carry a tiny fraction (of the order of 1\%) of the outer core angular momentum. 
We may lack resolution to detect them further back in time, and their existence at older epochs cannot be ruled out solely on the basis of the good fit to LOD variations from 1950 onwards.

\subsection{On electromagnetic core-mantle coupling}
\label{sec: EM coupling}

Putting aside the fast torsional waves governed by (\ref{Taylor ZNZ}), we read equation (\ref{Taylor_init}) as an indication that $u_G$ and ${\bf u}_{NZ}$ have similar time-scales.
This equation however does not directly constrain the solid body rotation part of the core flow, for which $\partial (u_G/s)/ \partial s =0$.
The time evolution of this rotation is instead governed by the coupling mechanisms acting at the outer core boundaries.

The torque $\Gamma^o(t)$ acting between the core and the mantle is linearly related to the time derivative of the observed LOD. 
In a series of papers, \cite{holme98I,holme98II} found that core surface flow models exist that explain geomagnetic data between 1900 and 1980, and can also account for decadal changes in $\Gamma^o(t)$ through electromagnetic coupling if the conductance of the mantle 
\begin{equation}
G = \int_{c}^{c+\delta} \kappa_m dr
\end{equation}
\noindent
is $10^8$ S or greater
($\delta$ is the thickness of the conducting layer; for the figures given below, the conductivity of the mantle, as a function of radius, is chosen as $\kappa_m(r)=3000(c/r)^{30}$ S/m). 
We test here whether this result holds for our ensemble of flow models, while accounting for their uncertainties. 
Assuming that these are correct, we can follow a direct approach rather than the inverse approach of \cite{holme98I}. 
We associate each member $\mathbf{u}^p$ of the ensemble of core flows with a time series $\Gamma^p(t)$ of the electromagnetic torque acting at the CMB
(see Appendix \ref{sec: app-em-torque}). 
We find that the mantle conductance $G$ has to be about $7\times 10^8$ S to make the typical amplitude of $\Gamma^p(t)$ match the torque values inferred from LOD changes at decadal periods (about $10^{18}$ N.m peak to peak), with a correlation coefficient of 0.36 between observations and the ensemble average predictions. It seems difficult to reconcile the time variation of our flow model with an electromagnetic explanation for the core-mantle coupling if the mantle conductance is less than $3\times 10^8$ S.
The difference with the results of \cite{holme98II} comes from the longer characteristic times of our large scale flows, which are responsible for most of the electromagnetic torque (when ignoring time correlations of the SV model errors, in case $D_{10}$, we find $G\sim4\times 10^8$ S for the magnitude of the predicted torque to match that of $\Gamma^o$).
We also calculate the torque at interannual periods. The correlation coefficient between the observed torque and the ensemble average predictions is 0.56. A large conductance $G\sim3\times10^9$ S is required to account for the amplitude of the observed torque, at these periods. 

The phase of the predicted series fits relatively well that of the geodetic series over both period ranges.
However, the conductances required to make electromagnetic coupling a viable mechanism at respectively 6 and 25 years periods are strongly at variance. 
This does not come as a surprise since the solid body rotation $t_{1}^0$ accounts for a significant portion of both the electromagnetic torque and the LOD changes. 
Assuming that fluctuations of  $t_{1}^0$ were linearly related to the variation of angular momentum at periods from 6 to 60 years, 
they could not also be linearly related also to the evolution of its time derivative (and hence of $\Gamma^o$) over the same period range.
The high value of $G$ that is required for electromagnetic core-mantle coupling at either period would certainly hinder the propagation of torsional waves across the core \citep{dumberry08,gillet10}.

\begin{figure}
\centerline{
 \noindent\includegraphics[width=30pc]{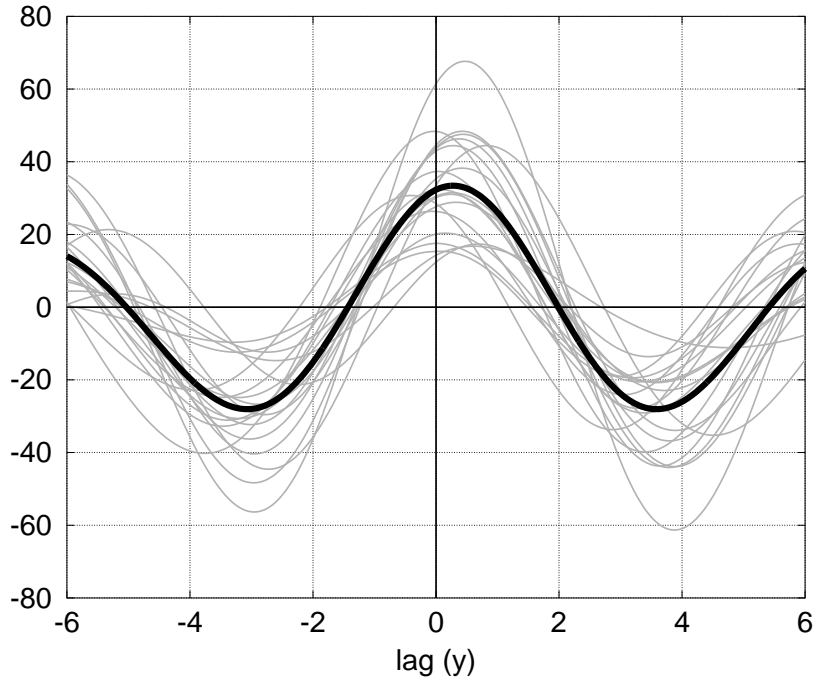}}
 \caption{Cross-correlation (non normalized) between the filtered LOD geodetic observations and predictions from the flow models, for all ensemble members (grey) and for their mean (black).
Positive delays correspond to observations ahead of predictions.}
 \label{fig: lag LOD}
 \end{figure}

Finally, we have also calculated the cross-correlation between observed and predicted LOD changes from our ensemble of core flows (figure \ref{fig: lag LOD}).
We find that the correlation is maximum for a delay between observed and predicted values of $\tau = 0.26 \pm 0.29$ year. 
The obtained value is the ensemble average of time lags, and the error is estimated as the standard deviation within the ensemble of time lags.
The positive value of the delay means observations are, in average, ahead of predictions. 
Since diffusion in the mantle may cause a negative delay, we estimate that the lag most probably lays between -0.3 and 0 year (within 2 standard deviations) ; 
its estimation may help constrain the electrical conductivity of the lowermost mantle.

\subsection{Dynamics of the equatorial region}
\label{sec: eq}

\begin{figure}
\centerline{
 \noindent\includegraphics[width=20pc]{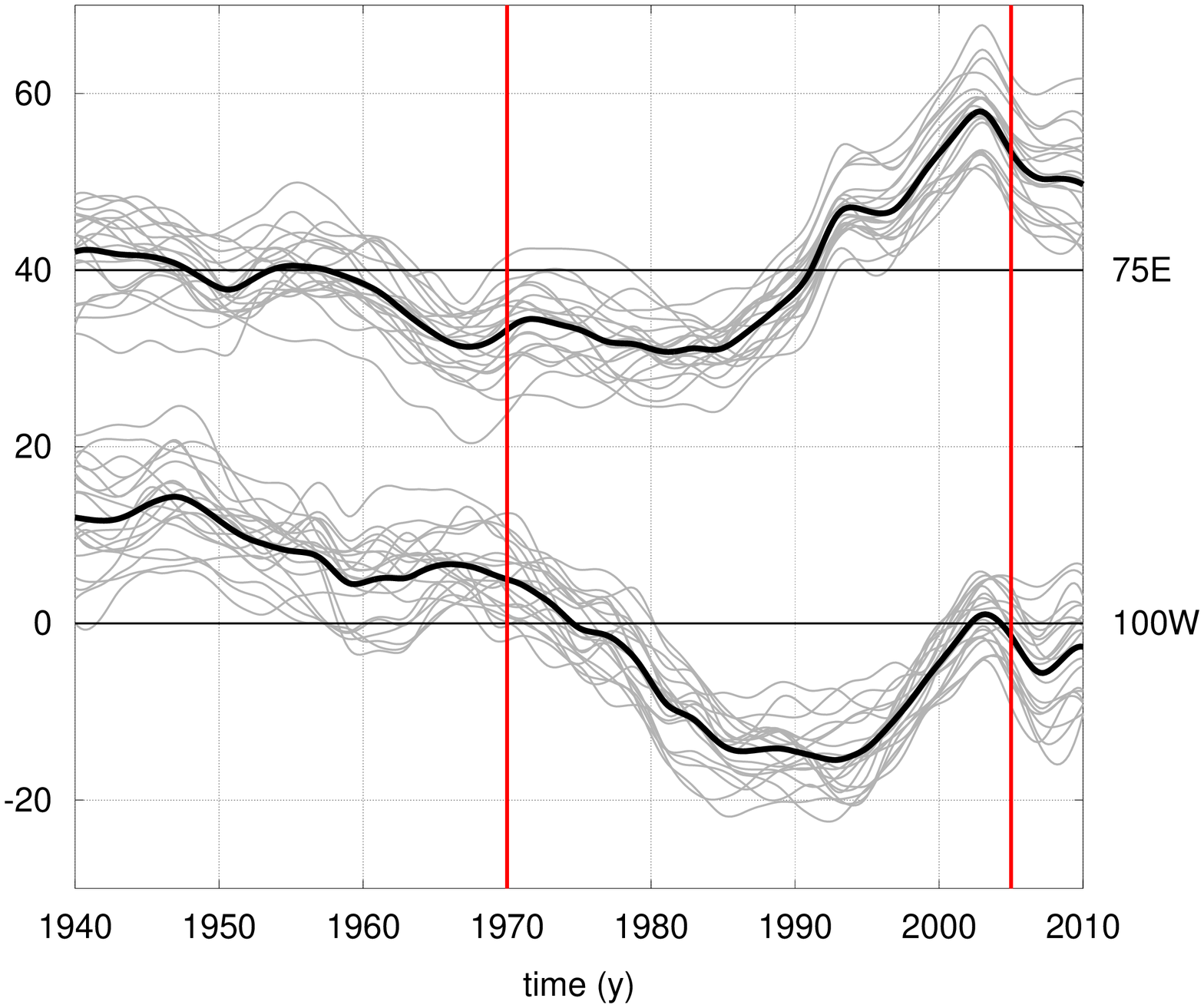}
 \noindent\includegraphics[width=20pc]{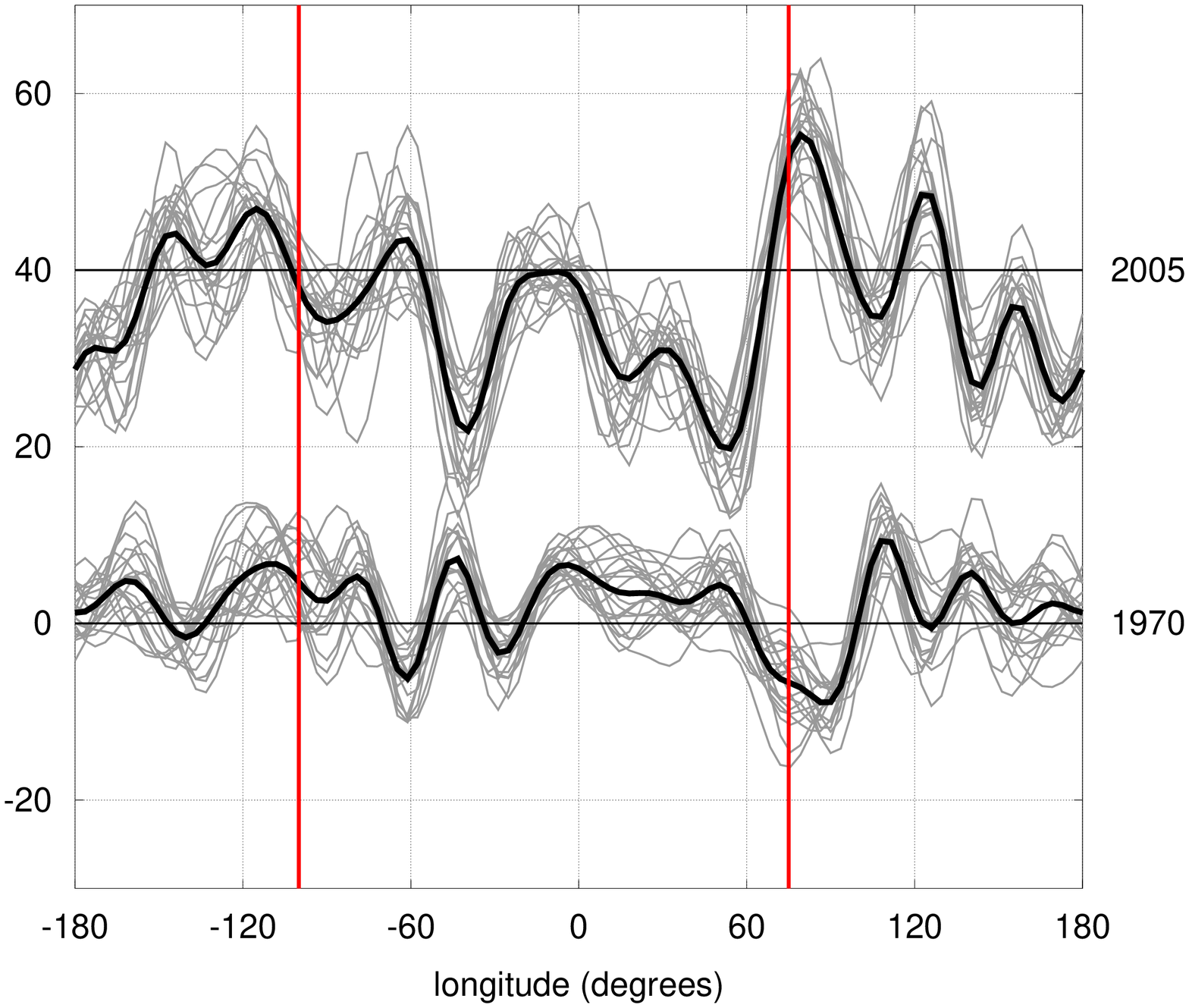}}
\centerline{
 \noindent\includegraphics[width=20pc]{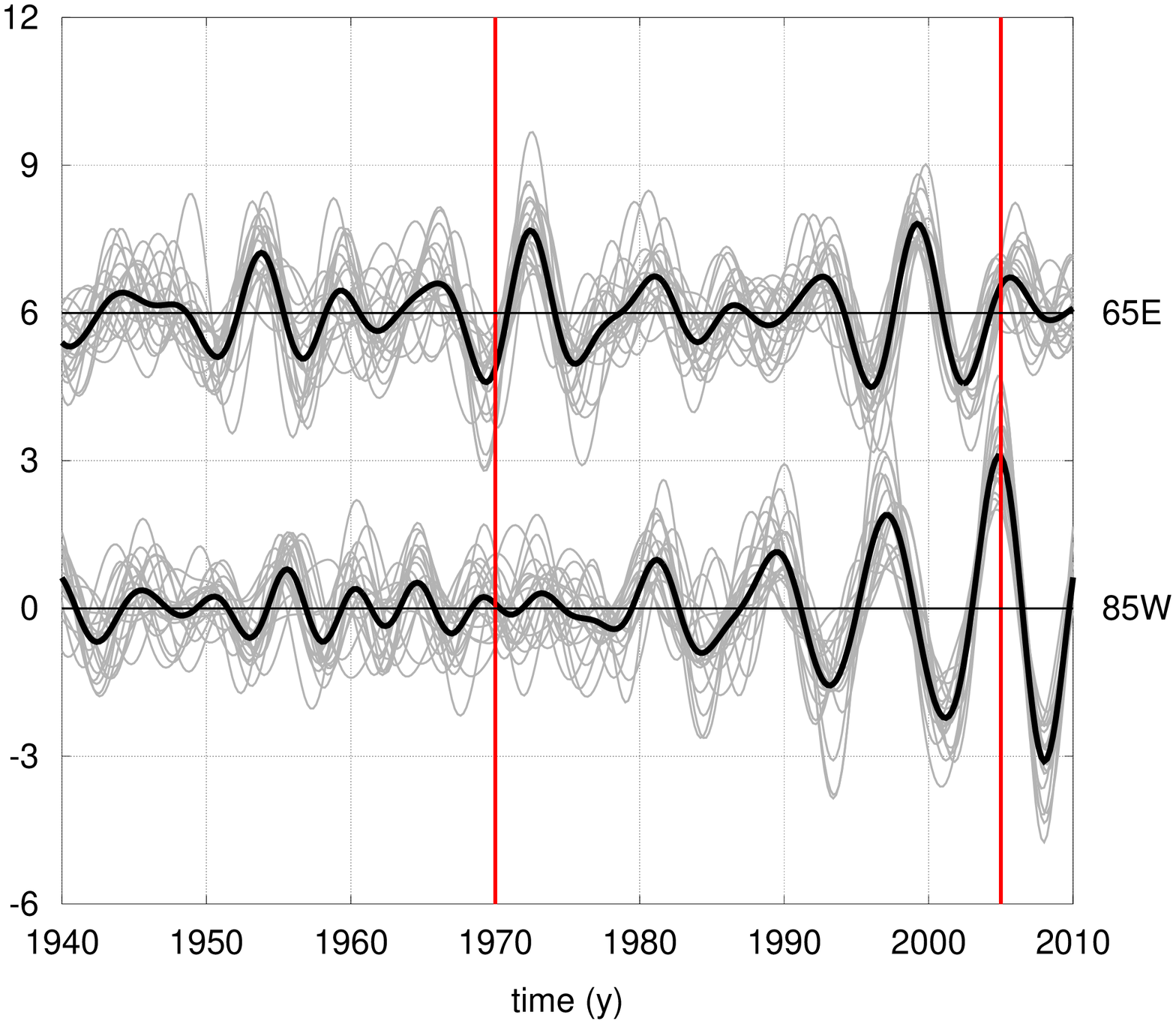}
 \noindent\includegraphics[width=20pc]{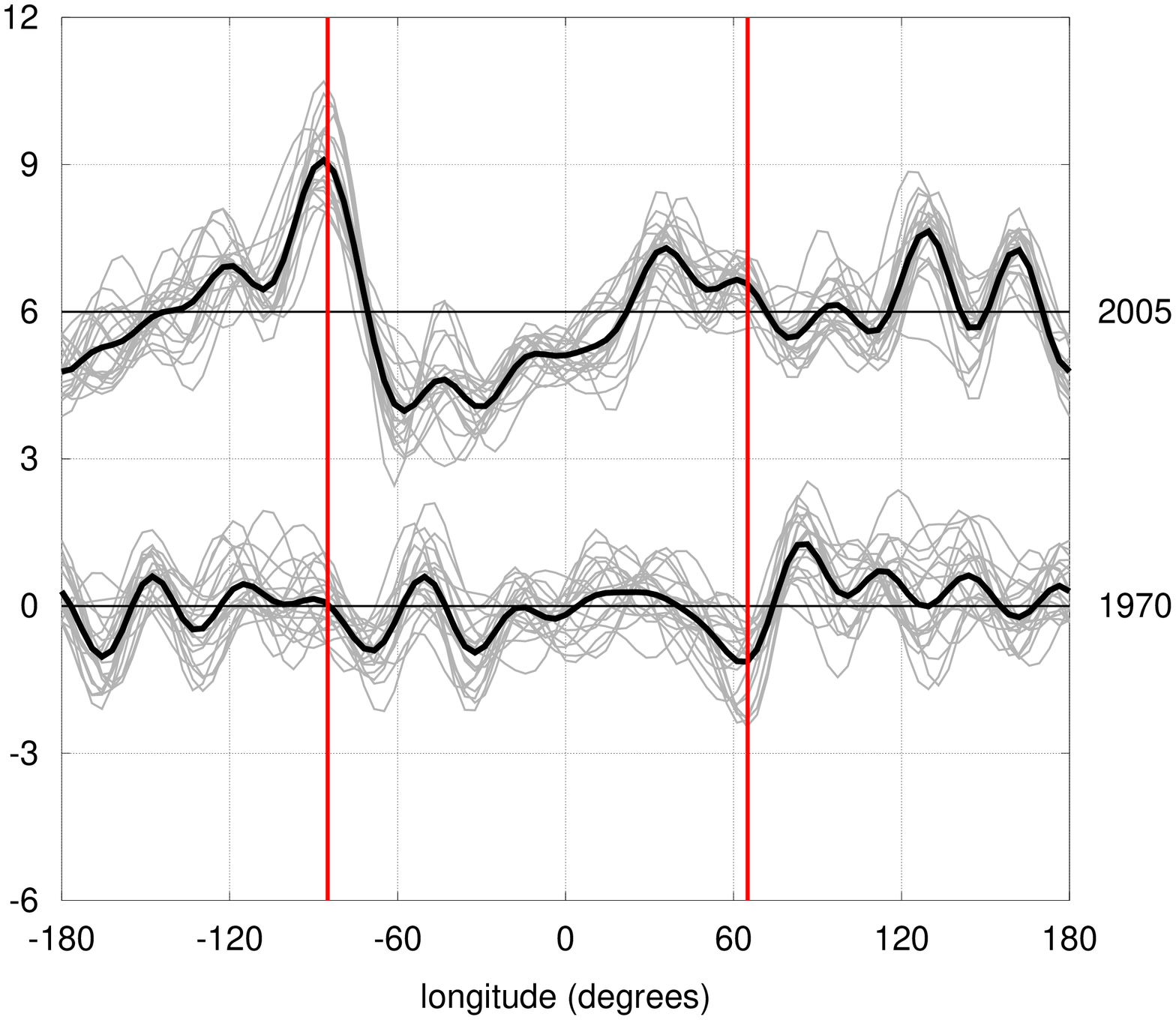}}
 \caption{Non-zonal azimuthal flow $u_{\phi}-u_G$ at the equator (truncated at spherical harmonic degree 14; in km/year): in grey the ensemble of realizations, in black the ensemble average. 
Left: time series at two different longitudes. 
Right: azimuthal profiles at two different epochs.
Top: flow anomaly with respect to the stationary flow. 
Bottom: flow anomaly band-pass filtered between 4 and 9.5 years. 
On each plot, the two ensembles of profiles are shifted with respect to one another. 
The red vertical lines on the left (resp. right) panels refer to the epochs (resp. the longitudes) presented on the right (resp. left) panels. 
}
 \label{fig: Uphi eq profiles}
 \end{figure}

We have observed in \S\ref{sec: geometry} and \S\ref{sec: TW} a remarkable morphology of the azimuthal velocity in the equatorial belt at various time-scales: the time averaged velocity, the decadal flow changes, as well as the interannual geostrophic motions all show a minimum in amplitude at about 10$^{\circ}$\, latitude (see figures \ref{fig: t-ave flow} and \ref{fig: TW}, bottom). 
Decadal, non-zonal (i.e. non-axisymmetric), azimuthal jets in the equatorial region do not seem directly related to $u_{\phi}$ at higher latitudes, even though they are linked to $u_{\theta}$ at latitudes around 10$^{\circ}$ through mass conservation (see for instance around longitudes 60$^{\circ}$W, 70$^{\circ}$E or 130$^{\circ}$E in Figure \ref{fig: t-ave flow}, bottom left). 
These low latitude features and their time evolution are consistently replicated within the flow ensemble (see figure \ref{fig: Uphi eq profiles}, top). 

\begin{figure}
\centerline{
 \noindent\includegraphics[width=30pc]{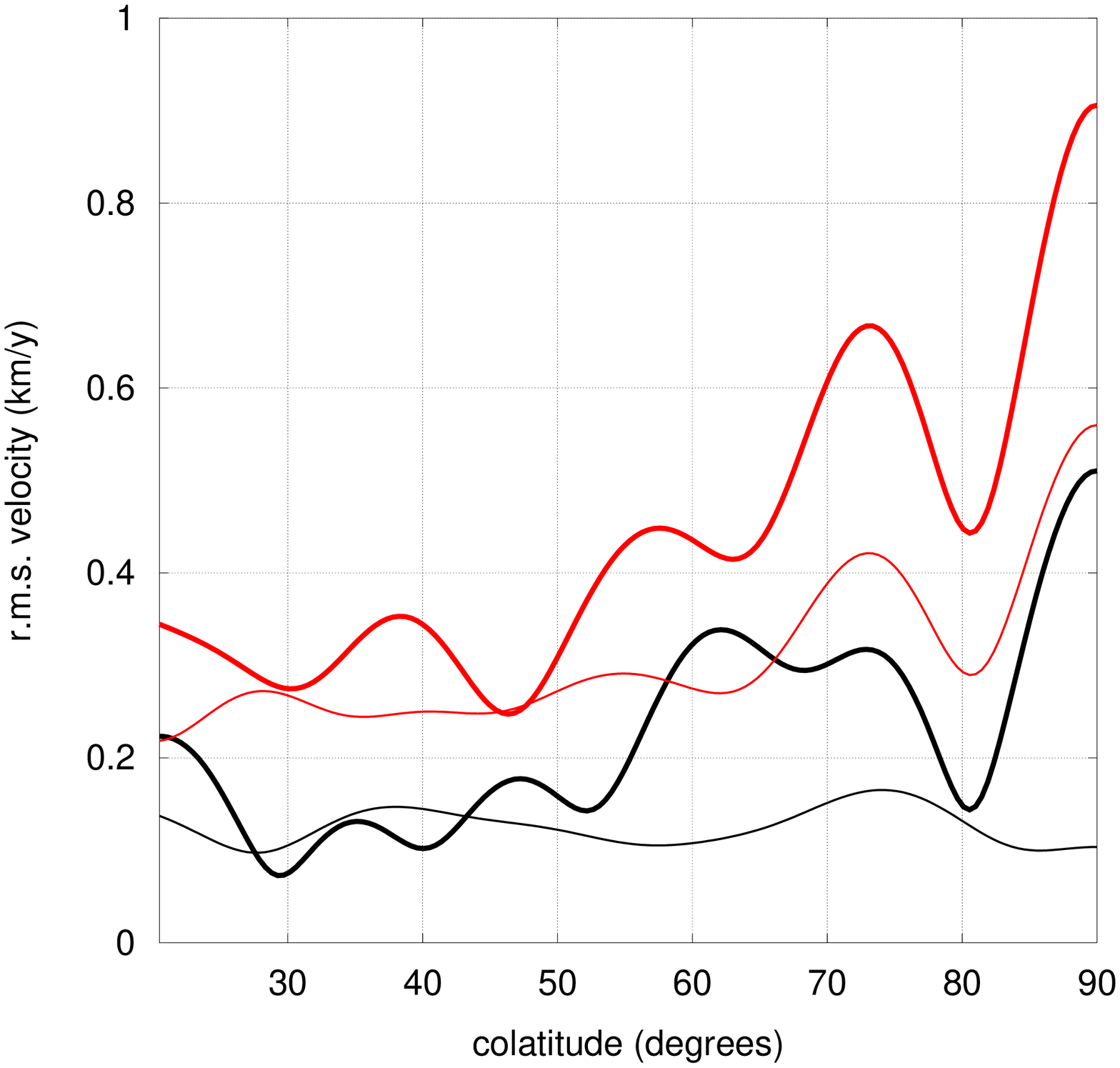}}
 \caption{As a function of colatitude, root-mean-square of the interannual (band-pass filtered between 4 and 9.5 years) azimuthal flow. 
For the zonal velocity (black): $\displaystyle\sqrt{\frac{1}{(t_b-t_a)}\int_{t_a}^{t_b} u_{G}(\theta,t)^2 dt}$.
For the non-zonal velocity (red): 
$\displaystyle\sqrt{\frac{1}{(t_b-t_a)}\int_{t_a}^{t_b}\frac{1}{2\pi}\oint (u_{\phi}(\theta,\phi,t)-u_{G}(\theta,t))^2 d\phi dt}$.
Two time intervals are considered: $[t_a,t_b]=1965-1975$ (thin lines) and 2000--2010 (thick lines).  
All profiles are for the ensemble average flow truncated at spherical harmonic degree 14.}
 \label{fig: std U 6y}
 \end{figure}

At interannual periods, non-zonal azimuthal motions have a minimum in amplitude at 10$^{\circ}$\, latitude, as illustrated with the latitudinal profiles of the rms of $u_{\phi}$ in figure \ref{fig: std U 6y}. 
Longitudinal profiles of $u_{\phi}$ at the equator present localized structures, with particularly intense flows around longitude 85$^{\circ}$ W at recent epochs (see figure \ref{fig: Uphi eq profiles}, bottom). 
Their peak-to-peak amplitude is much larger than that found for the geostrophic waves discussed in section \S\ref{sec: TW}.
Even though the largest of these jets below 10$^{\circ}$\, can be traced back only to the mid-1990's, be they either axisymmetric (zonal) or else non-zonal (see figures \ref{fig: TW} and \ref{fig: Uphi eq profiles} bottom), we cannot exclude that such active low-latitude dynamics was present at earlier times, on account of the improved resolution of magnetic field models constructed from satellite observations.
To our knowledge, the intriguing velocity minimum at 10$^{\circ}$ latitude has not been reported from geodynamo simulations. 
In our opinion, further theoretical and numerical studies of this particular region are called for.

\section{Discussion}
\label{sec: discussion}

\subsection{Core flow time changes}

We have presented an attempt to consistently reconstruct time-changes of core motions. 
These are primarily associated with disturbances of the westward eccentric gyre identified by \cite{pais08}. 
We obtain a weaker temporal variability compared to previous studies \citep[e.g.][]{amit2006time}.
Indeed, we find the gyre to be largely steady over 1940--2010. 

Our model does not perfectly account for the low frequency changes of the dipole term $g_1^0$ and of a few other low degree coefficients. 
This seems to be a result of the relatively large SV model errors, associated with our flow, affecting the slow variations of these coefficients. 
This indicates it may be necessary to model core motions as a perturbation centered on a non-zero background flow (the equivalent of the climatic mean for the oceans dynamics).
We must acknowledge, however, that there may be contributions to these slow, large-scale, field changes that are neglected in our quasi-geostrophic frozen-flux model.
For example, one consequence of quasi-geostrophy is that the longitudinal average of the meridional flow is zero.
A contrasting view has been offered by \cite{Buffett:2014qf} who calculated waves at the top of the core assuming that it is stably stratified. 
He found that the zonal flow is coupled to an axisymmetric meridional flow that causes the dipole magnetic field to fluctuate with a period of about 60 years. 
On the other hand, our quasi-geostrophic flows are able to reproduce much of the dipole decay observed in recent decades, when the impact of SV modeling errors is less pronounced, through non-zonal meridional flows acting on the longitudinally asymmetric core field.

\subsection{Limits on the time resolution of core flow models}

Information on the time variability of SV model errors amounts to a time-dependent constraint on core flow calculations.  
Our investigations show that consistently accounting for time-correlated SV model errors has very encouraging consequences regarding the rapid flow changes that may be inferred from satellite magnetic data.   
We found that the dispersion within our ensemble of flow solutions at sub-decadal periods was significantly reduced at recent epochs, indicating that a more detailed picture of the rapid core flow changes is emerging from the availability of continuous satellite data.

Our confidence in the estimated flow changes is supported by their ability to reproduce LOD variations at both decadal and interannual periods.  
Consistent modeling of interannual and decadal LOD changes was possible here for the first time thanks to our modeling of time variable SV model errors.
In previous core flow reconstructions investigating this issue \citep{wardinski04,gillet10}, there was no constraint on the flow changes between epochs  (i.e. $\tau_u\rightarrow 0$) and no time-correlated SV model error  was considered ($\rho_{loc}=\delta$).   As a result, decadal LOD changes were found to be significantly over-estimated.

We identify the approximately 6 years periodic LOD signal with torsional waves in the outer core. 
Our analysis is compatible with a modulation of their amplitude 
inside the time interval covered by our study, with particularly intense velocities at latitudes below 40$^{\circ}$\,  after 1995. 
These torsional waves may be excited by the interaction of non-zonal motions with the magnetic field inside the core.
Both the geostrophic flow and the non-zonal azimuthal velocity present a minimum in amplitude at 10$^{\circ}$\, latitude.  
The understanding of the interaction between zonal and non-zonal motions in the equatorial region requires further theoretical and numerical investigations. 

Torsional waves are not the main source of the intense and rapidly changing SV patterns recently identified near the equator from the analysis of satellite measurements during the past 15 years \citep{olsen07,chulliat2014geomagnetic}.
Rather, enhanced non-zonal flow at low latitudes, particularly around 90$^{\circ}$ W as well as 60$^{\circ}$ and 120$^{\circ}$ E, appear to be responsible.  

\subsection{Possible additions to the procedure}
\label{sec: extra}

Determining the posterior core flow probability density given the information contained in the SV data, as attempted here, not only provides an estimate of the uncertainty on the core motions, but also provides an opportunity to construct flow models subject to extra constraints. Indeed, this can be formalized as an optimization problem of the form
\begin{eqnarray} 
\left\{
\begin{array}{rcl}
\displaystyle
{\bf X} &= &\left<{\bf X}\right>+\delta{\bf X}\\
\displaystyle
0 & = & {\sf G}{\bf X}
\end{array}
\right.
\label{extra constraint}
\end{eqnarray}
with $E\left(\delta{\bf X}\delta{\bf X}^T\right)$ given by the posterior covariance matrix for the flow model errors.

A first instance of an extra constraint is $\displaystyle\frac{\partial(u_G/s)}{\partial s}=0$ at $s=c$, which is the boundary condition for torsional waves under the insulating mantle hypothesis \citep[see][]{schaeffer2012reflection,TOG8Jault13}. 
Similarly, one could produce flow solutions satisfying the non-penetration condition $\displaystyle u_s=0$ at the cylindrical surface tangent to the inner core. 

We have estimated the modeling error ${\bf E^m}$ from the unresolved magnetic field $\delta{\bf B}$.
Alternatively, one could also consider estimating directly ${\bf E^m}$, using `augmented state' data assimilation schemes \citep[see for instance][]{evensen2003ensemble}. 
In this framework, time covariances of SV model errors can be accounted for by advecting ${\bf E^m}$ with a stochastic equation that is consistent with their correlation functions, as presented in Figure \ref{fig: localization}. 

The stochastic approach employed throughout this study can be modified with the incorporation of deterministic elements involving either 3D geodynamo simulations \citep{aubert2014earth} or QG models \citep[such as that of][]{canet09}.
For instance the information about time covariances governed by a stochastic process could be combined with spatial covariances inferred from geodynamo simulations.
Another improvement may be to estimate simultaneously the velocity and magnetic fields \citep[as in][]{lesur2015geomagnetic} using sequential data assimilation.

The flow model presented here, together with its associated covariance matrix, is readily available from the address http://isterre.fr/recherche/equipes/geodynamo/themes-de-recherche/article/analyse-de-donnees-geomagnetiques?lang=fr.


%
%
%
%
%
%
%

\subsection*{Acknowledgments}

Computations presented in this paper were performed at the Service Commun de Calcul Intensif de l'Observatoire de Grenoble (SCCI).
We thank Vincent Lesur for his deep review of the methodological aspects of our manuscript, and an anonymous referee for his remarks on the analysis of torsional waves. 
We thank the International Space Science Institute for providing support for the workshops of international team no. 241. 
This work was partially funded by the French Centre National d'\'Etudes Spatiales (CNES) for the preparation and the exploitation of the Swarm mission of ESA. This work has also been supported by the French `Agence Nationale de la Recherche' under the grant ANR-2011-BS56-011.

\appendix

\section{Use of cross-covariances in problems with time-correlated errors: a tutorial example}
\label{sec: tutorial}

Let first focus on the case of a process sampled by two observations ${\sf y}_1$ and ${\sf y}_2$, associated with errors ${\sf z}_1$ and ${\sf z}_2$ (considered as random variables of variance $E\left({\sf z}_1^2\right) = E\left({\sf z}_2^2\right) = \sigma^2_{\sf z}$, with $E\left(\dots\right)$ the statistical expectation).
What are the consequences of ${\sf z}_1$ and ${\sf z}_2$ being correlated?
Noting $E\left({\sf z}_1{\sf z}_2\right)=r\sigma^2_{\sf z}$ ($r$ the correlation coefficient), the covariance matrix of the vector ${\bf z}=[{\sf z}_1,\, {\sf z}_2]^T$, 
\begin{eqnarray}
\displaystyle
{\sf C}_{\sf zz}= E\left({\bf z}{\bf z}^T\right)=\sigma_{\sf z}^2
\left[
\begin{array}{cc}
1  & r \\
r  & 1 
\end{array}
\right]\,,
\label{covar 2x2}
\end{eqnarray}
has two eigen vectors,
\begin{eqnarray}
\displaystyle
\left\{
\begin{array}{rl}
\displaystyle{\sf e}_a=  & ({\sf z}_2-{\sf z}_1)/\sqrt{2} \\
\displaystyle{\sf e}_b=  & ({\sf z}_2+{\sf z}_1)/\sqrt{2} 
\end{array}
\right.\,.
\end{eqnarray}
Their associated variances are 
\begin{eqnarray}
\displaystyle
\left\{
\begin{array}{l}
\displaystyle\sigma_a^2=E\left({\sf e}_a{\sf e}_a^T\right)=\sigma_{\sf z}^2(1-r) \\
\displaystyle\sigma_b^2=E\left({\sf e}_b{\sf e}_b^T\right)=\sigma_{\sf z}^2(1+r) 
\end{array}
\right.\,.
\end{eqnarray}
In the case of correlated noise ($r\neq 0$), we see that the process is sampled with variance of the error larger (resp. smaller) than $\sigma_{\sf z}^2$ in the direction ${\sf y}_2+{\sf y}_1$ (resp. ${\sf y}_2-{\sf y}_1$).
In other words, by allowing cross-covariances we decrease by a factor of $(1-r)$ the variance of the error on the difference ${\sf y}_2-{\sf y}_1$, and increase by a factor of $(1+r)$ the variance of the error on the average $({\sf y}_2+{\sf y}_1)/2$. 

We now further illustrate this idea with a one-dimensional toy-model.
Consider a true series $\varphi^t(t)$, polluted by a noise $\zeta(t)$, generating an observed series $\varphi^o=\varphi^t+\zeta$. 
All series are sampled at discrete epochs entering the vector ${\bf t}$, producing the vectors ${\bf y}^o$, ${\bf y}^t$ and  ${\bf z}$ for respectively the observed series, the true series and the noise.
We calculate a regression (or fit, or analysis) ${\bf y}^a$, sampled at the same epochs ${\bf t}$ by considering the information about the statistics of ${\bf y}$ and ${\bf z}$ contained in the covariance matrices $\displaystyle {\sf C}_{\sf zz}=E\left({\bf z}{\bf z}^T\right)$ of the noise and $\displaystyle {\sf C}_{\sf yy}=E\left({\bf y}^t{{\bf y}^t}^T\right)$ of the sampled series. 
This can be set up as an inverse problem, where  ${\bf y}^a$ is obtained as the best linear unbiased estimate, or BLUE \citep[e.g.,][]{rasmussen06}
\begin{eqnarray}
{\bf y}^a={\sf C}_{\sf yy}\left[{\sf C}_{\sf yy}+{\sf C}_{\sf zz}\right]^{-1}{\bf y}^o\,.
\label{eq:BLUE}
\end{eqnarray}

\begin{figure}
\centerline{
\noindent\includegraphics[width=30pc]{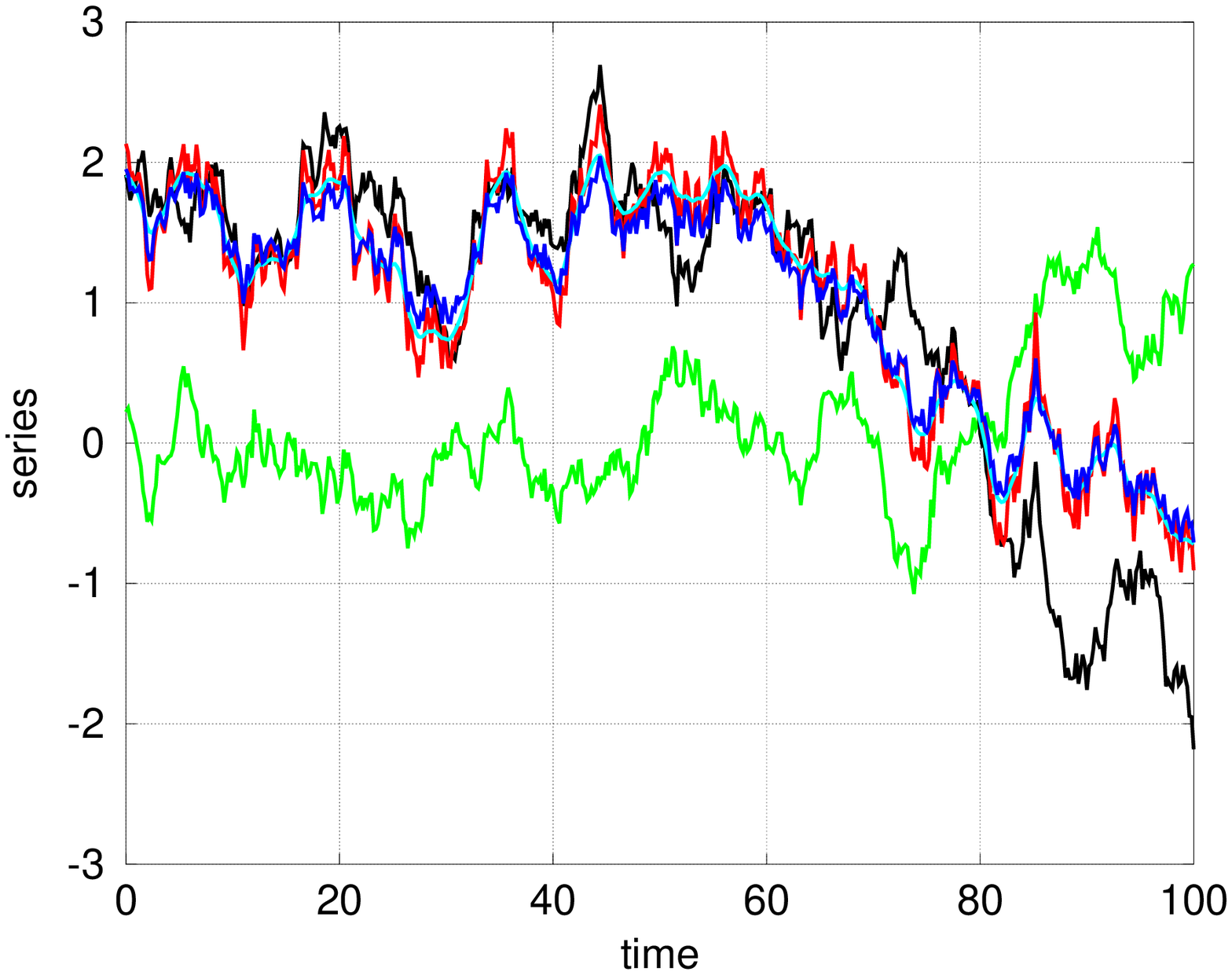}}
\centerline{
\noindent\includegraphics[width=30pc]{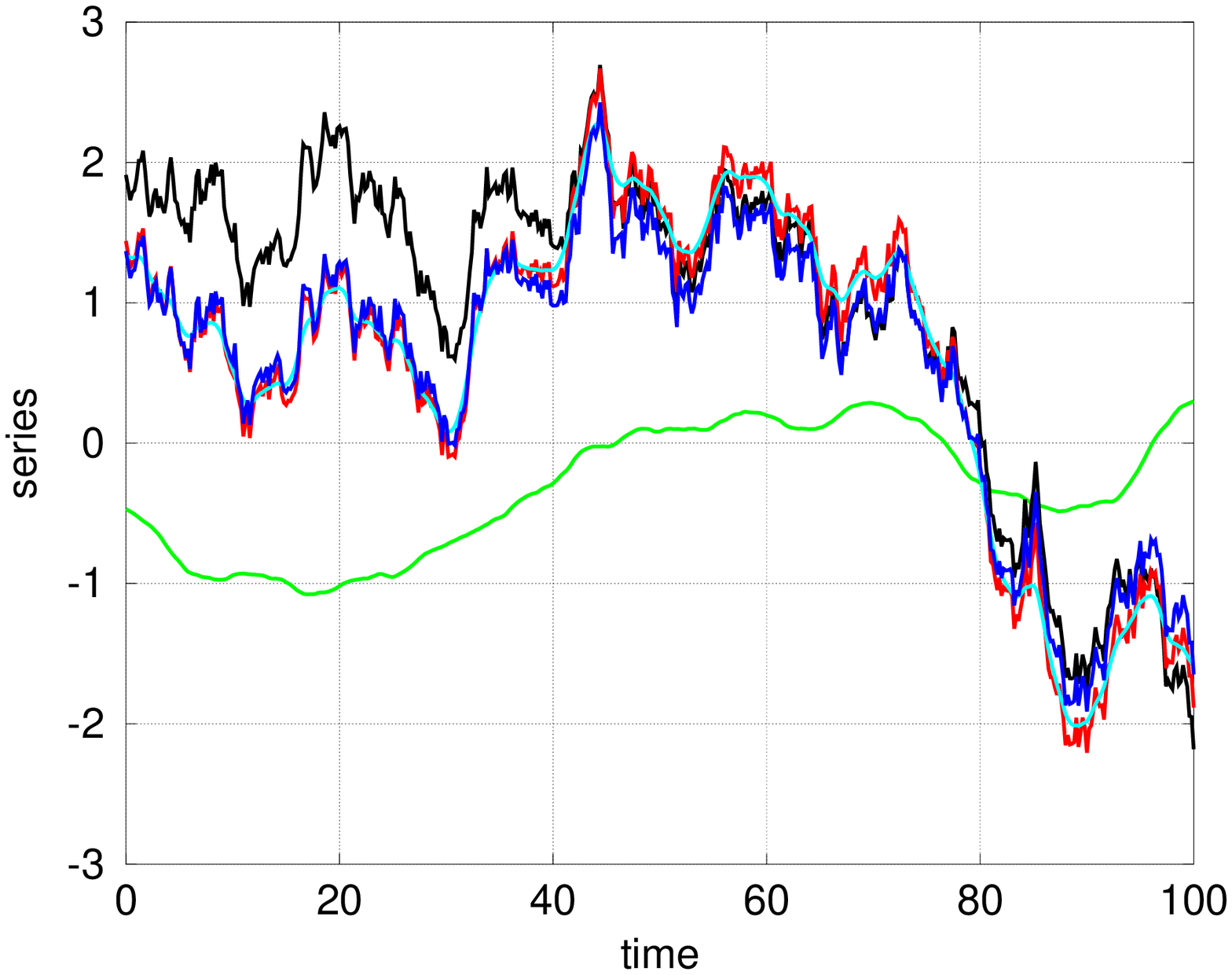}}
 \caption{Tutorial examples in cases $A$ (top) and $B$ (bottom):
true series (black), noise (green), data (red), BLUE with correlated errors (blue), and BLUE with uncorrelated errors (cyan) -- see text of Appendix \ref{sec: tutorial} for details.}
\label{fig: tutorial}
 \end{figure}

We present in Figure \ref{fig: tutorial} below two examples where ${\bf y}^t$ results from a process defined by equation (\ref{eq:stoch-AR1}), with the correlation time $\tau_{\sf u}$ replaced by $\tau_{\sf y}=100$ (dimensionless units).
In case $A$, the noise ${\bf z}$ also results from a process defined by equation (\ref{eq:stoch-AR1}), with $\tau_{\sf u}$ replaced by $\tau_{\sf z}=10$. 
In case $B$, the noise ${\bf z}$ results instead from a process defined by equation (\ref{eq:stoch-AR2}), with $\tau_{0}$ replaced by $\tau_{\sf z}=10$.
We set the variances $\sigma^2_{\sf y}=4$ and $\sigma^2_{\sf z}=0.25$.  
Both the true series and the noise present similar differentiability properties in case $A$, whereas the noise is smoother than the true series in case $B$.
In both cases we obtain two estimates of ${\bf y}^a$ using equation (\ref{eq:BLUE}): one with the correct matrix ${\sf C}_{\sf zz}$, and another one where we forgo the time correlation of the noise (${\bf z}$ is treated as a white noise, i.e. ${\sf C}_{\sf zz}$ is diagonal).

The results from several analyses are presented in Figure \ref{fig: tutorial}. 
We see that omitting cross-covariances in the statistics of the noise leads to an analyzed estimate ${\bf y}^a$ much smoother than the true series, but biased at some epochs (as a result of the time-correlated noise).  
On the contrary, when considering cross-covariances of the time-correlated noise we are able to retrieve part of the high frequency content of the true series; furthermore, the analysis is biased towards zero in comparison with the noisy observations (but not always when compared with the true series).
We finally see that the high frequency content is better retrieved in case $B$ than in case $A$. 
We attribute this to the fact that both the noise and the series display sharp time changes in case $A$, while in case $B$ the noise is assumed to be smoother.

\section{Calculation of the stream function $\psi$}
\label{sec: app-psi}

Within the incompressible QG approximation (see section \ref{sec:def-QG} and equation (\ref{ue psi})), the horizontal flow at the CMB is related to the stream function $\psi(\theta,\phi)$ through 
\begin{equation}
{\bf u}_h(\theta,\phi)=\frac{-1}{c^2\cos^2\theta}{\bf r}\times\nabla_h\psi\,.
\label{uh psi}
\end{equation}
From (\ref{uh psi}) the expression for the horizontal divergence gives 
\begin{equation}
\nabla_h\cdot{\bf u}_h
= -\sum_{\ell} \sum_{m=-\ell}^{\ell} \frac{\ell(\ell+1)}{c} s_{\ell m} Y_{\ell m}
=\frac{2}{c^3\cos^3\theta}\frac{\partial\psi}{\partial\phi}
\,,
\label{divh psilm}
\end{equation}
while the zonal velocity is 
\begin{equation}
u_{G}
=\sum_{\ell}t_{\ell 0}\frac{dP_{\ell 0}}{d\theta}
= -\frac{1}{c^2\cos^2\theta}\frac{d\psi_{G}}{d\theta}\,.
\label{uzon psizon}
\end{equation}
Writing 
\begin{equation}
\psi(\theta,\phi)=\sum_{\ell,m}\psi_{\ell m}Y_{\ell m}(\theta,\phi)\,,
\end{equation}
we use (\ref{divh psilm}) and relate the non-zonal coefficients $\psi_{\ell m}$ to the poloidal coefficients $s_{\ell m}$,
\begin{equation}
\forall m\neq 0, \; \; \; \sum_\ell\psi_{\ell m} Y_{\ell m}
= c^2 \cos^3\theta \sum_{\ell} \frac{\ell(\ell+1)}{2m} \mathrm{i} s_{\ell m} Y_{\ell m}.
\end{equation}
From (\ref{uzon psizon}) and the recurrence rules for Legendre polynomials, we obtain the following relation between the zonal coefficients $\psi_{\ell 0}$ and the toroidal coefficients $t_{\ell 0}$:
\begin{equation}
\psi_{\ell 0}= c^2\left(
 \frac{(\ell-1)(\ell-2)}{(2\ell-1)(2\ell-3)} t_{\ell-2,0} +
 \frac{2\ell^2+2\ell-3}{(2\ell-1)(2\ell+3)} t_{\ell 0} +
 \frac{(\ell+3)(\ell+2)}{(2\ell+5)(2\ell+3)} t_{\ell+2,0} 
\right)\,.
\end{equation}

\section{Estimation of the electromagnetic torque}
\label{sec: app-em-torque}

Following the theory of \cite{Roberts:1972uq}, the calculation relies on the linearization of both the induction equation and the expression for the axial magnetic torque acting on the solid mantle:
\begin{equation}
\Gamma= \mathbf{1}_z \cdot \int \mathbf{r} \times \left ( \mathbf{j} \times \mathbf{B} \right ) dV ,
\end{equation}
where the magnetic field $\mathbf{B}$ is obtained by downward extrapolation of the field known at the Earth's surface without considering any induction effect in the mantle. The electrical currents $\mathbf{j}$ are obtained from the surface electric field $\mathbf{E}_H$ at the CMB and hence from the field $\mathbf{u}B_r$ by continuity of $\mathbf{E}_H$ across the CMB.
Consistently with the equation (\ref{eq:ind_rad}) for the poloidal field, we neglect here the diffusion of toroidal magnetic field from the core.

%
%
%
%
%
%
%
%
%

\bibliographystyle{agufull08}

\end{document}